\documentclass[%
 reprint,
 amsmath,amssymb,
 aps,prx
]{revtex4-2}

\usepackage{mathdots}
\usepackage{graphicx}
\usepackage{dcolumn}
\usepackage{bm}
\usepackage{simplewick}
\usepackage[dvipsnames]{xcolor}
\usepackage{braket,cancel}
\definecolor{lgray}{gray}{0.4}
\definecolor{cor}{rgb}{0.94, 0.41, 0.35}
\definecolor{mag}{rgb}{0.7451, 0.2039, 0.3333}
\usepackage[colorlinks=true,urlcolor=mag,linkcolor=mag,
citecolor=mag,pdfpagelabels=true,hypertexnames=true,
plainpages=false,naturalnames=false]{hyperref}
\newcommand{\x}{{\boldsymbol x}}
\newcommand{\y}{{\boldsymbol y}}

\def\k{{\boldsymbol k}}
\newcommand{\q}{{\boldsymbol q}}
\newcommand{\p}{{\boldsymbol p}}

\def\bea{\begin{eqnarray}}
\def\eea{\end{eqnarray}}
\def\be{\begin{equation}}
\def\ee{\end{equation}}
\def\ba{\begin{array}}
\def\ea{\end{array}}
\def\nn{\nonumber}

\newcommand{\de}{{\rm d}}

\usepackage{tikz}
\usetikzlibrary{calc}
\usetikzlibrary{arrows,patterns}
\usetikzlibrary{shapes.misc}
\tikzset{cross/.style={cross out, draw=black, minimum size=2*(#1-\pgflinewidth), inner sep=0pt, outer sep=0pt},
cross/.default={1pt}}

\begin{document}

\title{Confronting infrared divergences in de Sitter: loops, logarithms\\ 
and the stochastic formalism}

\author{Gonzalo A. Palma$^a$}

\author{Spyros Sypsas$^{b,c}$}

\author{Danilo Tapia$^a$}

\affiliation{
$^a${\it Grupo de Cosmolog\'ia, Departamento de F\'isica, FCFM, Universidad de Chile, Blanco Encalada 2008, Santiago, Chile}
\\
$^b${\it Centro de Ciencias Exactas, Facultad de Ciencias, Universidad del B\'io-B\'io, Chill\'an, Chile}
\\
$^c${\it High Energy Physics Research Unit, Faculty of Science, Chulalongkorn University, Bangkok 10330, Thailand}
}

\begin{abstract}

A well-established result in quantum field theory in four-dimensional de Sitter space is that the vacuum state of a massless scalar field breaks the de Sitter isometry group, leading to time-dependent (secular) growth in correlation functions computed in inflationary coordinates. This behavior is widely believed to extend to more general theories involving light scalar fields with weak non-derivative interactions. In such cases, secular growth is thought to be further amplified by loop corrections, and the stochastic formalism is often regarded as the appropriate framework to resum these infrared effects. In this article we challenge this prevailing view. A crucial distinction must be made between two cases: a massless scalar field protected by a shift symmetry, and a light scalar without such a symmetry. In the former, the shift symmetry enforces derivative interactions, yielding observables in which secular growth plays no physical role. In the latter, although correlation functions develop infrared divergences in the massless limit, they remain fully invariant under the de Sitter isometry group. We analyze the structure of these divergences arising from loop integrals and show that, in the soft-momentum limit, they do not alter the time dependence of tree-level correlators. In fact, using a de Sitter-invariant renormalization scheme based on Wilson’s axioms for integration, these divergences can be systematically removed order by order. We therefore conclude that neither massless nor light scalar fields in de Sitter space exhibit genuine secular growth. We further discuss the implications of these findings for the validity and scope of the stochastic approach to inflation.

\end{abstract}

                          
\maketitle


\section{Introduction}

Quantum field theory (QFT) in de Sitter (dS) space is central to cosmology. Yet, the theoretical computation of observables, for instance, $n$-point correlation functions, comes with formidable technical challenges~\cite{Tsamis:1993ub,Brunier:2004sb,Onemli:2002hr,Miao:2005am,Sloth:2006az,Sloth:2006nu,Riotto:2008mv,Riotto:2011sf,Seery:2007we,Bartolo:2007ti,Rajaraman:2010zx,Xue:2011hm,Akhmedov:2013xka,Akhmedov:2017ooy,Akhmedov:2019cfd,Akhmedov:2024npw,Baumgart:2019clc,Baumgart:2020oby,Sasaki:1992ux,Bilandzic:2007nb,Weinberg:2005vy,Weinberg:2006ac,Senatore:2009cf,Senatore:2012nq,Afshordi:2000nr,Braglia:2025cee,Braglia:2025qrb,Kong:2024lac,Mirbabayi:2019qtx,Negro:2024iwy,Kamenshchik:2024ybm,Kamenshchik:2025ses}. Accounting for the effects of multi-particle states and loop corrections on observables requires solving difficult integrals and regularizing complicated divergent contributions. These difficulties, in particular, have obscured the handling of infrared (IR) divergences arising from loops; see~\cite{Seery:2010kh,Akhmedov:2013vka,Hu:2018nxy} for reviews.

Key to the debate surrounding IR divergences is the following result~\cite{Allen:1985ux}: It is impossible to define a normalizable inner product for massless scalar fields while keeping the full de Sitter symmetries intact. This finding has led to the conclusion that the presence of a massless scalar in a fixed dS background inevitably breaks the dS isometries to the 3D Euclidean group of translations and rotations. This symmetry breaking results from the IR behavior of the scalar field which manifests itself as time-dependent secular growth in correlation functions~\cite{Linde:1982uu,Vilenkin:1983xp,Starobinsky:1982ee,Vilenkin:1982wt}.  

This state of affairs has laid support to methods for handling IR divergences within the context of more general theories involving light scalar fields with weak non-derivative interactions, including approaches such as stochastic inflation~\cite{Starobinsky:1986fx,Starobinsky:1994bd,Gorbenko:2019rza}, which offers a framework for performing IR resummation~\cite{Tsamis:2005hd,Finelli:2008zg,Cespedes:2023aal,Kitamoto:2018dek,Honda:2023unh,Serreau:2013psa}. In particular, the abandonment of dS invariance has led to the widespread use of IR comoving cutoffs to regularize the contributions of long-wavelength modes to correlation functions in perturbation theory. To summarize, three interrelated statements play a crucial role in discussing the IR behavior of QFTs in dS: (a) Secular growth in correlation functions of light scalars arises from the breaking of dS symmetry; (b) The regularization of loop integrals requires comoving IR cutoffs explicitly breaking dS isometries; (c) The stochastic formalism provides a method for resumming secular growth originating from loops. 

The aim of the present article is to challenge this perspective on all three fronts. We do so by examining the impact of loop corrections on the infrared behavior of equal-time $n$-point correlation functions computed in momentum space, for a scalar field $\varphi(x)$ with non-derivative interactions governed by the action
\be \label{intro:action-basic}
S = - \int d^4 x \sqrt{-g} \left[ \frac{1}{2} (\partial \varphi)^2 + \mathcal{V}(\varphi) \right],
\ee
where $g_{\mu \nu}$ describes a fixed de Sitter background. Since our motivation is rooted in phenomenology relevant to cosmic inflation, we adopt a foliation of de Sitter using spatially flat cosmological coordinates, in which the line element takes the form
\be \label{intro:metric}
ds^2 = a^2(\tau)\left(-d\tau^2 + d\x^2\right),
\ee
with $a(\tau) = -1/ H \tau$ denoting the scale factor, $\tau \in (-\infty, 0)$ conformal time, and $H$ the Hubble expansion rate setting the de Sitter radius $R_H=H^{-1}$. Our focus is on fairly generic potentials $\mathcal{V}(\varphi)$ that satisfy $\mathcal{V}(\varphi) \ll H^4$ for the entire field range of interest, and hence only mildly break the shift symmetry $\varphi \to \varphi + c$ manifest in the free theory. 

In this setup, the Fourier modes of the scalar field $\varphi$ evolve analogously to those of inflationary scalar fluctuations in a quasi-de Sitter background. Due to the rapid expansion of space, the wavelengths of these fluctuations are stretched from subhorizon to superhorizon scales. At leading order—i.e., neglecting the potential $\mathcal{V}(\varphi)$—the field behaves as massless and its Fourier modes freeze after crossing the Hubble horizon. However, as the wavelength of a mode continues to grow, the non-linearities introduced by $\mathcal{V}(\varphi)$ begin to play an increasingly important role, modifying the statistical properties of $\varphi$ on length-scales much larger than $R_H$.

\subsection{de Sitter invariance of observables}

The scalar field theory~\eqref{intro:action-basic}, together with the coordinate system~\eqref{intro:metric}, provide a concrete framework for analyzing the behavior of correlation functions for both massless and light scalar fields. Reference~\cite{Allen:1985ux} was the first to observe that it is impossible to define a normalizable inner product for a massless scalar in de Sitter space (i.e., $\mathcal{V}(\varphi) = 0$) unless the isometries are broken. A direct consequence is that the two-point function in configuration space, $\langle \varphi(x) \varphi(x') \rangle$, for a free massless scalar field $\varphi(x)$, includes contributions that break de Sitter boosts. Specifically, it contains a de Sitter-invariant piece as well as a term proportional to $\ln a(\tau) + \ln a(\tau')$.

However, a truly massless field can only be realized if a shift symmetry is present to protect a flat direction in field space. In the absence of such a symmetry, a mass term is generated perturbatively, even if ${\cal V}''(0) = 0$ at tree level. Therefore, the action governing a genuinely massless field must be invariant under constant shifts, $\varphi \to \varphi + c$. This invariance, unlike in~\eqref{intro:action-basic}, restricts the allowed interactions to derivative couplings involving gradients of $\varphi$. As a result, physical observables must also respect the shift symmetry. In particular, the relevant two-point function, appearing in the constructions of observables, must be the de Sitter-invariant quantity
\be
G_{\mu \nu'} (x,x') \equiv \partial_{\mu} \partial_{\nu'} \langle \varphi(x) \varphi(x') \rangle.
\ee
These conditions ensure that the secular growth observed in $\langle \varphi(x) \varphi(x') \rangle$ does not propagate into physical observables. Consequently, despite the absence of a normalizable inner product, observables in a massless theory remain invariant under the full de Sitter group~\cite{Creminelli:2011mw} and do not exhibit secular growth~\cite{Tolley:2001gg}.

Theories with non-derivative interactions that break the shift symmetry---such as the one given in~\eqref{intro:action-basic}---are more subtle. In such cases, interactions inevitably introduce a constant physical infrared length scale, $2\pi \Lambda_{\rm IR}^{-1} \gg R_H$, beyond which the theory becomes strongly nonlinear. $2\pi \Lambda_{\rm IR}^{-1}$ must be a physical length (as opposed to a comoving length that stretches along with the expanding space); otherwise, a Fourier mode with comoving momentum $k \ll \Lambda_{\rm IR}$ would never enter the nonlinear regime where the interactions from $\mathcal{V}(\varphi)$ dominate. The scale $\Lambda_{\rm IR}$ thus effectively acts as the infrared cutoff of the theory in physical-momentum space. As a result, it is $\Lambda_{\rm IR}$---rather than the Hubble scale $H$---that provides the appropriate criterion for distinguishing between light and massive scalar field theories. For superhorizon physical momenta $p \ll H$, a massive field with mass parameter $m \ll H$ is described by a mode function whose amplitude decays as $\sim (p/H)^{\Delta}$, where $\Delta = m^2 / 3 H^2$ is the conformal weight of the mode function. To assess how efficiently a mode decays in the interval $\Lambda_{\rm IR} \ll p \ll H$, it is useful to introduce the scale
\be \label{mir}
m_{\rm IR}^2 \equiv \frac{3 H^2}{ 2 \ln(H / \Lambda_{\rm IR}) },
\ee
such that a scalar field is considered massive (i.e., its decay is efficient) if the mass parameter in the quadratic Lagrangian satisfies $m \gg m_{\rm IR}$. In this regime, perturbation theory remains well controlled: mode functions decay rapidly for physical momenta in the range $\Lambda_{\rm IR} \ll p \ll m$, ensuring that correlation functions are free of infrared divergences at both tree and loop levels.

Conversely, theories with $m \ll m_{\rm IR}$ describe effectively massless fields. In this case, mode functions do not decay in the interval $\Lambda_{\rm IR} \ll p \ll H$ because they are largely insensitive to the mass parameter, while for $p \sim \Lambda_{\rm IR}$, nonlinear interactions dominate, rendering the quadratic mass term operationally irrelevant. Nevertheless, a fiducial mass of order $m_{\rm IR}$ can still be introduced as a regulator in perturbative computations~\cite{Arai:2011dd}, capturing the effects of $\Lambda_{\rm IR}$ and thus the breaking of the shift symmetry. For external momenta $p \gg \Lambda_{\rm IR}$, a mass regulator $m \simeq m_{\rm IR}$ effectively controls IR divergences. In this regime, observable correlation functions remain de Sitter invariant, but exhibit divergences that scale as $H^2 / m_{\rm IR}^2$~\cite{Allen:1987tz,Huenupi:2024ksc}, or equivalently---using Eq.~\eqref{mir}---as $\ln(H / \Lambda_{\rm IR})$~\cite{Huenupi:2024ksc}.

All in all, observables must remain de Sitter invariant regardless of the value of the mass parameter $m$, both in the exactly massless case $m = 0$ and in the formal massless limit $m = m_{\rm IR} \to 0$, where $\mathcal{V}(\varphi) \neq 0$ is allowed. The crucial question, then, is whether IR divergences---which scale as $H^2 / m_{\rm IR}^2 \sim \ln(H/\Lambda_{\rm IR})$---can be renormalized in such a way that observables (i.e., correlation functions) become independent of the fiducial mass parameter $m_{\rm IR}$, which is set by the physical IR cutoff scale $\Lambda_{\rm IR}$ of the theory. We will argue that this is indeed the case: IR divergences arising from loop corrections can be handled using standard effective field theory methods, in which the effects from non-perturbative scales are systematically encoded through effective operators valid at perturbative scales. This procedure yields finite, de Sitter-invariant correlation functions in momentum space, valid for external momenta above the threshold $\Lambda_{\rm IR}$~\footnote{A physical IR momentum-cutoff has been advocated in Refs.~\cite{Palma:2017lww,Chen:2018brw,Chen:2018uul,Palma:2023idj,Palma:2023uwo,Huenupi:2024ksc,Burgess:2009bs,Burgess:2010dd,Negro:2024bbf,Urakawa:2009gb,Urakawa:2009my,Tanaka:2013caa}. Computations of correlators via the wavefunction of the universe approach, also support a physical IR regulator~\cite{Anninos:2014lwa,Melville:2021lst,Lee:2023jby,Huenupi:2024ztu}. In addition, a de Sitter invariant approach is compatible with the bootstrap program~\cite{Arkani-Hamed:2018kmz}. The boostless bootstrap~\cite{Pajer:2020wxk,Pimentel:2022fsc} on the other hand, departs from the dS isometries in an effort to capture slow-roll effects during inflation, or incorporate effects stemming from the finite duration of inflation~\cite{Wang:2022eop}. Since we are considering rigid dS, these latter cases are not relevant for our purpose. Finally, it would be interesting to compare with lower dimensional exactly solvable setups like for example the one considered in~\cite{Anninos:2024fty}.}.

At the core of our approach lies the simple observation that IR divergences introduced via an infinitesimal mass $m_{\rm IR}$ (or, equivalently, via a strong-coupling scale $\Lambda_{\rm IR}$) are mathematically equivalent to de Sitter-invariant IR divergences arising in loop integrals that satisfy the so-called Wilson's axioms, which define momentum-space integration as a unique operation. In an arbitrary spatial dimension $d$, these axioms take the form~\cite{Wilson:1972cf}:
\bea
\int \!\! d^d k \, [\alpha f(\k) + \beta g(\k)] &=& \alpha \!\! \int \!\! d^d k \, f(\k) + \beta \!\!  \int \!\! d^d k \, g(\k), \qquad \label{linearity} \\
\int \!\! d^d k \, f(\alpha \k) &=& \frac{1}{\alpha^d}  \int \!\! d^d k \, f(\k), \label{int-scaling} \\
\int \!\! d^d k \, f(\k + \q) &=&   \int \!\! d^d k \, f(\k), \label{translation-invariance}
\eea
where $f(\k)$ and $g(\k)$ are arbitrary functions of the momentum $\k$, and $\alpha$, $\beta$ are constants. Axiom~\eqref{linearity} expresses the linearity of the integration operation, Axiom~\eqref{int-scaling} reflects the scaling behavior under dilations, and Axiom~\eqref{translation-invariance} enforces translation invariance in momentum space. These axioms underpin dimensional regularization and, as we will show, provide a consistent framework for isolating both UV and IR divergences in the computation of correlation functions for light scalar fields in de Sitter space.

\subsection{No secular growth in a nutshell}

This article explores a wide range of interrelated issues concerning the computation of $n$-point correlation functions for light scalar fields in de Sitter spacetime. For readers not wishing to engage with the full technical details, we sketch here the core argument against the appearance of secular growth in correlation functions.

\begin{figure}[t]
\begin{center}
\includegraphics[scale=0.45]{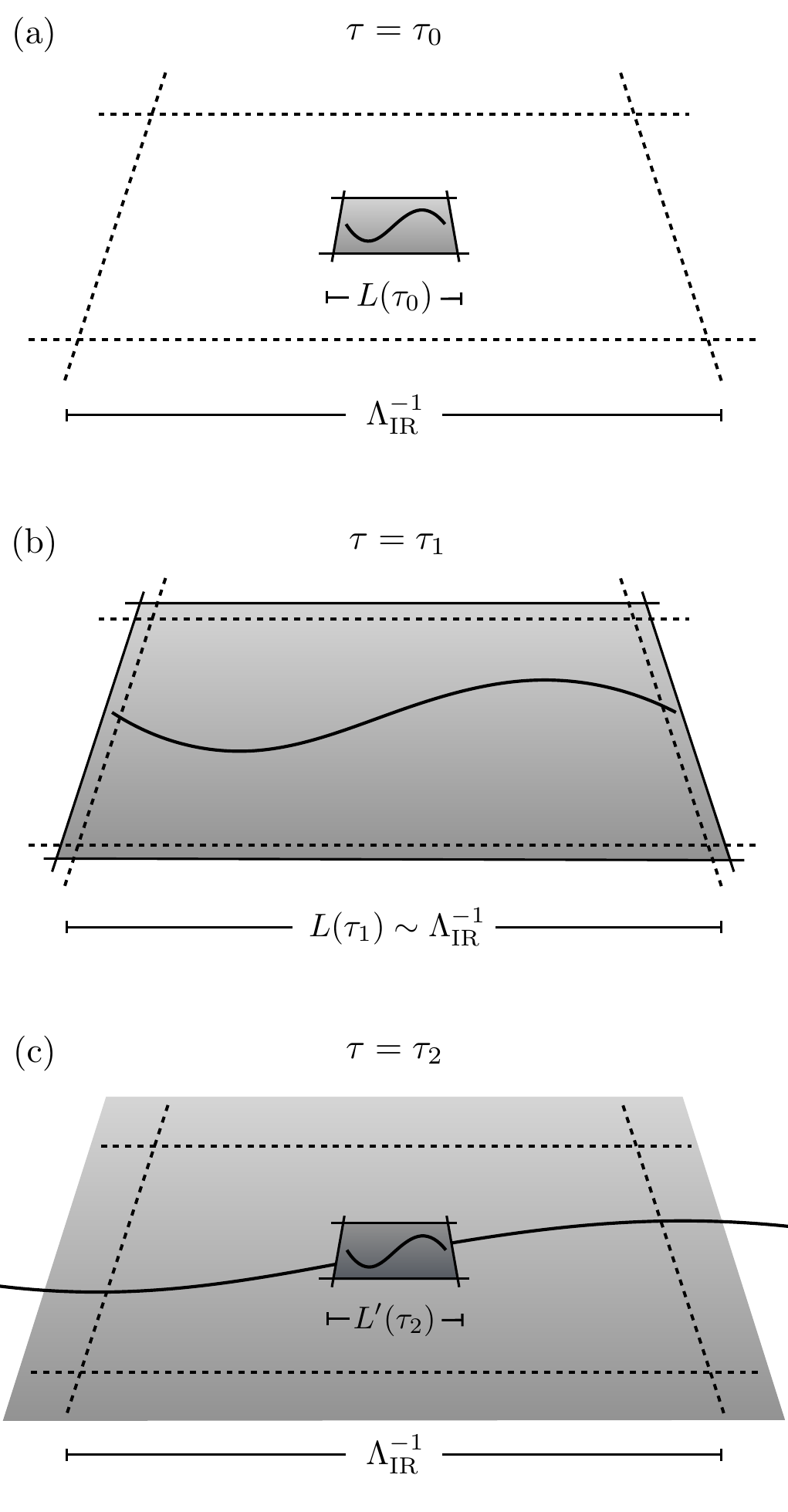}
\caption{(a) Initially, at $\tau = \tau_0$, the universe's size is $L(\tau_0) = L \times a(\tau_0)$. Modes of wavelengths longer than $L(\tau_0)$ are in their IR-safe configuration, inherited from the pre-inflationary era. (b) Eventually, at some time $\tau = \tau_1$, the universe's size $L(\tau_1) = L \times a(\tau_1)$ becomes of order $\Lambda_{\rm IR}^{-1}$, the lengthscale at which $\varphi_{\rm IR}$'s longest modes reach equilibrium. (c) After the longest modes reach equilibrium, the evolution of modes with comoving wavelengths $L' \ll \Lambda_{\rm IR}^{-1} / a(\tau_2)$ is drastically different from those of comoving wavelengths $L$: At lengths smaller than $\Lambda_{\rm IR}^{-1}$ the universe is effectively de Sitter, and the boundary doesn't influence the evolution of modes.
 }
\label{fig_02}
\end{center}
\end{figure}

To begin, consider the case where the leading-order dynamics in Eq.~(\ref{intro:action-basic}) correspond to an exactly massless scalar field, i.e., with $\mathcal{V} = 0$. In this case, according to the prevailing view, the de Sitter isometries are broken due to the presence of a comoving length scale $L$ that imposes a maximum wavelength for fluctuations (or equivalently, an infrared comoving momentum cutoff $k_{\rm IR} = 2\pi / L$). One can visualize this scale $L$ as defining an arbitrary comoving patch of physical size $L(\tau) = L a(\tau)$, located anywhere in space. This patch expands with the universe and contains fluctuations of all wavelengths, including the longest allowed comoving mode. A preferred time $\tau_0$ then exists when the patch size satisfies $L(\tau_0) \sim H^{-1}$, marking the moment the universe becomes populated with superhorizon fluctuations [see FIG.\ref{fig_02}(a)].

The setup just outlined may be interpreted as describing a universe that transitioned from a radiation-dominated era to an inflating phase at time $\tau_0$. Then, fluctuations with comoving wavelengths larger than $L$ are necessarily infrared safe, and therefore the comoving cutoff $k_{\rm IR}$ serves as an absolute lower bound on the momenta of fluctuations, including those appearing in loop integrals. As a result, correlation functions inevitably develop secular growth, which is further amplified by loop corrections. However, once non-derivative interactions encoded in $\mathcal{V}(\varphi)$ are turned on, they introduce a physical infrared scale $\Lambda_{\rm IR}$, beyond which strong nonlinear effects become relevant. As the patch grows and $L(\tau) = L a(\tau)$ reaches the physical scale $\Lambda_{\rm IR}^{-1}$, the longest fluctuations enter the strongly nonlinear regime and are expected to reach equilibrium [see FIG.~\ref{fig_02}(b)].

Beyond this point, fluctuations at shorter wavelengths—well below $\Lambda_{\rm IR}^{-1}$—are dynamically shielded from the long-wavelength sector that has reached equilibrium [see FIG.~\ref{fig_02}(c)]. The comoving patch size $L$ no longer plays any physical role, and the associated comoving cutoff $k_{\rm IR}$ becomes irrelevant to the dynamics. Instead, fluctuations on smaller scales, such as those with comoving wavelength $L' \ll \Lambda_{\rm IR}^{-1} / a(\tau_2)$, evolve in an effectively de Sitter background where the boundary plays no role. In this regime, it is the physical scale $\Lambda_{\rm IR}$ that determines the infrared structure of the theory. Since secular growth relies on the presence of a comoving IR cutoff, and such a cutoff no longer governs the dynamics, we conclude that no genuine secular growth can arise in this new universe.

The remainder of this introduction outlines our approach and key results before turning to the technical development in the main body of the paper.

\subsection{Overview and outlook}

The main technical goal of this article is to analyze the impact of loop corrections on equal-time $n$-point correlation functions in theories describing interacting light scalar fields. We begin in Section~\ref{sec:iso-vacua} with a review of basic aspects of scalar field theories in de Sitter spacetime and the emergence of secular growth in two-point correlation functions, focusing on the special case of free massless scalars. We also discuss how IR divergences arise in the limit $m \to 0$ for light scalar theories in which the shift symmetry is broken. This discussion sets the stage for analyzing more realistic models with nonlinear interactions.

To compute the effects of interactions on observables, we will adopt the Schwinger--Keldysh formalism, which will be reviewed in Section~\ref{sec:S-K}. This formalism enables a diagrammatic expansion of $n$-point correlation functions, allowing us to visualize the impact of interactions perturbatively, order by order. As is standard in perturbation theory, the formalism requires separating the theory into free and interacting sectors. As anticipated in the introduction, we take the free sector to be massless. That is, any term of the form $\frac{m^2}{2} \varphi^2$ present in the free theory must involve a mass parameter satisfying $m \ll m_{\rm IR}$, where $m_{\rm IR}$ is the infrared mass scale introduced in~\eqref{mir}. This ensures that the theory remains effectively massless down to the IR momentum scale $\Lambda_{\rm IR}$, below which it becomes strongly nonlinear. Interactions will be modeled by a scalar potential $\mathcal{V}(\varphi)$, which we expand in a Taylor series as
\be \label{intro:pot-taylor}
\mathcal{V}(\varphi) = \sum_{n = 2}^{\infty} \frac{\lambda_n}{n!} \varphi^n .
\ee
Each term proportional to $\lambda_n$ in~\eqref{intro:pot-taylor} corresponds to an $n$-leg vertex that serves as a building block for diagrams representing correlation functions. The Schwinger--Keldysh formalism allows us to express equal-time $n$-point functions in Fourier space as a sum over such diagrams, with the external momenta propagating through the diagram's internal structure, including virtual processes and loops.

The reader may have noticed that the expansion in Eq.~\eqref{intro:pot-taylor} includes a quadratic term of the form $\frac{1}{2} \lambda_2 \varphi^2$, which could, in principle, be incorporated into the free part of the theory governing the linear dynamics of fluctuations. What then should our stance be regarding this term when organizing the theory into free and interacting components? At first glance, it may seem natural to incorporate the term $\frac{1}{2} \lambda_2 \varphi^2$ into the free sector of the theory and work with massive linear mode functions characterized by a mass parameter $m^2 = \lambda_2$. In this setup, if $\lambda_2 \ll m_{\rm IR}^2$, the theory can be treated as effectively massless, whereas if $\lambda_2 \gg m_{\rm IR}^2$, the free theory is massive and, by extension, IR safe. 

However, this is not the perspective we adopt in the present work. Instead, the term $\frac{1}{2} \lambda_2 \varphi^2$ is treated as part of the interaction sector, regardless of the value of $\lambda_2$. In fact, it is entirely feasible to consider scenarios in which $\lambda_2 \gg H^2$, yet the evolution of modes remains effectively massless for a significant period after horizon crossing. Consider, for example, the axionic potential
\be \label{example-pot-cos}
\mathcal{V}(\varphi) = A^4 \left[ 1 - \cos(\omega_{\varphi} \varphi) \right],
\ee
with $A \ll H$. This potential satisfies $\mathcal{V}(\varphi)/H^4 \ll 1$ for all values of $\varphi$, ensuring weak interactions throughout field space. Now, if $\omega_{\varphi} \gg H$, then superhorizon Fourier modes will induce $\varphi$-fluctuations of amplitude $\sim H$, allowing the field to explore a large portion of the potential and traverse several of its hills and valleys. This behavior arises despite the potentially large value of $\lambda_2$, which in this case is given by $\lambda_2 = A^4 \omega_{\varphi}^2$ and may significantly exceed both $m_{\rm IR}^2$ and $H^2$. In such a scenario, interpreting $\lambda_2$ as a mass parameter for the free theory is misleading. The higher-order terms in the expansion of the potential~\eqref{example-pot-cos} contribute equally to the nonlinear dynamics, implying that the relaxation of fluctuations after horizon crossing is governed not by $\lambda_2/H^2$, but by the small ratio $A^4 / H^4$.

With this example in mind, we will assume that the potential satisfies $\mathcal{V}(\varphi) / H^4 \ll 1$ over the range of field fluctuations relevant to the correlation functions under consideration. Under this assumption, the theory should be organized such that $\lambda_2$ remains part of the interaction sector, regardless of the value of the mass parameter. The free theory is then described by a massless scalar field down to the momentum scale $\Lambda_{\rm IR}$.

The splitting of the theory just discussed will be particularly relevant for understanding Section~\ref{sec:logarithms}, which presents some of the key technical results of this article. There, we analyze the time dependence of equal-time $n$-point correlation functions in Fourier space in the superhorizon regime, where each external comoving momentum $\k_i$ corresponds to a wavelength much larger than the de Sitter radius $R_H$ (i.e., $|\k_i \tau| \ll 1$). We will show that, in this limit, the time dependence of correlation functions is governed by the number of interaction vertices rather than the number of external legs or loops. More precisely, in the superhorizon limit, a collection $D_T(\k_1, \cdots, \k_n; \tau)$, representing the sum of diagrams sharing a common topology $T$, takes the form
\bea 
D_{T} (\k_1, \cdots , \k_n; \tau) &=&   \delta^{(3)} (\boldsymbol{K}) \sum_{s} A^T_{s}(\k_1, \cdots , \k_n) \nn \\
&& \!\!\!\!\!\!\!\!\!\!\!\!\!\!\! \times \prod_{a=1}^{V}  \ln  \big[ -\tau f^T_{s,a}(\k_1, \cdots , \k_n)  \big] , \quad
\eea
where $\boldsymbol{K} \equiv \k_1 + \cdots + \k_n$ and $V$ is the number of vertices in the given topology. In this expression, the label $s$ indexes a set of amplitudes $A^T_s(\k_1, \cdots, \k_n)$, while $a$ runs from $1$ to $V$, labeling the time-dependent logarithmic terms. The functions $A^T_s(\k_1, \cdots, \k_n)$ and $f^T_{s,a}(\k_1, \cdots, \k_n)$ satisfy the scaling property
\bea
A^T_{s,a}(\alpha \k_1, \cdots, \alpha \k_n) &=& \alpha^{-3 (n-1)} A^T_{s,a}(\k_1, \cdots, \k_n), \quad \\
f^T_{s,a}(\alpha \k_1, \cdots, \alpha \k_n) &=& \alpha f^T_{s,a}(\k_1, \cdots, \k_n),
\eea
for a positive real number $\alpha$. Importantly, this structure holds independently of the number of loops. We will confirm this behavior through explicit examples involving nontrivial loop corrections in which external momenta propagate through the loops.

The relevance of this result should be clear: loop corrections to correlation functions cannot grow in time relative to tree-level contributions. This implies that if loop corrections are initially small, they remain small over time. In Section~\ref{sec:stoch}, we examine the implications of this result for the stochastic approach to inflation. As previously emphasized by other authors, matching the standard version of stochastic inflation to perturbation theory requires breaking de Sitter invariance, which in turn leads to time-dependent loop corrections. However, since we find that loop corrections remain de Sitter invariant and thus do not lead to secular growth, we conclude that the stochastic formalism requires revision. Building on previous work~\cite{Palma:2023idj,Palma:2023uwo}, in Section~\ref{sec:stoch} we revisit Starobinsky’s original derivation of the stochastic formalism and highlight that one of its key steps relies on an implicit assumption regarding how loop integrals are cut off in momentum space. We show that if loop integrals are regularized in a de Sitter-invariant manner, then the stochastic dynamics receives relevant corrections that modify the Fokker--Planck equation governing the evolution of fluctuation statistics in the long-wavelength regime.

We conclude in Section~\ref{sec:conclusions} with a summary of our findings and some final remarks.

\subsection{Conventions and notation}

We adopt units where $\hbar = 1$, and use a mostly-plus signature for the spacetime metric. Integrals in coordinate space are abbreviated as $\int d^3 x = \int_{\x}$, and those in momentum space as $\int d^3 k / (2\pi)^3 = \int_{\k}$. Fourier transforms are defined so that $\varphi(\x, \tau) = \int_{\k} \tilde{\varphi}(\k, \tau) e^{ i \k \cdot \x}$. Throughout most of this article we work in conformal time $\tau$, but in the final sections, where the stochastic formalism is examined, we switch to cosmic time, related to $\tau$ via $t = H^{-1} \ln (-1 / H \tau)$.


\section{de Sitter isometries and massless vacua}
\label{sec:iso-vacua}

\begin{figure}[t]
\begin{center}
\includegraphics[scale=0.55]{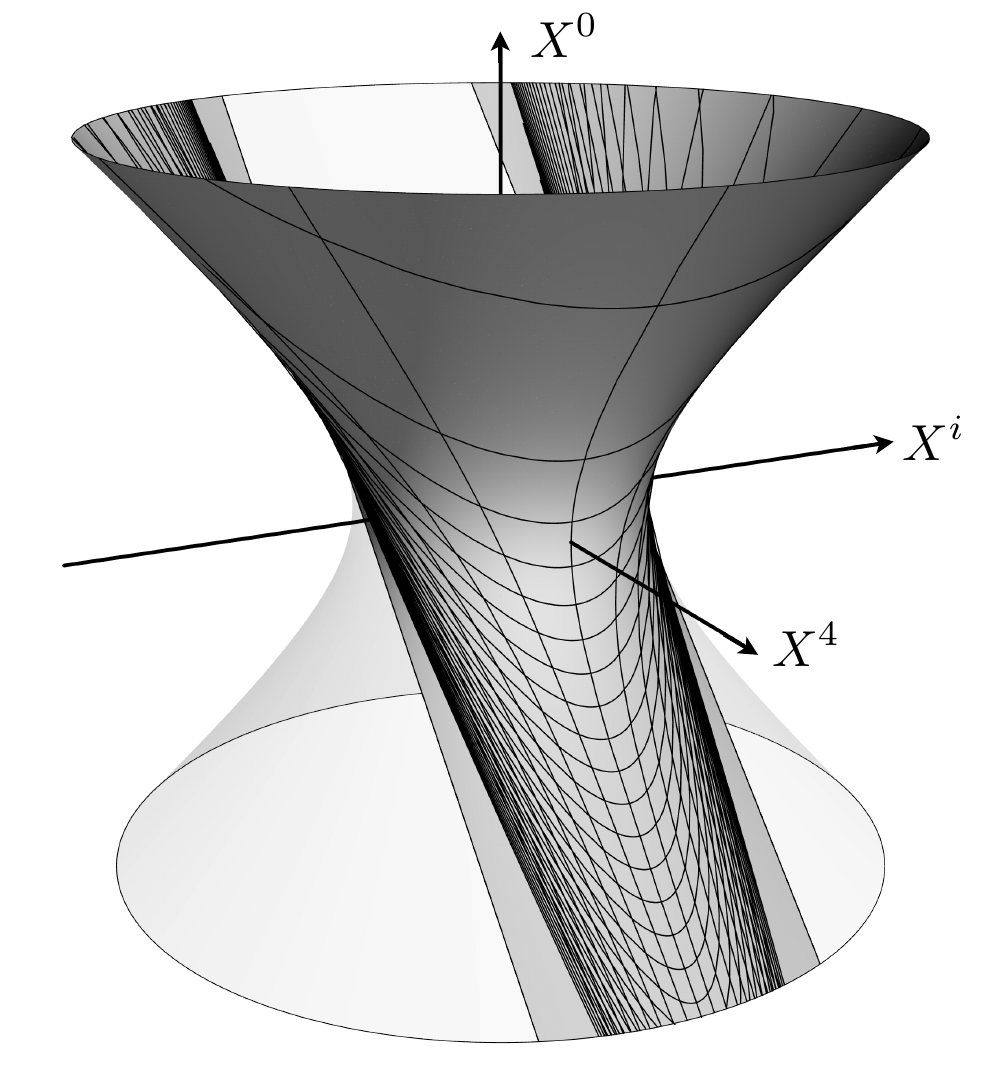}
\caption{A representation of de Sitter space covered by conformal inflationary coordinates. These coordinates cover only half of the full spacetime.}
\label{fig_01}
\end{center}
\end{figure}

A four-dimensional de Sitter spacetime~\cite{Spradlin:2001pw} can be visualized as a Lorentz-invariant hyperboloid embedded in five-dimensional Minkowski space (see FIG.~\ref{fig_01}). The hyperboloid is defined by the coordinates $X^0$ and $X^I$ with $I = 1,\dots,4$, satisfying the constraint
\be
- \left(X^0\right)^2 + \sum_{I=1}^{4} \left(X^I\right)^2 = R_H^2 , \label{eq-embedding}
\ee
where $R_H$ is the de Sitter radius. The conformal inflationary coordinates $(\x, \tau)$ used in the metric~(\ref{intro:metric}) are related to the embedding coordinates via the relations:
\bea
X^0 &=&   \frac{1}{2} a(\tau) H \big( H^{-2} +   \x^2 - \tau^2 \big) , \label{coord-param-1} \\
X^4 &=&  \frac{1}{2} a(\tau) H \big( H^{-2} -  \x^2 + \tau^2 \big) , \label{coord-param-2} \\
X^i &=& a(\tau) x^i , \label{coord-param-3}
\eea
with $i = 1,2,3$. Recall that $H = R_H^{-1}$ is the Hubble expansion rate associated with spatially flat slices, and $a(\tau) = -1/(H \tau)$ is the corresponding scale factor. FIG.~\ref{fig_01} illustrates how the inflationary coordinates $(\tau, \x)$ cover part of the hyperboloid.
Surfaces of constant conformal time correspond to the intersection of the hyperboloid with hyperplanes defined by $X^0 + X^4 = a(\tau)/H$. These planes are tilted at $45^\circ$ with respect to the horizontal. It is worth noting that these coordinates cover only half of de Sitter space. However, this restriction does not limit the scope of our analysis.

In addition to $E(3)$, the Euclidean group of translations and rotations acting on the spatial coordinates $\x$, the metric~(\ref{intro:metric}) is manifestly invariant under dilations:
\bea
\tau &\to& \bar \tau = e^{-\theta} \tau  , \label{additional-symmetry-1} \\
\x &\to& \bar \x = e^{- \theta} \, \x  . \label{additional-symmetry-2}
\eea
In ambient-space coordinates $(X^0,X^I)$, dilations are nothing but boosts of rapidity $\theta$ along the $X^4$-direction. de Sitter space admits additional isometries that are less obvious in these coordinates. These emerge from the invariance of the embedding relation~(\ref{eq-embedding}) under the full 5D Lorentz group, acting on the coordinates $(X^0, X^I)$ as $X \to \bar{X} = \Lambda X$. Consider the Lorentz transformation $\Lambda(\mathbf{b})$ parameterized by an arbitrary 3-vector $\mathbf{b}$, mixing boosts and rotations:
\be \label{Lorentz-b}
\Lambda ( {\bf b} ) \equiv \left(\begin{array}{ccccc}1 + \frac{1}{2} {\bf b}^2  & b_1 & b_2 & b_3 & - \frac{1}{2} {\bf b}^2  
\\
b_1 & 1 & 0 & 0 & -b_1 
\\
b_2 & 0 & 1 & 0 & -b_2 
\\
b_3 & 0 & 0 & 1 & - b_3
\\
\frac{1}{2} {\bf b}^2 & b_1 & b_2 & b_3 & 1- \frac{1}{2} {\bf b}^2 \end{array}\right) .
\ee
Applying this transformation to the parameterization~(\ref{coord-param-1})--(\ref{coord-param-3}) induces the following non-linear transformation on the cosmological coordinates $(\x, \tau)$:
\bea
\tau \to \bar \tau &=& \frac{\tau}{1 - 2 \, {\bf b} \cdot \x +  {\bf b}^2 \, (\x^2-\tau^2)}  ,  \label{NL-transf-1} \\
x^i \to  \bar x^i &=&  \frac{x^i - b^i \, (\x^2-\tau^2)}{1 - 2 \, {\bf b} \cdot \x +  {\bf b}^2 \, (\x^2-\tau^2)} .
 \label{NL-transf-2}
 \eea
Thus, the full 10-dimensional de Sitter isometry group, when expressed in cosmological coordinates, consists of: $E(3)$, the group of spatial rotations and translations, the non-linear transformations~(\ref{NL-transf-1})--(\ref{NL-transf-2}), which can be recognized as the special conformal transformation, and the dilation symmetry~(\ref{additional-symmetry-1})--(\ref{additional-symmetry-2}), altogether forming the dS group $SO(4,1)$.
Note that the non-linear transformations associated with $\Lambda(\mathbf{b})$ typically map points outside the inflationary patch covered by $(\tau, \x)$. This limitation is inherited by the transformations~(\ref{NL-transf-1})--(\ref{NL-transf-2}). In contrast, the $E(3)$ subgroup and the dilation symmetry preserve the patch.

To conclude this discussion, observe that the hyperboloid (\ref{eq-embedding}) can be mapped into a sphere in a five-dimensional Euclidean space by performing a Wick rotation via $X^I \to X_E^I = X^I$ and $X^0 \to X_E^5 = i X^0$. Working with the Euclidean version of de Sitter allows one compute the geodesic distance between pair of points $X^I$ and ${X'}^I$ on the sphere. This is simply given by
\be \label{geodesic-distance}
D = R_H \arccos \left( Z \right),
\ee
where $Z = H^2 \sum_{I=1}^{5} X_E^I {X_E'}^I $. Wick rotating back to the Lorentzian signature, one finds that the geodesic distance between two points on the hyperboloid is still given by (\ref{geodesic-distance}) but with quadratic form $Z$ given by
\be
  Z =   H^2 \left( \sum_{I=1}^{4} X^I {X'}^I -   X^0 {X'}^0  \right) .
\ee
By further substituting the embedding relations~(\ref{coord-param-1})--(\ref{coord-param-3}) into this expression for $Z$, one finds the more convenient form in terms of the points $(\x, \tau)$ and $(\x', \tau')$ expressed in conformal cosmological coordinates:
\be \label{Q-def}
Z =1 - \frac{|\x - \x'|^2 - (\tau - \tau')^2}{2 \tau \tau'}.
\ee
Note that $D$ can be either purely real or imaginary, depending on whether $|Z| < 1$ or $|Z| > 1$ respectively, implying that $D^2$ can be positive (for points with space-like separation) or negative (for points with time-like separation).

\subsection{Vacuum state for massless quanta} 

A scalar field $\varphi(\x, \tau)$ in de Sitter space, expressed in inflationary coordinates, may be regarded as a function of the five-dimensional embedding coordinates $(X^0, X^I)$ via the identification~\eqref{coord-param-1}--\eqref{coord-param-3}. Such a field is expected to transform as a scalar under the full set of de Sitter isometries. Consequently, the two-point function $\langle \Omega | \varphi(\x, \tau)\, \varphi(\x', \tau') | \Omega \rangle$, computed with respect to a de Sitter-invariant vacuum state $|\Omega\rangle$, is also expected to respect de Sitter isometries and thus must depend only on the invariant distance $Z$. However, for a massless scalar field, the two-point correlation function is found to include a contribution that explicitly breaks de Sitter invariance~\cite{Allen:1985ux}. Let us examine this in more detail.

Inserting the metric~\eqref{intro:metric} into Eq.~\eqref{intro:action-basic}, the action for a canonical scalar field with potential $\mathcal{V}(\varphi)$ in de Sitter space, written in conformal cosmological coordinates, becomes
\be \label{action-cosmo-coord}
S = \int \! d^3 x d\tau \, a^4(\tau) \left[ \frac{ \dot \varphi ^2 -   ( \nabla \varphi )^2}{2 a^2(\tau)}  - \mathcal V (\varphi) \right],
\ee
where the dot denotes a derivative with respect to conformal time $\tau$. It follows that in the absence of interactions (i.e., for $\mathcal{V} = 0$), the equation of motion satisfied by the field operator $\varphi(\x, \tau)$ is
\be
\ddot{\varphi} - \frac{2}{\tau} \dot{\varphi} - \nabla^2 \varphi = 0.
\ee
In this case, the theory exhibits invariance under the shift symmetry $\varphi \to \varphi + c$, where $c$ is a constant.

The quantization of $\varphi$ can be carried out by expressing the field in Fourier space as
\be \label{field-Fourier}
\varphi(\x, \tau) = \int_{\bf k} \left[ f_k(\tau) \, \hat{a}_{\k} + f_k^*(\tau) \, \hat{a}_{-\k}^\dag \right] e^{i \k \cdot \x},
\ee
where $\hat{a}_{\k}^\dag$ and $\hat{a}_{\k}$ are the usual creation and annihilation operators, satisfying the commutation relation $[\hat{a}_{\k}, \hat{a}_{\k'}^\dag] = (2\pi)^3 \delta^{(3)}(\k - \k')$. The mode functions $f_k(\tau)$ satisfy the equation of motion
\be \label{mode-massless}
\ddot{f}_k - \frac{2}{\tau} \dot{f}_k + k^2 f_k = 0.
\ee
The solution to this equation must ensure that the field operator obeys the canonical commutation relation $[\varphi(\x, \tau), \Pi(\y, \tau)] = \delta^{(3)}(\x - \y)$, where the canonical momentum is given by $\Pi(\y, \tau) = a^2(\tau) \dot{\varphi}(\y, \tau)$. 

It is standard to show that the general solution for $f_k(\tau)$ consistent with the Bunch--Davies initial condition and the canonical quantization rules is
\be \label{fk-t}
f_k(\tau) = \frac{i H}{\sqrt{2 k^3}} \left(1 + i k \tau \right) e^{-i k \tau}.
\ee
The vacuum state $|\Omega\rangle$ is defined by $\hat{a}_{\k} |\Omega\rangle = 0$. To study the properties of this vacuum under the de Sitter isometries discussed earlier—or under a subgroup thereof—it suffices to examine the transformation properties of the two-point function
\be \label{2-point-G}
G(|\x - \x'| ; \tau, \tau') \equiv \langle \Omega | \varphi(\x, \tau) \, \varphi(\x', \tau') | \Omega \rangle.
\ee
Inserting the expansion~\eqref{field-Fourier} into~\eqref{2-point-G}, we then obtain $
G(|\x - \x'| ; \tau, \tau') = \int_{\bf k} f_k(\tau) f_k^*(\tau') e^{i \k \cdot (\x - \x')}$. After performing the angular integration over the momentum direction, this expression yields
\be \label{2point-0}
G(|\x - \x'| ; \tau, \tau') = \int_0^{\infty} \frac{dk}{k} \, F(k \tau, k \tau') \, \frac{\sin(k |\x - \x'|)}{k |\x - \x'|},
\ee
where we have defined
\be
F(k \tau, k \tau') \equiv \frac{1}{2\pi^2} k^3 f_k(\tau) f_k^*(\tau').
\ee
From~\eqref{fk-t}, it is clear that $F(k \tau, k \tau')$ is regular at $k = 0$. This implies that the integral in~\eqref{2point-0} is logarithmically divergent in the infrared, i.e., as $k \to 0$.

Before evaluating the integral in~\eqref{2point-0} explicitly, it is helpful to first consider the implications of logarithmic divergences in integrals of this type. To illustrate the point, let us examine a function $I(s)$ defined by the integral
\be \label{def-I-s}
I(s) = \int_0^\infty \frac{dk}{k} \, f(s k),
\ee
where we assume that $f(sk)$ decays sufficiently rapidly as $k \to \infty$ to ensure convergence in the ultraviolet, and that $f(0) \neq 0$ is finite. Now consider evaluating $I(s)$ at the rescaled argument $\bar{s} = e^{-\theta}$:
\be \label{def-I-s-2}
I(\bar{s}) = I(e^{-\theta} s) = \int_0^\infty \frac{dk}{k} \, f(e^{-\theta} s k).
\ee
We may Taylor expand the function $f(e^{-\theta} s k) = f(s k + s k \epsilon) = \sum_n \frac{1}{n!} (s k \epsilon)^n f^{(n)} (sk) $, where $\epsilon = 1 - e^{-\theta}$ and $f^{(n)} (x)$ is the $n$th derivative of the function $f (x)$ with respect to $x$. Then, by performing integrations by parts, and assuming the fast decay of $f(s k)$ at large $k$, the integral yields the exact result
\be \label{def-I-s-3}
I(e^{-\theta} s) = I(s) + \theta f(0).
\ee
This demonstrates that, unless $f(s k)$ vanishes at $k = 0$, the function $I(s)$ is not invariant under rescalings of $s$. Notably, this behavior violates Wilson's second axiom~\eqref{int-scaling}. 

One might instead attempt to absorb the factor $e^{-\theta}$ into the integration variable $k$. However, doing so still yields the additional term $\theta f(0)$ on the right-hand side of~\eqref{def-I-s-3}. To see this explicitly, one can regularize the lower limit of the integral using a Heaviside function $\theta(k)$ (or a smooth approximation thereof). Under a rescaling of $k$, the derivative of this cutoff introduces a term proportional to $\theta \delta(k) f(k)$, which upon integration reproduces the same result.

Let us now examine how the previous analysis affects the behavior of $G(|\x - \x'|; \tau, \tau')$ under a dilation transformation of the form~\eqref{additional-symmetry-1}--\eqref{additional-symmetry-2}. Evaluating the two-point function~\eqref{2point-0} at the transformed coordinates $|\bar{\x} - \bar{\x}'| = e^{-\theta} |\x - \x'|$, $\bar{\tau} = e^{-\theta} \tau$, and $\bar{\tau}' = e^{-\theta} \tau'$, and using the result~\eqref{def-I-s-3} with $F(0,0) = H^2 / 4\pi^2$, we find:
\be \label{transf-G-x-tau-2}
G(|\bar{\x} - \bar{\x}'|; \bar{\tau}, \bar{\tau}') = G(|\x - \x'|; \tau, \tau') + \theta \frac{H^2}{4\pi^2}.
\ee
This result shows that the two-point function of a massless free scalar field is not invariant under dilations and, by extension, is not invariant under the full de Sitter group discussed earlier. Consequently, $G(|\x - \x'|; \tau, \tau')$ cannot be a function of the de Sitter-invariant quantity $Z$ defined in~\eqref{Q-def}.

A direct evaluation of the integral in~\eqref{2point-0} for $|\x - \x'| \neq 0$ yields:
\bea \label{2point-1}
G(|\x - \x'|; \tau, \tau') &=& \frac{H^2}{8 \pi^2} \bigg[ \frac{1}{1-Z} - \ln(1-Z) \nonumber \\
&& \hspace{-1.3cm} + \ln a(\tau) + \ln a(\tau') + c_{\infty} \bigg],
\eea
where the constant $c_{\infty}$, which is given by
\be
c_{\infty} = 2(1 - \gamma_E) - \ln 2 + 2 \lim_{k \to 0} \ln(H/k) ,
\ee
is a divergent contribution arising from the infrared limit of the integral. This result explicitly demonstrates the breaking of de Sitter symmetry at the level of the two-point function. The first line of~\eqref{2point-1}, usually referred to as the Hadamard form, depends only on the invariant quantity $Z$, while the second line contains symmetry-breaking terms that depend on the individual conformal times $\tau$ and $\tau'$. One can directly verify that~\eqref{transf-G-x-tau-2} holds by applying the rescalings $\tau \to \bar{\tau} = e^{-\theta} \tau$ and $\tau' \to \bar{\tau}' = e^{-\theta} \tau'$ to the result in~\eqref{2point-1}.

By evaluating~\eqref{2point-1} in the coincident-point limit $\x = \x'$ and $\tau = \tau'$, corresponding to $Z = 1$, the two-point correlation function becomes a function of time~\cite{Allen:1985ux}:
\be \label{secular-growth}
G^{(2)}(\tau) \equiv G(0; \tau) = \frac{H^2}{4 \pi^2} \ln \frac{a(\tau)}{a(\tau_0)},
\ee
where $\tau_0$ is a reference time that marks the onset of symmetry breaking and conveniently absorbs the divergent contributions arising from both $c_\infty$ and the limit $Z \to 1$.

The regularized and more manageable expression~\eqref{secular-growth} has led many authors to implement an infrared comoving cutoff $k_{\rm IR}$ directly in Eq.~\eqref{2point-0}. To be concrete, one can repeat the computation of the integral~\eqref{2point-0}, this time introducing an infrared comoving cutoff $k_{\rm IR}$ and an ultraviolet physical cutoff $\Lambda$. Since we are only interested in the superhorizon contributions to $G(\x, \x'; \tau)$, we may set $\Lambda = H$. With this choice, the evaluation of~\eqref{2point-0} reproduces the result~\eqref{secular-growth} in the coincident limit $|\x - \x'| \to 0$, with the identification $\tau_0 \simeq -k_{\rm IR}^{-1}$. From this perspective, the introduction of $k_{\rm IR}$ inevitably selects a preferred time slice, thereby breaking the de Sitter symmetries.


\subsection{Vacuum state for light quanta} 
\label{sec:light-quanta}

In a shift-symmetric theory with derivative interactions—such as a $P(X)$ theory—one typically expects strong coupling to emerge in the ultraviolet regime~\cite{Cheung:2007st}, even though infrared divergences may still be present. However, as discussed in the introduction, in QFTs with non-derivative interactions that break the shift symmetry of the massless theory, an infrared physical scale $\Lambda_{\rm IR}$ must arise, signaling the onset of strongly nonlinear dynamics. In this framework, the massless mode function~\eqref{fk-t} can only be trusted for physical momenta $p = -k \tau$ greater than $\Lambda_{\rm IR}$. Below this scale, the theory becomes strongly coupled, and the computation of observables requires resumming nonlinear corrections to all orders.

Let us examine the computation of correlation functions taking into account the presence of $\Lambda_{\rm IR}$. Given that this cutoff scale acts as a common yardstick for physical momenta belonging to the same time-slice, it is convenient to specialize this discussion to equal-time 2-point correlation functions $G (|\x-\x'| ; \tau ) \equiv \langle \Omega |  \varphi (\x , \tau)  \varphi (\x' , \tau)   | \Omega \rangle$. As we shall see, the resulting expressions are de Sitter invariant, and hence, they can be trivially extended to represent 2-point functions correlating points at different time slices. 

Of course, our aim is to perform perturbation theory valid for momenta above the scale $\Lambda_{\rm IR}$, by splitting the theory (\ref{action-cosmo-coord}) into the free part (corresponding to the massless kinetic term) and the interaction part (corresponding to the potential). At zeroth order, this scheme implies that the equal-time $2$-point correlator corresponds to the following integral summing modes with physical momentum $p = k / a(\tau) > \Lambda_{\rm IR}$:
\be \label{G-sin-k}
G (|\x - \x' |; \tau ) =  \!\! \int_{0}^{\infty} \! \frac{dk}{k}  F(k\tau)  \frac{\sin ( k | \x - \x' |) }{k | \x - \x' |} ,
\ee
where this time we have defined 
\be
F(k \tau) \equiv \frac{1}{2 \pi^2} k^3  | f_{k} (\tau)  |^2 \theta \left( k |\tau| - \Lambda_{\rm IR} / H \right) .
\ee
In this expression, $\theta(x)$ may be taken as the usual Heaviside step function, or as a smooth function keeping the transition to far-infrared momenta $p \ll \Lambda_{\rm IR}$ under control. Before explicitly solving this integral, notice that, because this time the integrand is regular in the entire integration domain, there are no subtleties with the validity of Wilson's second axiom (\ref{int-scaling}). Therefore, the implementation of the dilation transformation (\ref{additional-symmetry-1})-(\ref{additional-symmetry-2}) immediately results in
\be \label{transf-G-x-tau-3}
G (| \bar \x - \bar \x' |; \bar \tau)  =  G (| \x - \x' |; \tau)  ,
\ee
which may be obtained directly with a change of integration variable.
In fact, one is allowed to perform the change of variables $k \to p = - k \tau H$ which results in the following de Sitter invariant integral
\be \label{G-sin-p}
G (|\x - \x' |; \tau ) \! = \! \frac{H^2}{4 \pi^2} \!\! \int_{\Lambda_{\rm IR}}^{\infty} \!\! \frac{dp}{p} \! \left[ 1 + \! \frac{p^2}{H^2} \! \right] \!\! \frac{\sin (\frac{p}{H} \sqrt{2 (1-Z)}  ) }{\frac{p}{H} \sqrt{2 (1-Z)} } ,
\ee
where we used the form of the massless modes (\ref{fk-t}). 
This integral is explicitly invariant under de Sitter transformations and it is logarithmically sensitive to $\Lambda_{\rm IR}$. Direct integration of (\ref{G-sin-p}) then leads to the following result valid in the limit $\Lambda_{\rm IR}/H \to 0$:
\be \label{2point-2}
\!\!G (| \x - \x' |; \tau)\! = \frac{H^2}{8 \pi^2} \bigg[ \frac{1}{1-Z} \! - \ln (1-Z)  + 2 \ln \frac{H}{\Lambda_{\rm IR}}  \bigg] ,
\ee
where we have disregarded order 1 terms. This result is explicitly de Sitter invariant, while it diverges logarithmically as $\frac{H^2}{4 \pi^2} \ln (  H/ \Lambda_{\rm IR} )$, in the limit $\Lambda_{\rm IR}\to0$.

Notice that the coincident point limit may be obtained by evaluating the previous result at $Z \to 1$. This limit is of course divergent with the divergence coming from the UV end of the integral. However, given that the correlator is still invariant under the de Sitter isometries, the resulting expression must be a constant. To capture the divergences with the help of cutoffs, one may go back to (\ref{G-sin-p}) and directly evaluate it in the limit $\x\to\x'$. Then, by regularizing the integral with an additional physical UV cutoff $\Lambda_{\rm UV}$, one obtains~\cite{Palma:2017lww,Chen:2018uul,Chen:2018brw}
\be \label{G-0-p}
G (0; \tau ) = \frac{H^2}{4 \pi^2}\int_{\Lambda_{\rm IR}}^{\Lambda_{\rm UV}} \frac{dp}{p} \left(1 + p^2/H^2 \right) ,
\ee
which is a constant with divergencies coming from both the IR and UV parts of the integral. This result will be relevant in later sections when we deal with loop corrections to correlation functions.

\subsection{Removing zero modes}

There is an alternative way to derive~\eqref{2point-2} that incorporates the infrared physical cutoff scale $\Lambda_{\rm IR}$—induced by shift-symmetry-breaking interactions—directly at the level of the free theory. This consists of introducing a mass term $\frac{1}{2} m_{\rm IR}^2 \varphi^2$ in the action~\eqref{action-cosmo-coord}. With this modification, the equation of motion for the mode function $f_k(\tau)$ becomes
\be \label{mode-massive-IR}
\ddot{f}_k - \frac{2}{\tau} \dot{f}_k + \left( k^2 + \frac{m_{\rm IR}^2}{H^2 \tau^2} \right) f_k = 0.
\ee
The solution to this equation that satisfies the Bunch--Davies initial condition is
\be \label{mode-f-massive}
f_k(\tau) = \frac{i}{2} \sqrt{\frac{\pi}{H}} (-H \tau)^{3/2} H_\nu^{(1)}(-k \tau),
\ee
where $H_\nu^{(1)}(x)$ is the Hankel function of the first kind, and the index is $\nu = \sqrt{9/4 - m_{\rm IR}^2/H^2}$. Inspecting this solution shows that, for small mass $m_{\rm IR} \ll H$ and in the long-wavelength limit $|k \tau| \ll 1$, the mode function squared scales as $\sim (-k \tau)^{2 m_{\rm IR}^2 / 3 H^2}$. By definition, at physical momentum $p = -H k \tau$ of order $\Lambda_{\rm IR}$, the amplitude of the mode functions has not yet decayed, allowing nonlinearities to dominate. While this argument does not determine $\Lambda_{\rm IR}$ directly—since it depends on the structure of the interactions—it allows one to infer the relation
\be \label{mir-2}
m_{\rm IR}^2 = \frac{3 H^2}{2 \ln(H / \Lambda_{\rm IR})},
\ee
already shown in the Introduction. Continuing, if one uses~\eqref{mode-f-massive} to compute the two-point function, then in the limit $m_{\rm IR}/H \to 0$ one finds
\be \label{2point-3}
G(|\x - \x'|; \tau) = \frac{H^2}{8 \pi^2} \left[ \frac{1}{1 - Z} - \ln(1 - Z) + 3 \frac{H^2}{m_{\rm IR}^2} \right],
\ee
where, again, we have omitted $\mathcal{O}(1)$ terms. Comparing this result with~\eqref{2point-2} confirms the expression~\eqref{mir-2}, and illustrates how the introduction of a small mass parameter $m_{\rm IR}$ effectively captures the impact of shift-symmetry-breaking interactions at the level of the free theory for light scalar fields.

Note that the introduction of the infrared mass $m_{\rm IR}$ in Eq.~(\ref{mode-massive-IR}) effectively removes the zero mode, thereby breaking the shift symmetry responsible for the transformation rule of Eq.~(\ref{transf-G-x-tau-2}) underlying secular growth. As shown in Ref.~\cite{Allen:1987tz}, the limit $m_{\rm IR} \to 0$ can be taken smoothly in the two-point function $G(|\x - \x'|; \tau)$, provided that the role of the zero mode is properly accounted for in the massless limit. In this limit, one recovers the standard result of de Sitter symmetry breaking given in Eq.~(\ref{2point-1}). This analysis supports our perspective. 
In this framework, our proposal amounts to screening the zero mode in the same way that any strongly nonlinear mode is treated—by applying a de Sitter invariant regulator and renormalizing the result. Our main claim is that the Hadamard form of the two-point function remains the physically meaningful object in both shift-symmetric and light-field theories.

\subsection{Dimensional regularization}

In the previous discussion we gave preeminence to the idea that non-derivative interactions inevitably introduce an infrared mass scale $\Lambda_{\rm IR}$ setting the length $\Lambda_{\rm IR}^{-1}$ at which long wavelength modes become highly non-linear. This length scale appeared regularizing the divergence of the 2-point correlation function of the scalar $\varphi$, now diverging logarithmically in the limit $\Lambda_{\rm IR}/H \to 0$. Let us review how this divergence appears within dimensional regularization~\cite{Huenupi:2024ksc}.

Dimensional regularization first requires one to extend the dimensionality of integrals, in this case from $3$ to $d$, with $d$ a continuous parameter. After obtaining a finite value in $d$ dimensions, one can analytically extend this result back to $d \to 3$. This step usually allows one to isolate logarithmic divergencies, if any. The 2-point correlation function of the scalar field $\varphi$ in dimension $d$ is given by:
\be \label{two-corr-equal-time-2}
G_{d} (|\x - \x' |; \tau ) \equiv \frac{1}{\tilde \mu^{d-3}}  \int_{\bf k} | f_{k} (\tau)  |^2  e^{ i \k \cdot ( \x - \x' )},
\ee
where now $\int_{\bf k} = \int \frac{d^d k}{(2 \pi)^d}$. Note that we have added the factor $\frac{1}{\tilde \mu^{d-3}}$ (with $\tilde \mu$ an arbitrary parameter with mass dimension 1) to keep the mass dimension of $G_{d} (|\x - \x' |; \tau )$ fixed to $2$. To integrate the previous expression we can write $\k \cdot ( \x - \x' ) = k |\x - \x'| \cos \theta$, where $\theta$ is the angle between $\k$ and $\x - \x'$. Expressing the integration variables in polar coordinates, the integration measure can be written as $
 d^d k =   dk d\theta k^{d-1}  \sin \theta d\Omega_{d-1} $, where $d\Omega_{d-1}$ is the solid angle measure of a $(d-1)$-sphere. Then, integrating all the angles and using $\Omega_{d-1} = \frac{2 \pi^{(d-1)/2}}{\Gamma[\frac{1}{2} (d-1)]}$ as the solid angle of the $2$-sphere, Eq.~(\ref{two-corr-equal-time-2}) becomes
\bea \label{twopoint-dim-1}
G_{d} (|\x - \x' |; \tau ) &=& \frac{2 \Omega_{d-1}}{\tilde \mu^{d-3}(2 \pi)^d}   \int_0^{\infty} dk \,  k^{d-1}  \nn \\ 
&&  \times | f_{k} (\tau)  |^2  \frac{\sin(k | \x - \x' | )}{k | \x - \x' |} .
\eea
Here, the mode function $f_k(\tau)$ must solve the massless equation of motion in $d$-dimensions. While there is no unique way of extending this equation away from $d=3$, we find that the scheme proposed by Melville and Pajer in Ref.~\cite{Melville:2021lst} is particularly simple to implement in the case of light fields. 

This scheme consists in restricting the mass parameter to remain proportional to $d^2-9$ in such a way as to ensure that the conformal weight of the mode function remains zero~\cite{Melville:2021lst, Lee:2023jby}. In this case, the equation of motion is:
\be  \label{fk-t-d}
\ddot f_k - \frac{d-1}{\tau} \dot f_k + \left( k^2 + \frac{d^2-9}{4 \tau^2} \right) f_k = 0 .
\ee
The solution to this equation, after imposing Bunch-Davies initial conditions, is found to be:
\be   \label{fk-t-d-sol}
f_k(\tau) = (- H \tau)^{(d-3)/2} \frac{i H}{\sqrt{2 k^3} } \left[ 1 + i k \tau \right] e^{- i k \tau } .
\ee
Because of the analytical form of this solution, it should be clear that this implementation makes momentum-integrals much simpler than in other cases, alleviating intermediate steps prior to the limit $d \to 3$. With the mode function (\ref{fk-t-d-sol}), the equal-time 2-point function reads
\bea \label{twopoint-dim-2}
G_{d} (|\x - \x' |; \tau ) &=& \frac{H^2 \Omega_{d-1}}{\tilde \mu^{d-3}(2 \pi)^d}   \int_0^{\infty} \frac{dk}{k} \,   (- H \tau k)^{d-3}  \nn \\ 
&&  \times \left[ 1 + k^2 \tau^2 \right]  \frac{\sin(k | \x - \x' | )}{k | \x - \x' |} .
\eea
Crucially, the lower end of this integral converges for $d>3$, and so, the full integral is finite as long as $d>3$, with a piece proportional to $\frac{1}{d-3}$ capturing the IR divergence. Similarly, our analysis that led to (\ref{def-I-s-2}) ensures that for $d>3$ the result will respect Wilson's second axiom (\ref{int-scaling}), further making it invariant under dilations.

Performing the integral explicitly for $d>3$ and analytically continuing the result for all values of $d$, one finds:
\bea
G_{d} (|\x - \x' |; \tau ) &=&       H^2 \pi^{-\frac{d+1}{2}} \sin\left(\text{\scriptsize $\frac{d \pi}{2}$}\right) 2^{\frac{3(1-d)}{2}}   \left(\frac{H}{\mu}\right)^{d-3} \nn \\
&&\times  \frac{2 (1-Z) \Gamma[d-4] - \Gamma[d-2]}{(1-Z)^{\frac{d-1}{2}}  \Gamma[\frac{1}{2}(d-1)]} , \qquad
\eea
where $1-Z = |\x - \x'|^2 / 2 \tau^2$ is the de Sitter invariant combination introduced earlier. Next, performing the limit $d\to 3$ one finds:
\be \label{2point-dim}
\!\!G (| \x - \x' |; \tau )\! = \frac{H^2}{8 \pi^2} \bigg[ \frac{1}{1-Z} \! - \ln (1-Z)  + \frac{2}{d-3} \bigg] .
\ee
Once more, here we are disregarding order 1 terms, among which there is a contribution proportional to $\ln (\mu / H)$, as expected in dimensional regularization. The result (\ref{2point-dim}) is consistent with the previous de Sitter invariant results (\ref{2point-2}) and (\ref{2point-3}). This is not a coincidence. Notice that the scheme (\ref{fk-t-d}) breaks the shift symmetry of the massless case away from $d=3$, in favor of keeping a vanishing conformal weight. Thus, at least in the computation of 2-point functions, this scheme ensures capturing the de Sitter invariant result, eliminating the unphysical secular growth. 

Before concluding this section, it is instructive to examine the particular case of the 2-point correlation function at coincident points within the present scheme. If one where to impose the limit $|\x - \x'| \to 0$ (or equivalently $Z \to 1$) directly in (\ref{2point-dim}), in addition to the term proportional to $\frac{1}{d-3}$, one finds another diverging piece contributed from both $1/(1-Z)$ and $\ln (1-Z)$. To make sense of this UV divergence, it is convenient to give a step back and perform the limit in (\ref{twopoint-dim-2}) before integrating. One obtains
\be \label{two-corr-equal-time-6}
G_{d} (0; \tau ) =  \frac{H^2 \Omega_{d-1}}{\tilde \mu^{d-3}(2 \pi)^d}  \int_0^{\infty} \frac{dk}{k} (- H \tau k)^{d-3} \left[ 1 + k^2 \tau^2  \right]  .
\ee
Given that (\ref{two-corr-equal-time-6}) is a polynomial integral, then (thanks to Wilson's integration axioms just mentioned) in dimensional regularization it yields vanishing result $G_{d} (0, \tau ) = 0$, independently of $d$ \cite{Lee:2023jby}.

\subsection{Dealing with IR-divergent integrals}
\label{sec:div-ints-Wilson}

After discussing various ways to compute the $2$-point correlation function, it is convenient to pause to identify an important element regarding the computation of IR-divergent integrals. Thanks to the result (\ref{def-I-s-3}), we have learned that the appearance of secular growth is computationally linked to the fact that massless mode functions $f_k (\tau)$ reach the zero mode limit $k=0$ with a non-vanishing value of the combination $ k^3 f_{k} (\tau)   f_{k}^* (\tau)$. On the other hand, it is clear that this behavior breaks down in the presence of non-derivative interactions. In this case, the validity of the free-field description to arbitrary large wavelengths is guarantied to become invalid, and one has to regularize computations with the help of an IR physical cutoff scale $\Lambda_{\rm IR}$. 

We would like to point out that, as far as the computation of equal-time correlation functions is concerned, one may simply adopt the validity of Wilson's axioms from the start, independently of the behavior of the mode functions being integrated. As already noticed, Wilson's axioms provide a unique definition of integration, dictating the allowed operations that can be performed with integrals. In particular, it ensures the validity of changes of integration coordinates without yielding anomalous results such as (\ref{def-I-s-3}). In the rest of this article, we adopt the axioms to analyze the computation of $n$-point functions of light scalars in de Sitter. Adopting this perspective will allow us to dissect the the structure of loop integrals and perform explicit computations of correlation functions keeping the physical property of de Sitter invariance intact.


\section{Schwinger-Keldysh formalism}
\label{sec:S-K}

Having discussed the computation of 2-point correlation functions in Section~\ref{sec:iso-vacua}, we turn our attention to more general $n$-point functions.
An equal-time $n$-point correlation function $G^{(n)}(\x_1 , \cdots \x_n ; \tau) \equiv \langle \varphi (\x_1 , \tau) \cdots \varphi (\x_n , \tau) \rangle$ can be written as a sum of terms involving all possible products of connected correlation functions with $n$ or fewer points. These connected correlation functions $G_c^{(n)}(\x_1 , \cdots \x_n ; \tau)$ are, by definition, the components that cannot be factored into products of lower-point correlation functions, and can be expressed as
\be \label{G-n-coordinate}
 G_c^{(n)} \! = \!\! \int_{{\bf k}_1}  \cdots \! \int_{{\bf k}_n} \tilde G^{(n)}_c ( {\k}_1 , \cdots \! , {\k}_n ; \tau )  e^{ i \sum_i \! \k_i \cdot \x_i} ,
\ee
where $ \tilde G^{(n)}_c ( {\k}_1 , \cdots \! , {\k}_n ; \tau )$ is the connected $n$-point correlation function in momentum space. The Schwinger-Keldysh formalism~\cite{Calzetta:1986ey} allows for the direct computation of $ \tilde G^{(n)}_c ( {\k}_1 , \cdots \! , {\k}_n , \tau )$ with the help of diagrams following simple Feynman rules. Before summarizing these rules it is useful to define the so called Wightman function $G(k,\tau_a,\tau_b)$, which is the basic building block allowing the definition of propagators:
\be \label{G-basic-prop}
G (k , \tau_a , \tau_b) \equiv f_k(\tau_a) f^*_k( \tau_b) .
\ee
More explicitly, this function is given by
\be \label{Wightman-function}
 G (k , \tau_a , \tau_b) = \frac{H^2}{2 k^3} (1 + i k \tau_a) (1 - i k \tau_b) e^{- i k (\tau_a - \tau_b)} .
\ee
Note that $G (k , \tau_a , \tau_b)$ is nothing but the Fourier transform of the two point function $G (|\x-\x'| ; \tau , \tau')$ introduced in (\ref{2-point-G}).

\subsection{Feynman rules}
\label{sec:Feynman}

We now list the Feynman rules for the theory (\ref{action-cosmo-coord}). A detailed derivation can be found in Ref.~\cite{Chen:2017ryl}. Our starting point is the Taylor expansion of the potential $\mathcal V (\varphi)$:
\be \label{taylor-bare-potential}
\mathcal V (\varphi) = \sum_n \frac{\lambda_n}{n!} \varphi^n .
\ee
A single term of the expansion, proportional to $\lambda_n$, defines two classes of $n$-legged vertices, hereby distinguished by black and white solid dots:
\def\nvertexb{\tikz[baseline=-0.6ex,scale=1.8, every node/.style={scale=1.4}]{
\coordinate (v1) at (0ex,0ex);
\coordinate (phi1) at (-3ex,2ex);
\coordinate (phi2) at (-4ex,0ex);
\coordinate (phi3) at (-3ex,-2ex);
\coordinate (phi4) at (3ex,2ex);
\coordinate (phi5) at (4ex,0ex);
\coordinate (phi6) at (3ex,-2ex);
\draw[thick] (v1) -- (phi1);
\draw[thick] (v1) -- (phi2);
\draw[thick] (v1) -- (phi3);
\draw[thick] (v1) -- (phi4);
\draw[thick] (v1) -- (phi5);
\draw[thick] (v1) -- (phi6);
\filldraw[color=black, fill=black, thick] (v1) circle (0.5ex);
\node[anchor=south] at ($(v1)+(0,-2.5ex)$) {\scriptsize{$\tau_a$}};
\node[anchor=south] at ($(v1)+(0,+1.0ex)$) {\scriptsize{$\cdots$}};
}
}
\def\nvertexw{\tikz[baseline=-0.6ex,scale=1.8, every node/.style={scale=1.4}]{
\coordinate (v1) at (0ex,0ex);
\coordinate (phi1) at (-3ex,2ex);
\coordinate (phi2) at (-4ex,0ex);
\coordinate (phi3) at (-3ex,-2ex);
\coordinate (phi4) at (3ex,2ex);
\coordinate (phi5) at (4ex,0ex);
\coordinate (phi6) at (3ex,-2ex);
\draw[thick] (v1) -- (phi1);
\draw[thick] (v1) -- (phi2);
\draw[thick] (v1) -- (phi3);
\draw[thick] (v1) -- (phi4);
\draw[thick] (v1) -- (phi5);
\draw[thick] (v1) -- (phi6);
\filldraw[color=black, fill=white, thick] (v1) circle (0.5ex);
\node[anchor=south] at ($(v1)+(0,-2.5ex)$) {\scriptsize{$\tau_a$}};
\node[anchor=south] at ($(v1)+(0,+1.0ex)$) {\scriptsize{$\cdots$}};
}
}
\begin{flalign}
    & \nvertexb \qquad  \longrightarrow  \qquad V_{+} (\tau , {\bf K})  , \label{vertex-black} \\
    & \nvertexw \qquad  \longrightarrow  \qquad V_{-} (\tau , {\bf K})  , \label{vertex-white}
\end{flalign}
where ${\bf K} \equiv \sum_i \k_{i}$ is the sum of the momenta flowing into the vertex, and $V_{\pm} (\tau , {\bf K})$ stands for the following instruction involving the integral over the vertex time $\tau_a$:
\be \label{Feynman-vertices}
 V_{\pm} (\tau , {\bf K}) = \mp i \lambda_n \! \int^{\tau}_{- \infty} \!\!\!\!\!\! \de \tau_a \, a^4(\tau_a) (2 \pi)^3 \delta^{(3)} ({\bf K}) \bigg[ \cdots \bigg] ,
\ee
where $\delta^{(3)} ({\bf K})$ denotes a Dirac delta function enforcing conservation of momenta flowing through the vertex. Each vertex is characterized by a time integral from $-\infty$ up until the final time $\tau$ at which $n$-point correlation functions are evaluated. The square brackets on the right hand side indicates that any function of $\tau_a$ must be integrated in this way. 

In order to compute $n$-point correlation functions, vertices must be connected to a single surface (or boundary) labeled by $\tau$ via bulk-to-boundary propagators. This boundary represents the equal time surface at which the $n$-point function is evaluated. These propagators receive the following assignments:
\def\propbf{\tikz[baseline=-0.6ex,scale=1.8, every node/.style={scale=1.4}]{
\coordinate (tau) at (-3ex,0ex);
\coordinate (phi) at (3ex,0ex);
\draw[thick] (tau) -- (phi);
\filldraw[color=black, fill=black, thick] (tau) circle (0.5ex);
\node[anchor=south] at ($(tau)+(0,0.5ex)$) {\scriptsize{$\tau_a$}};
\pgfmathsetmacro{\arista}{0.06}
\filldraw[color=black, fill=white, thick] ($(phi)-(\arista,\arista)$) rectangle ($(phi)+(\arista,\arista)$);
\node[anchor=south] at ($(phi)+(0,0.5ex)$){\scriptsize{$\tau$}};
}
}
\def\propwf{\tikz[baseline=-0.6ex,scale=1.8, every node/.style={scale=1.4}]{
\coordinate (tau) at (-3ex,0ex);
\coordinate (phi) at (3ex,0ex);
\draw[thick] (tau) -- (phi);
\filldraw[color=black, fill=white, thick] (tau) circle (0.5ex);
\node[anchor=south] at ($(tau)+(0,0.5ex)$) {\scriptsize{$\tau_a$}};
\pgfmathsetmacro{\arista}{0.06}
\filldraw[color=black, fill=white, thick] ($(phi)-(\arista,\arista)$) rectangle ($(phi)+(\arista,\arista)$);
\node[anchor=south] at ($(phi)+(0,0.5ex)$){\scriptsize{$\tau$}};
}
}
\begin{flalign}
  \propbf   \;\;\;\;\;\;  &\longrightarrow  \;\;\;\;\;\;  G_{+}  (k, \tau_a, \tau) , \\
  \propwf  \;\;\;\;\;\;  &\longrightarrow  \;\;\;\;\;\;  G_{-}  (k, \tau_a, \tau) .
\end{flalign}
The analytical expressions for the quantities appearing at the right hand side of the previous assignments are given in terms of the Wightman function $G(k,\tau_a,\tau_b)$ introduced in (\ref{G-basic-prop}). These are: 
\bea
G_{+}  (k, \tau_a , \tau) &=& G^*  (k, \tau_a, \tau) , \label{G_+} \\ 
G_{-}  (k, \tau_a, \tau) &=& G  (k, \tau_a, \tau).
\eea
Notice that $G_{+} = G_{-}^*$.

Vertices can also be connected among themselves via internal, bulk-to-bulk propagators. Given that we have two classes of vertices, the rules contain four classes of internal propagators:
\def\propbb{\tikz[baseline=-0.6ex,scale=1.8, every node/.style={scale=1.4}]{
\coordinate (tau1) at (-3ex,0ex);
\coordinate (tau2) at (3ex,0ex);
\draw[thick] (tau1) -- (tau2);
\filldraw[color=black, fill=black, thick] (tau1) circle (0.5ex);
\node[anchor=south] at ($(tau1)+(0,0.5ex)$) {\scriptsize{$\tau_a$}};
\filldraw[color=black, fill=black, thick] (tau2) circle (0.5ex);
\node[anchor=south] at ($(tau2)+(0,0.5ex)$) {\scriptsize{$\tau_b$}};
}
}
\def\propww{\tikz[baseline=-0.6ex,scale=1.8, every node/.style={scale=1.4}]{
\coordinate (tau1) at (-3ex,0ex);
\coordinate (tau2) at (3ex,0ex);
\draw[thick] (tau1) -- (tau2);
\filldraw[color=black, fill=white, thick] (tau1) circle (0.5ex);
\node[anchor=south] at ($(tau1)+(0,0.5ex)$) {\scriptsize{$\tau_a$}};
\filldraw[color=black, fill=white, thick] (tau2) circle (0.5ex);
\node[anchor=south] at ($(tau2)+(0,0.5ex)$) {\scriptsize{$\tau_b$}};
}
}
\def\propbw{\tikz[baseline=-0.6ex,scale=1.8, every node/.style={scale=1.4}]{
\coordinate (tau1) at (-3ex,0ex);
\coordinate (tau2) at (3ex,0ex);
\draw[thick] (tau1) -- (tau2);
\filldraw[color=black, fill=black, thick] (tau1) circle (0.5ex);
\node[anchor=south] at ($(tau1)+(0,0.5ex)$) {\scriptsize{$\tau_a$}};
\filldraw[color=black, fill=white, thick] (tau2) circle (0.5ex);
\node[anchor=south] at ($(tau2)+(0,0.5ex)$) {\scriptsize{$\tau_b$}};
}
}
\def\propwb{\tikz[baseline=-0.6ex,scale=1.8, every node/.style={scale=1.4}]{
\coordinate (tau1) at (-3ex,0ex);
\coordinate (tau2) at (3ex,0ex);
\draw[thick] (tau1) -- (tau2);
\filldraw[color=black, fill=white, thick] (tau1) circle (0.5ex);
\node[anchor=south] at ($(tau1)+(0,0.5ex)$) {\scriptsize{$\tau_a$}};
\filldraw[color=black, fill=black, thick] (tau2) circle (0.5ex);
\node[anchor=south] at ($(tau2)+(0,0.5ex)$) {\scriptsize{$\tau_b$}};
}
}
\begin{flalign}
\propbb  \;\;\;\;\;\;  &\longrightarrow  \;\;\;\;\;\; G_{++} (k, \tau_a , \tau_b) , \\
\propww  \;\;\;\;\;\; &\longrightarrow  \;\;\;\;\;\; G_{--} (k, \tau_a , \tau_b) , \\
\propbw  \;\;\;\;\;\; &\longrightarrow \;\;\;\;\;\;  G_{+-} (k, \tau_a , \tau_b) ,\\
\propwb  \;\;\;\;\;\; &\longrightarrow \;\;\;\;\;\;  G_{-+} (k, \tau_a , \tau_b) .
\end{flalign}
The right hand side of the previous assignments can be written in terms of the Wightman function:
\bea
G_{++} (k, \tau_a , \tau_b)  &=&  G (k , \tau_a , \tau_b)  \theta (\tau_a - \tau_b) \nn \\
&& + G^* (k , \tau_a , \tau_b)  \theta (\tau_b - \tau_a),  \\
G_{--} (k, \tau_a , \tau_b) &=& G^* (k , \tau_a , \tau_b)  \theta (\tau_a - \tau_b) \nn \\
&& + G (k , \tau_a , \tau_b)  \theta (\tau_b - \tau_a), \\
G_{+-} (k, \tau_a , \tau_b)  &=& G^* (k , \tau_a , \tau_b)  , \\
G_{-+} (k, \tau_a , \tau_b) &=& G (k , \tau_a , \tau_b)  ,
\eea
where $\theta(x)$ denotes the standard Heaviside step function. Notice that $G_{++} = G_{--}^*$ and $G_{+-} = G_{-+}^*$.

The previous analytical assignments for vertices and propagators allow one to write the connected correlation function $ \tilde G^n_c ( {\k}_1 , \cdots , {\k}_n , \tau )$ in momentum space, as the sum of every diagram with $n$ external legs, truncated to the desired order (with respect to vertices and loops). In doing so, every internal momentum $\k$ flowing through internal propagators must be integrated with $\int \frac{d^3 k}{(2 \pi)^3}$. Additionally, a particular diagram must include an overall factor $1/S_D$ where $S_D$ is the symmetry factor of the diagram. Because the final result consists of a sum of diagrams running through all classes of vertices ---both black and white--- the resulting $n$-point function will necessarily be real.

To finish, notice that the existence of two classes of vertices (black and white) implies the proliferation of diagrams as the number of vertices increases. A collection of diagrams contributing to $ \tilde G^n_c ( {\k}_1 , \cdots , {\k}_n , \tau )$, obtained with $V$ vertices, and sharing the same topology, would consist in $2^V$ individual diagrams differing in the colors of their vertices. Therefore, with the purpose of discussing diagrams without specifying a particular class of vertex, it is useful to define the concept of a non-colored vertex as the one resulting from the formal sum of black and white vertices:
\def\nvertexn{\tikz[baseline=-0.6ex,scale=1.8, every node/.style={scale=1.4}]{
\coordinate (v1) at (0ex,0ex);
\coordinate (phi1) at (-3ex,2ex);
\coordinate (phi2) at (-4ex,0ex);
\coordinate (phi3) at (-3ex,-2ex);
\coordinate (phi4) at (3ex,2ex);
\coordinate (phi5) at (4ex,0ex);
\coordinate (phi6) at (3ex,-2ex);
\draw[thick] (v1) -- (phi1);
\draw[thick] (v1) -- (phi2);
\draw[thick] (v1) -- (phi3);
\draw[thick] (v1) -- (phi4);
\draw[thick] (v1) -- (phi5);
\draw[thick] (v1) -- (phi6);
\fill[white] (v1) circle (0.5ex);
\filldraw[pattern=north east lines, thick] (v1) circle (0.5ex);
\node[anchor=south] at ($(v1)+(0,-2.5ex)$) {\scriptsize{$\tau_a$}};
\node[anchor=south] at ($(v1)+(0,+1.0ex)$) {\scriptsize{$\cdots$}};
}
}
\be
\nvertexn = \nvertexb + \nvertexw
\ee
Given that propagators distinguish the color of each vertex, this element has the sole purpose of alleviating the representation of diagrams.

\subsection{Examples}
\label{SK-examples}

It will be useful to count with a few concrete examples of analytical expressions for diagrams contributing to general $n$-point correlation functions. 

To start with, consider the contribution to $ \tilde G^n_c ( {\k}_1 , \cdots , {\k}_n , \tau )$ coming from diagrams with a single vertex and $L$ loops. This contribution comes from a vertex of strength $\lambda_{n + 2L}$, and may be expressed as a single diagram with a single non-colored vertex representing the sum of two diagrams:
\def\Oconnectedloopszero{\tikz[baseline=-1.4ex]{
\coordinate (P) at (0,-3ex);
\coordinate (C) at (-7ex,4ex);
\coordinate (k1) at (-5.0ex,4.0ex);
\coordinate (k2) at (-2.0ex,4.0ex);
\coordinate (kn) at (5.0ex,4.0ex);
\pgfmathsetmacro{\arista}{0.1}
\draw[thick] (-7ex,4ex) -- (7ex,4ex);
\draw[thick] (0,-3ex) -- (k1);
\draw[thick] (0,-3ex) -- (k2);
\draw[thick] (0,-3ex) -- (kn);
\filldraw[color=black, fill=black, thick] (0ex,2ex) circle (0.1ex);
\filldraw[color=black, fill=black, thick] (1.2ex,2ex) circle (0.1ex);
\filldraw[color=black, fill=black, thick] (2.4ex,2ex) circle (0.1ex);
\node at ($(P) + (-1.ex,0)$) [anchor=east]{{$\tau'$}};
\filldraw[color=black, fill=white, thick] ($(k1)-(\arista,\arista)$) rectangle ($(k1)+(\arista,\arista)$)
node[anchor=south]{\footnotesize{$k_1$}};
\filldraw[color=black, fill=white, thick] ($(k2)-(\arista,\arista)$) rectangle ($(k2)+(\arista,\arista)$)
node[anchor=south]{\footnotesize{$k_2$}};
\filldraw[color=black, fill=white, thick] ($(kn)-(\arista,\arista)$) rectangle ($(kn)+(\arista,\arista)$)
node[anchor=south]{\footnotesize{$k_n$}};
\draw[thick,scale=3] (0,-1ex)  to[in=-100,out=-140,loop] (0,-1ex);
\draw[thick,scale=3] (0,-1ex)  to[in=-50,out=-90,loop] (0,-1ex);
\draw[thick,scale=3] (0,-1ex)  to[in=30,out=-10,loop] (0,-1ex);
\pgfmathsetmacro{\distanceone}{0.8}
\pgfmathsetmacro{\angleone}{315}
\pgfmathsetmacro{\distancetwo}{0.8}
\pgfmathsetmacro{\angletwo}{330}
\pgfmathsetmacro{\distancethree}{0.8}
\pgfmathsetmacro{\anglethree}{345}
\coordinate (Qone) at ($(P) + (\angleone:\distanceone)$);
\filldraw[color=black, fill=black, thick] (Qone) circle (0.1ex);
\coordinate (Qtwo) at ($(P) + (\angletwo:\distancetwo)$);
\filldraw[color=black, fill=black, thick] (Qtwo) circle (0.1ex) node[anchor=west]{\scriptsize{$L$}};
\coordinate (Qthree) at ($(P) + (\anglethree:\distancethree)$);
\filldraw[color=black, fill=black, thick] (Qthree) circle (0.1ex);
\fill[white] (P) circle (0.9ex);
\filldraw[pattern=north east lines, thick] (P) circle (0.9ex);
}
}
\def\Oconnectedloopsone{\tikz[baseline=-1.4ex]{
\coordinate (P) at (0,-3ex);
\coordinate (C) at (-7ex,4ex);
\coordinate (k1) at (-5.0ex,4.0ex);
\coordinate (k2) at (-2.0ex,4.0ex);
\coordinate (kn) at (5.0ex,4.0ex);
\pgfmathsetmacro{\arista}{0.1}
\draw[thick] (-7ex,4ex) -- (7ex,4ex);
\draw[thick] (0,-3ex) -- (k1);
\draw[thick] (0,-3ex) -- (k2);
\draw[thick] (0,-3ex) -- (kn);
\filldraw[color=black, fill=black, thick] (0ex,2ex) circle (0.1ex);
\filldraw[color=black, fill=black, thick] (1.2ex,2ex) circle (0.1ex);
\filldraw[color=black, fill=black, thick] (2.4ex,2ex) circle (0.1ex);
\node at ($(P) + (-1.ex,0)$) [anchor=east]{{$\tau'$}};
\filldraw[color=black, fill=white, thick] ($(k1)-(\arista,\arista)$) rectangle ($(k1)+(\arista,\arista)$)
node[anchor=south]{\footnotesize{$k_1$}};
\filldraw[color=black, fill=white, thick] ($(k2)-(\arista,\arista)$) rectangle ($(k2)+(\arista,\arista)$)
node[anchor=south]{\footnotesize{$k_2$}};
\filldraw[color=black, fill=white, thick] ($(kn)-(\arista,\arista)$) rectangle ($(kn)+(\arista,\arista)$)
node[anchor=south]{\footnotesize{$k_n$}};
\draw[thick,scale=3] (0,-1ex)  to[in=-100,out=-140,loop] (0,-1ex);
\draw[thick,scale=3] (0,-1ex)  to[in=-50,out=-90,loop] (0,-1ex);
\draw[thick,scale=3] (0,-1ex)  to[in=30,out=-10,loop] (0,-1ex);
\pgfmathsetmacro{\distanceone}{0.8}
\pgfmathsetmacro{\angleone}{315}
\pgfmathsetmacro{\distancetwo}{0.8}
\pgfmathsetmacro{\angletwo}{330}
\pgfmathsetmacro{\distancethree}{0.8}
\pgfmathsetmacro{\anglethree}{345}
\coordinate (Qone) at ($(P) + (\angleone:\distanceone)$);
\filldraw[color=black, fill=black, thick] (Qone) circle (0.1ex);
\coordinate (Qtwo) at ($(P) + (\angletwo:\distancetwo)$);
\filldraw[color=black, fill=black, thick] (Qtwo) circle (0.1ex) node[anchor=west]{\scriptsize{$L$}};
\coordinate (Qthree) at ($(P) + (\anglethree:\distancethree)$);
\filldraw[color=black, fill=black, thick] (Qthree) circle (0.1ex);
\filldraw[color=black, fill=black, thick] (P) circle (0.9ex);
}
}
\def\Oconnectedloopstwo{\tikz[baseline=-1.4ex]{
\coordinate (P) at (0,-3ex);
\coordinate (C) at (-7ex,4ex);
\coordinate (k1) at (-5.0ex,4.0ex);
\coordinate (k2) at (-2.0ex,4.0ex);
\coordinate (kn) at (5.0ex,4.0ex);
\pgfmathsetmacro{\arista}{0.1}
\draw[thick] (-7ex,4ex) -- (7ex,4ex);
\draw[thick] (0,-3ex) -- (k1);
\draw[thick] (0,-3ex) -- (k2);
\draw[thick] (0,-3ex) -- (kn);
\filldraw[color=black, fill=black, thick] (0ex,2ex) circle (0.1ex);
\filldraw[color=black, fill=black, thick] (1.2ex,2ex) circle (0.1ex);
\filldraw[color=black, fill=black, thick] (2.4ex,2ex) circle (0.1ex);
\node at ($(P) + (-1.ex,0)$) [anchor=east]{{$\tau'$}};
\filldraw[color=black, fill=white, thick] ($(k1)-(\arista,\arista)$) rectangle ($(k1)+(\arista,\arista)$)
node[anchor=south]{\footnotesize{$k_1$}};
\filldraw[color=black, fill=white, thick] ($(k2)-(\arista,\arista)$) rectangle ($(k2)+(\arista,\arista)$)
node[anchor=south]{\footnotesize{$k_2$}};
\filldraw[color=black, fill=white, thick] ($(kn)-(\arista,\arista)$) rectangle ($(kn)+(\arista,\arista)$)
node[anchor=south]{\footnotesize{$k_n$}};
\draw[thick,scale=3] (0,-1ex)  to[in=-100,out=-140,loop] (0,-1ex);
\draw[thick,scale=3] (0,-1ex)  to[in=-50,out=-90,loop] (0,-1ex);
\draw[thick,scale=3] (0,-1ex)  to[in=30,out=-10,loop] (0,-1ex);
\pgfmathsetmacro{\distanceone}{0.8}
\pgfmathsetmacro{\angleone}{315}
\pgfmathsetmacro{\distancetwo}{0.8}
\pgfmathsetmacro{\angletwo}{330}
\pgfmathsetmacro{\distancethree}{0.8}
\pgfmathsetmacro{\anglethree}{345}
\coordinate (Qone) at ($(P) + (\angleone:\distanceone)$);
\filldraw[color=black, fill=black, thick] (Qone) circle (0.1ex);
\coordinate (Qtwo) at ($(P) + (\angletwo:\distancetwo)$);
\filldraw[color=black, fill=black, thick] (Qtwo) circle (0.1ex) node[anchor=west]{\scriptsize{$L$}};
\coordinate (Qthree) at ($(P) + (\anglethree:\distancethree)$);
\filldraw[color=black, fill=black, thick] (Qthree) circle (0.1ex);
\filldraw[color=black, fill=white, thick] (P) circle (0.9ex);
}
}
\be \label{D-tree-level-1-vertex}
\Oconnectedloopszero \!\!\! = \!\!\!\Oconnectedloopsone \!\!\! + \!\!\! \Oconnectedloopstwo .
\ee
Using the Feynman rules already outlined, one finds that this contribution has the following analytical form:
\begin{widetext}
\be \label{n-point-one-vertex}
\Oconnectedloopszero \!\!\! =   (2\pi)^3 \delta^{(3)} (\boldsymbol{K}) \, 2\,\text{Im}\Bigg\{  \frac{\lambda_{n+2L}}{2^L L! H^4} 
     \int_{-\infty}^{\tau}  \frac{d \tau'}{{ \tau' }^{4}}G_+(\tau', \tau, k_1)\cdots G_+(\tau', \tau ,k_n)\left[\int_{\bf k} G( \tau', \tau',k)\right]^L\!\Bigg\}, 
\ee
\end{widetext}
where ${\bf K} = \sum_{i} \k_i$. Note the presence of the symmetry factor $2^L L!$, where $2^L$ arises from the contribution of each individual loop, accounting for the two indistinguishable ways loops can be formed with a single propagator, and $L!$ comes from the symmetry of the diagram under the interchange of identical loops. We will come back to this example in later sections.

As a second example, let us consider the most general diagram with two vertices contributing to a connected $n$-point function. Such diagrams will have $n_1$ external legs connected to the first vertex $\tau_1$ and $n_2$ external legs connected to the second vertex $\tau_2$, with $n_1 + n_2 = n$: 
\begin{widetext}
\def\diagramextwoverticesNoLabels{\tikz[baseline=-1.4ex]{
\coordinate (k1) at (-9ex,0ex);
\coordinate (k2) at (-3.ex,0ex);
\coordinate (k3) at (+3.ex,0ex);
\coordinate (k4) at (+9.ex,0ex);
\coordinate (t1) at (-4.5ex,-7ex);
\coordinate (t2) at (+4.5ex,-7ex);
\pgfmathsetmacro{\arista}{0.1}
\draw[thick] (-11ex,0ex) -- (11ex,0ex);
\draw[-,thick] (t1) -- (k1);
\draw[-,thick] (t1) -- (k2);
\draw[-,thick] (t2) -- (k3);
\draw[-,thick] (t2) -- (k4);
\draw[thick,scale=2] (t1) to [bend left=60] (t2);
\draw[thick,scale=2] (t1) to [bend left=-60] (t2);
\draw[thick,scale=3] (t1)  to[in=-90,out=-130,loop] (t1);
\draw[thick,scale=3] (t1)  to[in=-180,out=-220,loop] (t1);
\draw[thick,scale=3] (t2)  to[in=-90,out=-50,loop] (t2);
\draw[thick,scale=3] (t2)  to[in=0,out=40,loop] (t2);
\fill[white] (t1) circle (0.9ex);
\filldraw[pattern=north east lines, thick] (t1) circle (0.9ex);
\fill[white] (t2) circle (0.9ex);
\filldraw[pattern=north east lines, thick] (t2) circle (0.9ex);
\filldraw[color=black, fill=white, thick] ($(k1)-(\arista,\arista)$) rectangle ($(k1)+(\arista,\arista)$);
\filldraw[color=black, fill=white, thick] ($(k2)-(\arista,\arista)$) rectangle ($(k2)+(\arista,\arista)$);
\filldraw[color=black, fill=white, thick] ($(k3)-(\arista,\arista)$) rectangle ($(k3)+(\arista,\arista)$);
\filldraw[color=black, fill=white, thick] ($(k4)-(\arista,\arista)$) rectangle ($(k4)+(\arista,\arista)$);
\coordinate (Q) at ($(0ex,-7.0ex)$);
\coordinate (P) at ($(-8ex,-8.5ex)$);
\coordinate (K) at ($(8ex,-8.5ex)$);
\node at ($(P)$) [anchor=east]{{\footnotesize{$L_1$}}};
\filldraw[color=black, fill=black, thick] ($(P)+(-0.3ex,0.9ex)$) circle (0.1ex);
\filldraw[color=black, fill=black, thick] (P) circle (0.1ex);
\filldraw[color=black, fill=black, thick] ($(P)+(0.5ex,-0.8ex)$) circle (0.1ex);
\node at ($(K)$) [anchor=west]{{\footnotesize{$L_2$}}};
\filldraw[color=black, fill=black, thick] ($(K)+(0.3ex,0.9ex)$) circle (0.1ex);
\filldraw[color=black, fill=black, thick] (K) circle (0.1ex);
\filldraw[color=black, fill=black, thick] ($(K)+(-0.5ex,-0.8ex)$) circle (0.1ex);
\filldraw[color=black, fill=black, thick] ($(Q)+(0ex,1ex)$) circle (0.1ex);
\filldraw[color=black, fill=black, thick] (Q) circle (0.1ex);
\filldraw[color=black, fill=black, thick] ($(Q)+(0ex,-1ex)$) circle (0.1ex);
\node at ($(0ex,-9.5ex)$) [anchor=north]{\footnotesize{$L+1$}};
\node at ($(k1)+(0ex,0.5ex)$) [anchor=south]{\footnotesize{$\k_1$}};
\node at ($(k2)+(0ex,0.5ex)$) [anchor=south]{\footnotesize{$\k_{n_1}$}};
\node at ($(k3)+(0ex,0.5ex)$) [anchor=south]{\footnotesize{$\k_{n_1+1}$}};
\node at ($(k4)+(0ex,0.5ex)$) [anchor=south]{\footnotesize{$\k_{n}$}};
}
}
\def\diagramextwoverticesone{\tikz[baseline=-1.4ex]{
\coordinate (k1) at (-9ex,0ex);
\coordinate (k2) at (-3.ex,0ex);
\coordinate (k3) at (+3.ex,0ex);
\coordinate (k4) at (+9.ex,0ex);
\coordinate (t1) at (-4.5ex,-7ex);
\coordinate (t2) at (+4.5ex,-7ex);
\pgfmathsetmacro{\arista}{0.1}
\draw[thick] (-11ex,0ex) -- (11ex,0ex);
\draw[-,thick] (t1) -- (k1);
\draw[-,thick] (t1) -- (k2);
\draw[-,thick] (t2) -- (k3);
\draw[-,thick] (t2) -- (k4);
\draw[thick,scale=2] (t1) to [bend left=60] (t2);
\draw[thick,scale=2] (t1) to [bend left=-60] (t2);
\draw[thick,scale=3] (t1)  to[in=-90,out=-130,loop] (t1);
\draw[thick,scale=3] (t1)  to[in=-180,out=-220,loop] (t1);
\draw[thick,scale=3] (t2)  to[in=-90,out=-50,loop] (t2);
\draw[thick,scale=3] (t2)  to[in=0,out=40,loop] (t2);
\filldraw[color=black, fill=black, thick] (t1) circle (0.9ex);
\filldraw[color=black, fill=black, thick] (t2) circle (0.9ex);
\filldraw[color=black, fill=white, thick] ($(k1)-(\arista,\arista)$) rectangle ($(k1)+(\arista,\arista)$);
\filldraw[color=black, fill=white, thick] ($(k2)-(\arista,\arista)$) rectangle ($(k2)+(\arista,\arista)$);
\filldraw[color=black, fill=white, thick] ($(k3)-(\arista,\arista)$) rectangle ($(k3)+(\arista,\arista)$);
\filldraw[color=black, fill=white, thick] ($(k4)-(\arista,\arista)$) rectangle ($(k4)+(\arista,\arista)$);
\coordinate (Q) at ($(0ex,-7.0ex)$);
\coordinate (P) at ($(-8ex,-8.5ex)$);
\coordinate (K) at ($(8ex,-8.5ex)$);
\node at ($(P)$) [anchor=east]{{\footnotesize{$L_1$}}};
\filldraw[color=black, fill=black, thick] ($(P)+(-0.3ex,0.9ex)$) circle (0.1ex);
\filldraw[color=black, fill=black, thick] (P) circle (0.1ex);
\filldraw[color=black, fill=black, thick] ($(P)+(0.5ex,-0.8ex)$) circle (0.1ex);
\node at ($(K)$) [anchor=west]{{\footnotesize{$L_2$}}};
\filldraw[color=black, fill=black, thick] ($(K)+(0.3ex,0.9ex)$) circle (0.1ex);
\filldraw[color=black, fill=black, thick] (K) circle (0.1ex);
\filldraw[color=black, fill=black, thick] ($(K)+(-0.5ex,-0.8ex)$) circle (0.1ex);
\filldraw[color=black, fill=black, thick] ($(Q)+(0ex,1ex)$) circle (0.1ex);
\filldraw[color=black, fill=black, thick] (Q) circle (0.1ex);
\filldraw[color=black, fill=black, thick] ($(Q)+(0ex,-1ex)$) circle (0.1ex);
\node at ($(0ex,-9.5ex)$) [anchor=north]{\footnotesize{$L+1$}};
\node at ($(k1)+(0ex,0.5ex)$) [anchor=south]{\footnotesize{$\k_1$}};
\node at ($(k2)+(0ex,0.5ex)$) [anchor=south]{\footnotesize{$\k_{n_1}$}};
\node at ($(k3)+(0ex,0.5ex)$) [anchor=south]{\footnotesize{$\k_{n_1+1}$}};
\node at ($(k4)+(0ex,0.5ex)$) [anchor=south]{\footnotesize{$\k_{n}$}};
}
}
\def\diagramextwoverticestwo{\tikz[baseline=-1.4ex]{
\coordinate (k1) at (-9ex,0ex);
\coordinate (k2) at (-3.ex,0ex);
\coordinate (k3) at (+3.ex,0ex);
\coordinate (k4) at (+9.ex,0ex);
\coordinate (t1) at (-4.5ex,-7ex);
\coordinate (t2) at (+4.5ex,-7ex);
\pgfmathsetmacro{\arista}{0.1}
\draw[thick] (-11ex,0ex) -- (11ex,0ex);
\draw[-,thick] (t1) -- (k1);
\draw[-,thick] (t1) -- (k2);
\draw[-,thick] (t2) -- (k3);
\draw[-,thick] (t2) -- (k4);
\draw[thick,scale=2] (t1) to [bend left=60] (t2);
\draw[thick,scale=2] (t1) to [bend left=-60] (t2);
\draw[thick,scale=3] (t1)  to[in=-90,out=-130,loop] (t1);
\draw[thick,scale=3] (t1)  to[in=-180,out=-220,loop] (t1);
\draw[thick,scale=3] (t2)  to[in=-90,out=-50,loop] (t2);
\draw[thick,scale=3] (t2)  to[in=0,out=40,loop] (t2);
\filldraw[color=black, fill=black, thick] (t1) circle (0.9ex);
\filldraw[color=black, fill=white, thick] (t2) circle (0.9ex);
\filldraw[color=black, fill=white, thick] ($(k1)-(\arista,\arista)$) rectangle ($(k1)+(\arista,\arista)$);
\filldraw[color=black, fill=white, thick] ($(k2)-(\arista,\arista)$) rectangle ($(k2)+(\arista,\arista)$);
\filldraw[color=black, fill=white, thick] ($(k3)-(\arista,\arista)$) rectangle ($(k3)+(\arista,\arista)$);
\filldraw[color=black, fill=white, thick] ($(k4)-(\arista,\arista)$) rectangle ($(k4)+(\arista,\arista)$);
\coordinate (Q) at ($(0ex,-7.0ex)$);
\coordinate (P) at ($(-8ex,-8.5ex)$);
\coordinate (K) at ($(8ex,-8.5ex)$);
\node at ($(P)$) [anchor=east]{{\footnotesize{$L_1$}}};
\filldraw[color=black, fill=black, thick] ($(P)+(-0.3ex,0.9ex)$) circle (0.1ex);
\filldraw[color=black, fill=black, thick] (P) circle (0.1ex);
\filldraw[color=black, fill=black, thick] ($(P)+(0.5ex,-0.8ex)$) circle (0.1ex);
\node at ($(K)$) [anchor=west]{{\footnotesize{$L_2$}}};
\filldraw[color=black, fill=black, thick] ($(K)+(0.3ex,0.9ex)$) circle (0.1ex);
\filldraw[color=black, fill=black, thick] (K) circle (0.1ex);
\filldraw[color=black, fill=black, thick] ($(K)+(-0.5ex,-0.8ex)$) circle (0.1ex);
\filldraw[color=black, fill=black, thick] ($(Q)+(0ex,1ex)$) circle (0.1ex);
\filldraw[color=black, fill=black, thick] (Q) circle (0.1ex);
\filldraw[color=black, fill=black, thick] ($(Q)+(0ex,-1ex)$) circle (0.1ex);
\node at ($(0ex,-9.5ex)$) [anchor=north]{\footnotesize{$L+1$}};
\node at ($(k1)+(0ex,0.5ex)$) [anchor=south]{\footnotesize{$\k_1$}};
\node at ($(k2)+(0ex,0.5ex)$) [anchor=south]{\footnotesize{$\k_{n_1}$}};
\node at ($(k3)+(0ex,0.5ex)$) [anchor=south]{\footnotesize{$\k_{n_1+1}$}};
\node at ($(k4)+(0ex,0.5ex)$) [anchor=south]{\footnotesize{$\k_{n}$}};
}
}
\be \label{D-diagrams-2-vertices}
\diagramextwoverticesNoLabels  = 2 {\rm Re} \Bigg\{ \diagramextwoverticesone  +  \diagramextwoverticestwo \Bigg\} .
\ee
\end{widetext}
Such a diagram admits $L_1$ loops starting and ending on the first vertex $\tau_1$ and $L_2$ loops starting and ending on the second vertex $\tau_2$, together with $L$ loops formed by propagators joining together the two vertices $\tau_1$ and $\tau_2$. With this configuration, the interaction couplings setting the strength of the two vertices are $\lambda_{n_1 + 2 L_1 + L +1}$ and $\lambda_{n_2 + 2 L_2 + L+1}$ respectively. Using the Feynman rules outlined in the previous section, this translates to
\begin{widetext}
\def\diagramextwoverticesNoLabels{\tikz[baseline=-1.4ex]{
\coordinate (k1) at (-9ex,0ex);
\coordinate (k2) at (-3.ex,0ex);
\coordinate (k3) at (+3.ex,0ex);
\coordinate (k4) at (+9.ex,0ex);
\coordinate (t1) at (-4.5ex,-7ex);
\coordinate (t2) at (+4.5ex,-7ex);
\pgfmathsetmacro{\arista}{0.1}
\draw[thick] (-11ex,0ex) -- (11ex,0ex);
\draw[-,thick] (t1) -- (k1);
\draw[-,thick] (t1) -- (k2);
\draw[-,thick] (t2) -- (k3);
\draw[-,thick] (t2) -- (k4);
\draw[thick,scale=2] (t1) to [bend left=60] (t2);
\draw[thick,scale=2] (t1) to [bend left=-60] (t2);
\draw[thick,scale=3] (t1)  to[in=-90,out=-130,loop] (t1);
\draw[thick,scale=3] (t1)  to[in=-180,out=-220,loop] (t1);
\draw[thick,scale=3] (t2)  to[in=-90,out=-50,loop] (t2);
\draw[thick,scale=3] (t2)  to[in=0,out=40,loop] (t2);
\fill[white] (t1) circle (0.9ex);
\filldraw[pattern=north east lines, thick] (t1) circle (0.9ex);
\fill[white] (t2) circle (0.9ex);
\filldraw[pattern=north east lines, thick] (t2) circle (0.9ex);
\filldraw[color=black, fill=white, thick] ($(k1)-(\arista,\arista)$) rectangle ($(k1)+(\arista,\arista)$);
\filldraw[color=black, fill=white, thick] ($(k2)-(\arista,\arista)$) rectangle ($(k2)+(\arista,\arista)$);
\filldraw[color=black, fill=white, thick] ($(k3)-(\arista,\arista)$) rectangle ($(k3)+(\arista,\arista)$);
\filldraw[color=black, fill=white, thick] ($(k4)-(\arista,\arista)$) rectangle ($(k4)+(\arista,\arista)$);
\coordinate (Q) at ($(0ex,-7.0ex)$);
\coordinate (P) at ($(-8ex,-8.5ex)$);
\coordinate (K) at ($(8ex,-8.5ex)$);
\node at ($(P)$) [anchor=east]{{\footnotesize{$L_1$}}};
\filldraw[color=black, fill=black, thick] ($(P)+(-0.3ex,0.9ex)$) circle (0.1ex);
\filldraw[color=black, fill=black, thick] (P) circle (0.1ex);
\filldraw[color=black, fill=black, thick] ($(P)+(0.5ex,-0.8ex)$) circle (0.1ex);
\node at ($(K)$) [anchor=west]{{\footnotesize{$L_2$}}};
\filldraw[color=black, fill=black, thick] ($(K)+(0.3ex,0.9ex)$) circle (0.1ex);
\filldraw[color=black, fill=black, thick] (K) circle (0.1ex);
\filldraw[color=black, fill=black, thick] ($(K)+(-0.5ex,-0.8ex)$) circle (0.1ex);
\filldraw[color=black, fill=black, thick] ($(Q)+(0ex,1ex)$) circle (0.1ex);
\filldraw[color=black, fill=black, thick] (Q) circle (0.1ex);
\filldraw[color=black, fill=black, thick] ($(Q)+(0ex,-1ex)$) circle (0.1ex);
\node at ($(0ex,-9.5ex)$) [anchor=north]{\footnotesize{$L+1$}};
\node at ($(k1)+(0ex,0.5ex)$) [anchor=south]{\footnotesize{$\k_1$}};
\node at ($(k2)+(0ex,0.5ex)$) [anchor=south]{\footnotesize{$\k_{n_1}$}};
\node at ($(k3)+(0ex,0.5ex)$) [anchor=south]{\footnotesize{$\k_{n_1+1}$}};
\node at ($(k4)+(0ex,0.5ex)$) [anchor=south]{\footnotesize{$\k_{n}$}};
}
}
\bea \label{two-vertex-diagram}
\diagramextwoverticesNoLabels \!\!\!\!\!\!
&=& - \frac{\lambda_{n_1+2L_1+L+1} \lambda_{n_2+2L_2+L+1}}{(L+1)! L_1! L_2! 2^{L_1} 2^{L_2} H^8} (2\pi)^6 \int^{\tau}_{-\infty} \!\! \frac{d\tau_1}{\tau_1^4}  \int^{\tau}_{-\infty} \!\! \frac{d\tau_2}{\tau_2^4} \int_{{\bf q}_1} \cdots \int_{{\bf q}_{L+1}} \nn \\ [-55pt]  
&& \times
\left[\int_{{\bf l}_1} G( \tau_1, \tau_1,l_1)\right]^{L_1} \left[\int_{{\bf l}_2} G( \tau_2, \tau_2,l_2)\right]^{L_2}  \delta^{(3)} ({\bf K}_1 + {\bf Q}) \delta^{(3)} ({\bf K}_2 - {\bf Q})  \nn \\ 
 &&  \times 2 {\rm Re} \Bigg\{ \bigg( G_{+} (k_1 , \tau_1 , \tau) \cdots  G_{+} (k_{n_1} , \tau_1 , \tau)   
  G_{+} (k_{n_1+1} , \tau_2 , \tau) \cdots  G_{+} (k_{n} , \tau_2 , \tau)  \\ 
&& \times G_{++} (q_1 , \tau_1 , \tau_2) \cdots G_{++} (q_{L+1} , \tau_1 , \tau_2) \bigg)  - \bigg( G_{+} (k_1 , \tau_1 , \tau) \cdots  G_{+} (k_{n_1} , \tau_1 , \tau)   
 \nn \\
&& \times G_{-} (k_{n_1+1} , \tau_2 , \tau) \cdots  G_{-} (k_{n} , \tau_2 , \tau)  G_{+-} (q_1 , \tau_1 , \tau_2) \cdots G_{+-} (q_{L+1} , \tau_1 , \tau_2) \bigg) \Bigg\} , \nn  \qquad
\eea

\end{widetext}
where ${\bf K}_1 = \k_1 + \cdots + \k_{n_1}$ and ${\bf K}_2 = \k_{n_1 + 1} + \cdots + \k_{n}$ represent the total external momentum flowing into the vertices $\tau_1$ and $\tau_2$ respectively. In addition, ${\bf Q} = \q_1 + \cdots + \q_{L+1}$ is the total momentum carried by bulk-to-bulk propagators laying between $\tau_1$ and $\tau_2$. Notice that in the particular case where both vertices share the same number of legs together with the same number of external legs attached to them, that is $n_1= n_2$ and $L_1 = L_2$, then the expression above must be corrected to contain an additional factor $1/2$ due to the implied extra symmetry of such a configuration. 

Note that the product $\delta^{(3)} ({\bf K}_1 + {\bf Q}) \delta^{(3)} ({\bf K}_2 - {\bf Q})$ implies that, after integrating the momentum of any of the bulk-to-bulk propagators between $\tau_1$ and $\tau_2$, the diagram will be proportional to an overall delta function $\delta^{(3)} ({\bf K}_1 + {\bf K}_2)$, implying that the diagram conserves the total external momentum. It is direct to see that this property is preserved for any diagram, independently of the number of vertices or loops.

\subsection{Dilation symmetry of $n$-point functions} 
\label{sec:dilation-sym}

Having the Schwinger-Keldysh rules at our disposal, we are in a condition to check that equal-time correlation functions, computed to any desired order in perturbation theory, are dilation invariant. This will rely on the validity of Wilson's axioms which, as discussed in Section~\ref{sec:div-ints-Wilson}, must be in play in the case of theories with non-derivative interactions. Let us start by assessing how connected correlation functions in momentum space $ \tilde G^{(n)}_c ( {\k}_1 , \cdots , {\k}_n ; \tau )$ transform under dilations. For this, notice first that the reciprocal effect of the dilation transformation (\ref{additional-symmetry-2}) in momentum space, takes the form:
\be
\k_ i \to \bar \k_i = e^{\theta} \k_ i .
\ee
Then, a simple inspection of the various elements participating in the Feynman rules introduced in Section~\ref{sec:Feynman} reveals their scaling properties under the transformation (\ref{additional-symmetry-1}). For instance, by replacing $\tau = e^{\theta} \bar \tau$ in (\ref{vertex-black}) and (\ref{vertex-white}), it follows that 
\be \label{vertex-invariant}
V_{\pm} ( \tau , \sum_i \k_{i}) [\cdots] = V_{\pm} (  \bar \tau , \sum_i \bar \k_{i}) [\cdots] .
\ee
On the other hand, every propagator, either bulk-to-bulk or bulk-to-boundary, respects the rule:
\be \label{scaling-props}
G (k , \tau_a , \tau_b) = e^{ 3 \theta} G (\bar k , \bar \tau_a , \bar \tau_b) .
\ee
Regardless of the number of loops, in any term contributing to $ \tilde G^n_c ( {\k}_1 , \cdots , {\k}_n ; \tau )$, there will be as many momentum-integrals as internal lines. This means that the scaling of (\ref{scaling-props}) for internal lines is canceled by the change of variables $\int \frac{d^3 k}{(2 \pi)^3} = e^{-3 \theta} \int \frac{d^3 \bar k}{(2 \pi)^3}$ ensured by second Wilson's axiom (\ref{int-scaling}). This further implies that only external lines contribute to the scaling properties of $\tilde G^n_c ( {\k}_1 , \cdots , {\k}_n ; \tau )$ under dilations, which is found to be
\be \label{n-point-momentum-scaling}
\tilde G^{(n)}_c ( {\k}_1 , \cdots , {\k}_n ; \tau ) = e^{3 n \theta} \tilde G^{(n)}_c ( \bar {\k}_1 , \cdots , \bar {\k}_n ; \bar \tau ) .
\ee
We would like to emphasize, once again, the importance of second Wilson's axiom (\ref{int-scaling}) in order to reach this conclusion.  

Next, recall the relation between equal-time connected $n$-point correlation functions in coordinate and momentum space, given in Eq.~(\ref{G-n-coordinate}). Then, from the fact that under dilations one has $\int \frac{d^3 k}{(2 \pi)^3} = e^{-3 \theta} \int \frac{d^3 \bar k}{(2 \pi)^3}$ together with the property $\x\cdot \k = \bar \x \cdot \bar \k$, it immediately follows that
\be
G^{(n)}(\x_1 , \cdots \x_n ; \tau) = G^{(n)}(\bar \x_1 , \cdots \bar \x_n ; \bar \tau) , \label{G-n-dilation}
\ee
where $\bar \x_i = e^{-\theta} \x_i$. Thus, as anticipated, correlation functions in coordinates space are invariant under the dilation transformation (\ref{additional-symmetry-1})-(\ref{additional-symmetry-2}). 

This result imposes a strong restriction on the coordinate-dependence of $n$-point correlators. Recall that $\tilde G^{(n)}_c ( {\k}_1 , \cdots , {\k}_n ; \tau )$ must be proportional to an overall delta function $\delta^{3} (\k_1 + \cdots + \k_n)$. This means that $G^{(n)}(\x_1 , \cdots \x_n ; \tau)$ can only depend on spatial coordinates through the difference between pairs of positions $|\x_i - \x_j|$. In other words, correlation functions must be invariant under rotations and spatial translations. Then, by choosing $e^{\theta} = 1/a(\tau)$ in (\ref{G-n-dilation}) we end up with 
\be
G^{(n)}(\x_1 , \cdots \x_n ; \tau) = G^{(n)}(\x_1 a(\tau) , \cdots , \x_n a(\tau)  ; -H^{-1}) . 
\ee
This result further implies that $G^{(n)}(\x_1 , \cdots \x_n , \tau)$ must be a function of combinations $|(\x_i - \x_j) / \tau|^2$, which are de Sitter invariant. Note that this is true at any order in perturbation theory, independently of the number of loops. Furthermore, evaluating the $n$-point function at coincident point, one finally learns that:
\be \label{G-n-general}
G^{n}(\x , \cdots \x ; \tau) = G^{n}(0 , \cdots 0 ; - H^{-1}),
\ee
which is independent of time, regardless of the number of vertices and/or loops taking place in the diagrams contributing to the $n$-point function.

\subsection{Dilation symmetry and effective field theory}

A supporting viewpoint, complementing the analysis of the previous subsection, is offered by the Wilsonian approach to effective field theory (EFT).

Suppose we count with an EFT defined over a finite range of momenta, limited by IR and UV cutoffs $\Lambda_{\rm IR}$ and $\Lambda_{\rm UV}$, and allowing the computation of $n$-point correlation functions $\tilde G^{(n)}_c ( {\k}_1 , \cdots , {\k}_n ; \tau )$ in momentum space. One of the main features of such an EFT is that loop corrections are finite, simply because any integration of momentum must be performed between the two cutoffs $\Lambda_{\rm IR}$ and $\Lambda_{\rm UV}$. This is because loop contributions coming from momenta outside this range are already encoded in the Wilsonian coefficients of the EFT, which must be finite in order to yield finite observables.

A crucial aspect underlying this EFT approach, is that the computation of $n$-point correlation functions $\tilde G^{(n)}_c ( {\k}_1 , \cdots , {\k}_n ; \tau )$ in momentum space must coincide with the full computation outlined in the previous sections. If this was not the case, the EFT would not serve any purpose. Another way to say this, is that the results must be independent of the chosen values for $\Lambda_{\rm IR}$ and $\Lambda_{\rm UV}$ employed to perform integrals for internal momenta. For this to be possible, the Wilsonian coefficients appearing in the theory (for instance the couplings $\lambda_n$) must depend on the cutoffs in such a way to make the resulting $n$-point functions cutoff independent. The previous statement implies that $n$-point functions $\tilde G^{(n)}_c ( {\k}_1 , \cdots , {\k}_n ; \tau )$ computed within the EFT approach must preserve the scaling property (\ref{n-point-momentum-scaling}).

As already mentioned, the new Feynman rules derived within this EFT specify that internal momentum-integrals must be integrated within a finite range of momenta. This introduces an important criterion of how to choose the type of cutoffs employed to limit integrals. For instance, physical cutoffs are  explicitly invariant under dilations, implying that the scaling property (\ref{n-point-momentum-scaling}) is trivially preserved when working with the new rules. 

It is pertinent to mention that, from this perspective, it makes little difference what IR cutoff one chooses to work with. In the popular case of a comoving regulator, the dilation symmetry is broken. Since the theory is de Sitter invariant, the dilation symmetry must be restored in some way in order to preserve the scaling property (\ref{n-point-momentum-scaling}). The way to preserve this property is by having time-dependent Wilsonian coefficients, with the right time dependence to cancel out the choice of IR cutoff~\cite{Huenupi:2024ksc}. Failure to adopt this step leads to the appearance of secular growth from loops in correlation functions.

\subsection{Split propagators}
\label{sec:split-propagators}

The fact that the final sum of diagrams yields a real quantity implies that much of the information contained by a simple diagram does not have an observable impact. Consider the splitting of the Wightman function $G(k,\tau_a,\tau_b)$ defined in (\ref{G-basic-prop}) into real and imaginary parts:
\be \label{split-G}
G(k,\tau_a,\tau_b) = G_R(k,\tau_a,\tau_b) + i \, G_I(k,\tau_a,\tau_b) .
\ee
The functions $G_R(k,\tau_a,\tau_b)$ and $G_I(k,\tau_a,\tau_b)$ are given by the following expressions:
\bea
G_R(k,\tau_a,\tau_b) &=& \frac{H^2}{2 k^3} \Big[ k (\tau_a - \tau_b) \sin \big(k (\tau_a - \tau_b) \big) \nn \\
&& + (1 + k^2 \tau_a \tau_b )\cos \big(k (\tau_a - \tau_b) \big)\Big] , \qquad  \\
G_R (k,\tau_a,\tau_b) &=& \frac{H^2}{2 k^3} \Big[ k (\tau_a - \tau_b) \cos \big(k (\tau_a - \tau_b) \big) \nn \\
&& - (1 + k^2 \tau_a \tau_b )\sin \big(k (\tau_a - \tau_b) \big)\Big] . \qquad 
\eea
Notice that $G_R(k,\tau_a,\tau_b)$ and $G_I(k,\tau_a,\tau_b)$ are respectively even and odd functions under the interchange of $\tau_a$ and $\tau_b$. In the limit $k |\tau_a| \ll 1 $ and $k |\tau_b| \ll 1 $ these functions respectively become
\bea
G_R(k,\tau_a,\tau_b) &=& \frac{H^2}{2 k^3} \Big[ 1 + \cdots\Big] , \qquad  \label{prop-re}  \\
G_I (k,\tau_a,\tau_b) &=& - \frac{H^2}{2 k^3} \Big[ \frac{
k^3}{3}(\tau_a^3 - \tau_b^3)  + \cdots \Big] , \qquad   \label{prop-im} 
\eea
where the ellipses denote subleading terms with respect to $k \tau_a$ and $k \tau_b$. The splitting of $G(k,\tau_a,\tau_b)$ defined in (\ref{split-G}) leads to the splitting of all propagators into real and imaginary parts. For instance, bulk-to-bulk propagators are found to acquire the following forms:
\bea
G_{++}(\tau_a,\tau_b) &=& G_R(\tau_a,\tau_b) + i \, G_I(\tau_a,\tau_b) I (\tau_a , \tau_b) , \qquad \\ 
G_{--}(\tau_a,\tau_b) &=& G_R(\tau_a,\tau_b) - i \, G_I(\tau_a,\tau_b) I (\tau_a , \tau_b) , \qquad \\ 
G_{+-}(\tau_a,\tau_b) &=& G_R(\tau_a,\tau_b) - i \, G_I(\tau_a,\tau_b)  , \qquad \\ 
G_{-+}(\tau_a,\tau_b) &=& G_R(\tau_a,\tau_b) + i \, G_I(\tau_a,\tau_b)  , \qquad 
\eea
where we have defined the function $I (\tau_a , \tau_b)$ as:
\be
I (\tau_a , \tau_b) \equiv \theta(\tau_a - \tau_b) - \theta(\tau_b - \tau_a) .
\ee
Notice that $I ( \tau_a , \tau_b)$ is an odd function under the interchange of $\tau_a$ and $\tau_b$. With these definitions in mind, let us decompose the diagrammatic representation of propagators as
\def\propbb{\tikz[baseline=-0.6ex,scale=1.8, every node/.style={scale=1.4}]{
\coordinate (tau1) at (-2.5ex,0ex);
\coordinate (tau2) at (2.5ex,0ex);
\draw[thick] (tau1) -- (tau2);
\filldraw[color=black, fill=black, thick] (tau1) circle (0.5ex);
\node[anchor=south] at ($(tau1)+(0,0.5ex)$) {\scriptsize{$\tau_a$}};
\filldraw[color=black, fill=black, thick] (tau2) circle (0.5ex);
\node[anchor=south] at ($(tau2)+(0,0.5ex)$) {\scriptsize{$\tau_b$}};
}
}
\def\propww{\tikz[baseline=-0.6ex,scale=1.8, every node/.style={scale=1.4}]{
\coordinate (tau1) at (-2.5ex,0ex);
\coordinate (tau2) at (2.5ex,0ex);
\draw[thick] (tau1) -- (tau2);
\filldraw[color=black, fill=white, thick] (tau1) circle (0.5ex);
\node[anchor=south] at ($(tau1)+(0,0.5ex)$) {\scriptsize{$\tau_a$}};
\filldraw[color=black, fill=white, thick] (tau2) circle (0.5ex);
\node[anchor=south] at ($(tau2)+(0,0.5ex)$) {\scriptsize{$\tau_b$}};
}
}
\def\propbw{\tikz[baseline=-0.6ex,scale=1.8, every node/.style={scale=1.4}]{
\coordinate (tau1) at (-2.5ex,0ex);
\coordinate (tau2) at (2.5ex,0ex);
\draw[thick] (tau1) -- (tau2);
\filldraw[color=black, fill=black, thick] (tau1) circle (0.5ex);
\node[anchor=south] at ($(tau1)+(0,0.5ex)$) {\scriptsize{$\tau_a$}};
\filldraw[color=black, fill=white, thick] (tau2) circle (0.5ex);
\node[anchor=south] at ($(tau2)+(0,0.5ex)$) {\scriptsize{$\tau_b$}};
}
}
\def\propwb{\tikz[baseline=-0.6ex,scale=1.8, every node/.style={scale=1.4}]{
\coordinate (tau1) at (-2.5ex,0ex);
\coordinate (tau2) at (2.5ex,0ex);
\draw[thick] (tau1) -- (tau2);
\filldraw[color=black, fill=white, thick] (tau1) circle (0.5ex);
\node[anchor=south] at ($(tau1)+(0,0.5ex)$) {\scriptsize{$\tau_a$}};
\filldraw[color=black, fill=black, thick] (tau2) circle (0.5ex);
\node[anchor=south] at ($(tau2)+(0,0.5ex)$) {\scriptsize{$\tau_b$}};
}
}
\def\imbb{\tikz[baseline=-0.6ex,scale=1.8, every node/.style={scale=1.4}]{
\coordinate (tau1) at (-2.5ex,0ex);
\coordinate (tau2) at (2.5ex,0ex);
\draw[thick, dashed] (tau1) -- (tau2);
\filldraw[color=black, fill=black, thick] (tau1) circle (0.5ex);
\node[anchor=south] at ($(tau1)+(0,0.5ex)$) {\scriptsize{$\tau_a$}};
\filldraw[color=black, fill=black, thick] (tau2) circle (0.5ex);
\node[anchor=south] at ($(tau2)+(0,0.5ex)$) {\scriptsize{$\tau_b$}};
}
}
\def\rebb{\tikz[baseline=-0.6ex,scale=1.8, every node/.style={scale=1.4}]{
\coordinate (tau1) at (-2.5ex,0ex);
\coordinate (tau2) at (2.5ex,0ex);
\draw[thick, double] (tau1) -- (tau2);
\filldraw[color=black, fill=black, thick] (tau1) circle (0.5ex);
\node[anchor=south] at ($(tau1)+(0,0.5ex)$) {\scriptsize{$\tau_a$}};
\filldraw[color=black, fill=black, thick] (tau2) circle (0.5ex);
\node[anchor=south] at ($(tau2)+(0,0.5ex)$) {\scriptsize{$\tau_b$}};
}
}
\def\imww{\tikz[baseline=-0.6ex,scale=1.8, every node/.style={scale=1.4}]{
\coordinate (tau1) at (-2.5ex,0ex);
\coordinate (tau2) at (2.5ex,0ex);
\draw[thick, dashed] (tau1) -- (tau2);
\filldraw[color=black, fill=white, thick] (tau1) circle (0.5ex);
\node[anchor=south] at ($(tau1)+(0,0.5ex)$) {\scriptsize{$\tau_a$}};
\filldraw[color=black, fill=white, thick] (tau2) circle (0.5ex);
\node[anchor=south] at ($(tau2)+(0,0.5ex)$) {\scriptsize{$\tau_b$}};
}
}
\def\reww{\tikz[baseline=-0.6ex,scale=1.8, every node/.style={scale=1.4}]{
\coordinate (tau1) at (-2.5ex,0ex);
\coordinate (tau2) at (2.5ex,0ex);
\draw[thick, double] (tau1) -- (tau2);
\filldraw[color=black, fill=white, thick] (tau1) circle (0.5ex);
\node[anchor=south] at ($(tau1)+(0,0.5ex)$) {\scriptsize{$\tau_a$}};
\filldraw[color=black, fill=white, thick] (tau2) circle (0.5ex);
\node[anchor=south] at ($(tau2)+(0,0.5ex)$) {\scriptsize{$\tau_b$}};
}
}
\def\imbw{\tikz[baseline=-0.6ex,scale=1.8, every node/.style={scale=1.4}]{
\coordinate (tau1) at (-2.5ex,0ex);
\coordinate (tau2) at (2.5ex,0ex);
\draw[thick, dashed] (tau1) -- (tau2);
\filldraw[color=black, fill=black, thick] (tau1) circle (0.5ex);
\node[anchor=south] at ($(tau1)+(0,0.5ex)$) {\scriptsize{$\tau_a$}};
\filldraw[color=black, fill=white, thick] (tau2) circle (0.5ex);
\node[anchor=south] at ($(tau2)+(0,0.5ex)$) {\scriptsize{$\tau_b$}};
}
}
\def\rebw{\tikz[baseline=-0.6ex,scale=1.8, every node/.style={scale=1.4}]{
\coordinate (tau1) at (-2.5ex,0ex);
\coordinate (tau2) at (2.5ex,0ex);
\draw[thick, double] (tau1) -- (tau2);
\filldraw[color=black, fill=black, thick] (tau1) circle (0.5ex);
\node[anchor=south] at ($(tau1)+(0,0.5ex)$) {\scriptsize{$\tau_a$}};
\filldraw[color=black, fill=white, thick] (tau2) circle (0.5ex);
\node[anchor=south] at ($(tau2)+(0,0.5ex)$) {\scriptsize{$\tau_b$}};
}
}
\def\imwb{\tikz[baseline=-0.6ex,scale=1.8, every node/.style={scale=1.4}]{
\coordinate (tau1) at (-2.5ex,0ex);
\coordinate (tau2) at (2.5ex,0ex);
\draw[thick, dashed] (tau1) -- (tau2);
\filldraw[color=black, fill=white, thick] (tau1) circle (0.5ex);
\node[anchor=south] at ($(tau1)+(0,0.5ex)$) {\scriptsize{$\tau_a$}};
\filldraw[color=black, fill=black, thick] (tau2) circle (0.5ex);
\node[anchor=south] at ($(tau2)+(0,0.5ex)$) {\scriptsize{$\tau_b$}};
}
}
\def\rewb{\tikz[baseline=-0.6ex,scale=1.8, every node/.style={scale=1.4}]{
\coordinate (tau1) at (-2.5ex,0ex);
\coordinate (tau2) at (2.5ex,0ex);
\draw[thick, double] (tau1) -- (tau2);
\filldraw[color=black, fill=white, thick] (tau1) circle (0.5ex);
\node[anchor=south] at ($(tau1)+(0,0.5ex)$) {\scriptsize{$\tau_a$}};
\filldraw[color=black, fill=black, thick] (tau2) circle (0.5ex);
\node[anchor=south] at ($(tau2)+(0,0.5ex)$) {\scriptsize{$\tau_b$}};
}
}
\bea
\propbb &=& \rebb + \imbb , \qquad \quad \\
\propww &=& \reww + \imww ,  \qquad \quad \\
\propbw &=& \rebw + \imbw , \qquad \quad\\
\propwb &=& \rewb + \imwb , \qquad \quad
\eea
where double lines stand for the real part and dashed lines denote their imaginary part. Similarly, bulk-to-boundary propagators are found to be given by:
\bea
G_{+} ( k, \tau_a , \tau)  &=& G_R (k , \tau_a , \tau) -  i G_I (k , \tau_a , \tau)  , \\
G_{-} (k, \tau_a , \tau)  &=& G_R (k , \tau_a , \tau) + i G_I (k , \tau_a , \tau)  . \quad
\eea
The diagrammatic representations for these bulk-to-boundary split propagators are
\def\propbf{\tikz[baseline=-0.6ex,scale=1.8, every node/.style={scale=1.4}]{
\coordinate (tau) at (-2.5ex,0ex);
\coordinate (phi) at (2.5ex,0ex);
\draw[thick] (tau) -- (phi);
\filldraw[color=black, fill=black, thick] (tau) circle (0.5ex);
\node[anchor=south] at ($(tau)+(0,0.5ex)$) {\scriptsize{$\tau_a$}};
\pgfmathsetmacro{\arista}{0.06}
\filldraw[color=black, fill=white, thick] ($(phi)-(\arista,\arista)$) rectangle ($(phi)+(\arista,\arista)$);
\node[anchor=south] at ($(phi)+(0,0.5ex)$){\scriptsize{$\tau$}};
}
}
\def\propwf{\tikz[baseline=-0.6ex,scale=1.8, every node/.style={scale=1.4}]{
\coordinate (tau) at (-2.5ex,0ex);
\coordinate (phi) at (2.5ex,0ex);
\draw[thick] (tau) -- (phi);
\filldraw[color=black, fill=white, thick] (tau) circle (0.5ex);
\node[anchor=south] at ($(tau)+(0,0.5ex)$) {\scriptsize{$\tau_a$}};
\pgfmathsetmacro{\arista}{0.06}
\filldraw[color=black, fill=white, thick] ($(phi)-(\arista,\arista)$) rectangle ($(phi)+(\arista,\arista)$);
\node[anchor=south] at ($(phi)+(0,0.5ex)$){\scriptsize{$\tau$}};
}
}
\def\propbfim{\tikz[baseline=-0.6ex,scale=1.8, every node/.style={scale=1.4}]{
\coordinate (tau) at (-2.5ex,0ex);
\coordinate (phi) at (2.5ex,0ex);
\draw[thick, dashed] (tau) -- (phi);
\filldraw[color=black, fill=black, thick] (tau) circle (0.5ex);
\node[anchor=south] at ($(tau)+(0,0.5ex)$) {\scriptsize{$\tau_a$}};
\pgfmathsetmacro{\arista}{0.06}
\filldraw[color=black, fill=white, thick] ($(phi)-(\arista,\arista)$) rectangle ($(phi)+(\arista,\arista)$);
\node[anchor=south] at ($(phi)+(0,0.5ex)$){\scriptsize{$\tau$}};
}
}
\def\propwfim{\tikz[baseline=-0.6ex,scale=1.8, every node/.style={scale=1.4}]{
\coordinate (tau) at (-2.5ex,0ex);
\coordinate (phi) at (2.5ex,0ex);
\draw[thick, dashed] (tau) -- (phi);
\filldraw[color=black, fill=white, thick] (tau) circle (0.5ex);
\node[anchor=south] at ($(tau)+(0,0.5ex)$) {\scriptsize{$\tau_a$}};
\pgfmathsetmacro{\arista}{0.06}
\filldraw[color=black, fill=white, thick] ($(phi)-(\arista,\arista)$) rectangle ($(phi)+(\arista,\arista)$);
\node[anchor=south] at ($(phi)+(0,0.5ex)$){\scriptsize{$\tau$}};
}
}
\def\propbfre{\tikz[baseline=-0.6ex,scale=1.8, every node/.style={scale=1.4}]{
\coordinate (tau) at (-2.5ex,0ex);
\coordinate (phi) at (2.5ex,0ex);
\draw[thick, double] (tau) -- (phi);
\filldraw[color=black, fill=black, thick] (tau) circle (0.5ex);
\node[anchor=south] at ($(tau)+(0,0.5ex)$) {\scriptsize{$\tau_a$}};
\pgfmathsetmacro{\arista}{0.06}
\filldraw[color=black, fill=white, thick] ($(phi)-(\arista,\arista)$) rectangle ($(phi)+(\arista,\arista)$);
\node[anchor=south] at ($(phi)+(0,0.5ex)$){\scriptsize{$\tau$}};
}
}
\def\propwfre{\tikz[baseline=-0.6ex,scale=1.8, every node/.style={scale=1.4}]{
\coordinate (tau) at (-2.5ex,0ex);
\coordinate (phi) at (2.5ex,0ex);
\draw[thick, double] (tau) -- (phi);
\filldraw[color=black, fill=white, thick] (tau) circle (0.5ex);
\node[anchor=south] at ($(tau)+(0,0.5ex)$) {\scriptsize{$\tau_a$}};
\pgfmathsetmacro{\arista}{0.06}
\filldraw[color=black, fill=white, thick] ($(phi)-(\arista,\arista)$) rectangle ($(phi)+(\arista,\arista)$);
\node[anchor=south] at ($(phi)+(0,0.5ex)$){\scriptsize{$\tau$}};
}
}
\bea
\propbf &=& \propbfre + \propbfim , \qquad \quad \\
\propwf &=& \propbfre + \propwfim .  \qquad \quad 
\eea

As shown in Ref.~\cite{Palma:2023idj} the appearance of real and imaginary propagators in diagrams is restricted to satisfy the following rules (see Appendix~\ref{app:split-prop} for a refined derivation of these statements):
\begin{itemize}
\item[I.] Every vertex must have at least one imaginary propagator attached to it. 

\item[II.] It is impossible to form a closed loop only with imaginary propagators.

\end{itemize}
Rule I implies that in splitting propagators into real and imaginary parts, the lowest number of imaginary propagators that a diagram of $V$ vertices can have is precisely $V$. On the other hand, rule II implies that at least one external leg must consist in an imaginary propagator. The following diagram offers an example:
\def\diagramextwoverticesNoLabels{\tikz[baseline=-1.4ex,scale=1.8, every node/.style={scale=1.4}]{
\coordinate (k1) at (-3ex,0ex);
\coordinate (k2) at (-1.ex,0ex);
\coordinate (k3) at (+3.ex,0ex);
\coordinate (t1) at (-2ex,-4ex);
\coordinate (t2) at (+2ex,-4ex);
\pgfmathsetmacro{\arista}{0.06}
\draw[thick] (-4ex,0ex) -- (4ex,0ex);
\draw[-,thick] (t1) -- (k1);
\draw[-,thick] (t1) -- (k2);
\draw[-,thick] (t2) -- (k3);
\draw[thick,scale=2] (t1) to [bend left=45] (t2);
\draw[thick,scale=2] (t1) to [bend left=-45] (t2);
\filldraw[color=black, fill=white, thick] ($(k1)-(\arista,\arista)$) rectangle ($(k1)+(\arista,\arista)$);
\filldraw[color=black, fill=white, thick] ($(k2)-(\arista,\arista)$) rectangle ($(k2)+(\arista,\arista)$);
\filldraw[color=black, fill=white, thick] ($(k3)-(\arista,\arista)$) rectangle ($(k3)+(\arista,\arista)$);
\fill[white] (t1) circle (0.45ex);
\filldraw[pattern=north east lines, thick] (t1) circle (0.45ex);
\fill[white] (t2) circle (0.45ex);
\filldraw[pattern=north east lines, thick] (t2) circle (0.45ex);
}
}
\def\diagramextwoverticesReImSplit{\tikz[baseline=-1.4ex,scale=1.8, every node/.style={scale=1.4}]{
\coordinate (k1) at (-3ex,0ex);
\coordinate (k2) at (-1.ex,0ex);
\coordinate (k3) at (+3.ex,0ex);
\coordinate (t1) at (-2ex,-4ex);
\coordinate (t2) at (+2ex,-4ex);
\pgfmathsetmacro{\arista}{0.06}
\draw[thick] (-4ex,0ex) -- (4ex,0ex);
\draw[-,thick,dashed] (t1) -- (k1);
\draw[-,thick,double] (t1) -- (k2);
\draw[-,thick,dashed] (t2) -- (k3);
\draw[thick,scale=2,double] (t1) to [bend left=45] (t2);
\draw[thick,scale=2,double] (t1) to [bend left=-45] (t2);
\filldraw[color=black, fill=white, thick] ($(k1)-(\arista,\arista)$) rectangle ($(k1)+(\arista,\arista)$);
\filldraw[color=black, fill=white, thick] ($(k2)-(\arista,\arista)$) rectangle ($(k2)+(\arista,\arista)$);
\filldraw[color=black, fill=white, thick] ($(k3)-(\arista,\arista)$) rectangle ($(k3)+(\arista,\arista)$);
\fill[white] (t1) circle (0.45ex);
\filldraw[pattern=north east lines, thick] (t1) circle (0.45ex);
\fill[white] (t2) circle (0.45ex);
\filldraw[pattern=north east lines, thick] (t2) circle (0.45ex);
}
}
\def\diagramextwoverticesTwoImExt{\tikz[baseline=-1.4ex,scale=1.8, every node/.style={scale=1.4}]{
\coordinate (k1) at (-3ex,0ex);
\coordinate (k2) at (-1.ex,0ex);
\coordinate (k3) at (+3.ex,0ex);
\coordinate (t1) at (-2ex,-4ex);
\coordinate (t2) at (+2ex,-4ex);
\pgfmathsetmacro{\arista}{0.06}
\draw[thick] (-4ex,0ex) -- (4ex,0ex);
\draw[-,thick,double] (t1) -- (k1);
\draw[-,thick,dashed] (t1) -- (k2);
\draw[-,thick,dashed] (t2) -- (k3);
\draw[thick,scale=2,double] (t1) to [bend left=45] (t2);
\draw[thick,scale=2,double] (t1) to [bend left=-45] (t2);
\filldraw[color=black, fill=white, thick] ($(k1)-(\arista,\arista)$) rectangle ($(k1)+(\arista,\arista)$);
\filldraw[color=black, fill=white, thick] ($(k2)-(\arista,\arista)$) rectangle ($(k2)+(\arista,\arista)$);
\filldraw[color=black, fill=white, thick] ($(k3)-(\arista,\arista)$) rectangle ($(k3)+(\arista,\arista)$);
\fill[white] (t1) circle (0.45ex);
\filldraw[pattern=north east lines, thick] (t1) circle (0.45ex);
\fill[white] (t2) circle (0.45ex);
\filldraw[pattern=north east lines, thick] (t2) circle (0.45ex);
}
}
\def\diagramextwoverticesThreeIm{\tikz[baseline=-1.4ex,scale=1.8, every node/.style={scale=1.4}]{
\coordinate (k1) at (-3ex,0ex);
\coordinate (k2) at (-1.ex,0ex);
\coordinate (k3) at (+3.ex,0ex);
\coordinate (t1) at (-2ex,-4ex);
\coordinate (t2) at (+2ex,-4ex);
\pgfmathsetmacro{\arista}{0.06}
\draw[thick] (-4ex,0ex) -- (4ex,0ex);
\draw[-,thick,double] (t1) -- (k1);
\draw[-,thick,double] (t1) -- (k2);
\draw[-,thick,dashed] (t2) -- (k3);
\draw[thick,scale=2,double] (t1) to [bend left=45] (t2);
\draw[thick,scale=2,dashed] (t1) to [bend left=-45] (t2);
\filldraw[color=black, fill=white, thick] ($(k1)-(\arista,\arista)$) rectangle ($(k1)+(\arista,\arista)$);
\filldraw[color=black, fill=white, thick] ($(k2)-(\arista,\arista)$) rectangle ($(k2)+(\arista,\arista)$);
\filldraw[color=black, fill=white, thick] ($(k3)-(\arista,\arista)$) rectangle ($(k3)+(\arista,\arista)$);
\fill[white] (t1) circle (0.45ex);
\filldraw[pattern=north east lines, thick] (t1) circle (0.45ex);
\fill[white] (t2) circle (0.45ex);
\filldraw[pattern=north east lines, thick] (t2) circle (0.45ex);
}
}
\def\diagramextwoverticesFourIm{\tikz[baseline=-1.4ex,scale=1.8, every node/.style={scale=1.4}]{
\coordinate (k1) at (-3ex,0ex);
\coordinate (k2) at (-1.ex,0ex);
\coordinate (k3) at (+3.ex,0ex);
\coordinate (t1) at (-2ex,-4ex);
\coordinate (t2) at (+2ex,-4ex);
\pgfmathsetmacro{\arista}{0.06}
\draw[thick] (-4ex,0ex) -- (4ex,0ex);
\draw[-,thick,double] (t1) -- (k1);
\draw[-,thick,double] (t1) -- (k2);
\draw[-,thick,dashed] (t2) -- (k3);
\draw[thick,scale=2,dashed] (t1) to [bend left=45] (t2);
\draw[thick,scale=2,double] (t1) to [bend left=-45] (t2);
\filldraw[color=black, fill=white, thick] ($(k1)-(\arista,\arista)$) rectangle ($(k1)+(\arista,\arista)$);
\filldraw[color=black, fill=white, thick] ($(k2)-(\arista,\arista)$) rectangle ($(k2)+(\arista,\arista)$);
\filldraw[color=black, fill=white, thick] ($(k3)-(\arista,\arista)$) rectangle ($(k3)+(\arista,\arista)$);
\fill[white] (t1) circle (0.45ex);
\filldraw[pattern=north east lines, thick] (t1) circle (0.45ex);
\fill[white] (t2) circle (0.45ex);
\filldraw[pattern=north east lines, thick] (t2) circle (0.45ex);
}
}
\def\diagramextwoverticesFiveIm{\tikz[baseline=-1.4ex,scale=1.8, every node/.style={scale=1.4}]{
\coordinate (k1) at (-3ex,0ex);
\coordinate (k2) at (-1.ex,0ex);
\coordinate (k3) at (+3.ex,0ex);
\coordinate (t1) at (-2ex,-4ex);
\coordinate (t2) at (+2ex,-4ex);
\pgfmathsetmacro{\arista}{0.06}
\draw[thick] (-4ex,0ex) -- (4ex,0ex);
\draw[-,thick,double] (t1) -- (k1);
\draw[-,thick,dashed] (t1) -- (k2);
\draw[-,thick,double] (t2) -- (k3);
\draw[thick,scale=2,double] (t1) to [bend left=45] (t2);
\draw[thick,scale=2,dashed] (t1) to [bend left=-45] (t2);
\filldraw[color=black, fill=white, thick] ($(k1)-(\arista,\arista)$) rectangle ($(k1)+(\arista,\arista)$);
\filldraw[color=black, fill=white, thick] ($(k2)-(\arista,\arista)$) rectangle ($(k2)+(\arista,\arista)$);
\filldraw[color=black, fill=white, thick] ($(k3)-(\arista,\arista)$) rectangle ($(k3)+(\arista,\arista)$);
\fill[white] (t1) circle (0.45ex);
\filldraw[pattern=north east lines, thick] (t1) circle (0.45ex);
\fill[white] (t2) circle (0.45ex);
\filldraw[pattern=north east lines, thick] (t2) circle (0.45ex);
}
}
\def\diagramextwoverticesSixIm{\tikz[baseline=-1.4ex,scale=1.8, every node/.style={scale=1.4}]{
\coordinate (k1) at (-3ex,0ex);
\coordinate (k2) at (-1.ex,0ex);
\coordinate (k3) at (+3.ex,0ex);
\coordinate (t1) at (-2ex,-4ex);
\coordinate (t2) at (+2ex,-4ex);
\pgfmathsetmacro{\arista}{0.06}
\draw[thick] (-4ex,0ex) -- (4ex,0ex);
\draw[-,thick,dashed] (t1) -- (k1);
\draw[-,thick,double] (t1) -- (k2);
\draw[-,thick,double] (t2) -- (k3);
\draw[thick,scale=2,double] (t1) to [bend left=45] (t2);
\draw[thick,scale=2,dashed] (t1) to [bend left=-45] (t2);
\filldraw[color=black, fill=white, thick] ($(k1)-(\arista,\arista)$) rectangle ($(k1)+(\arista,\arista)$);
\filldraw[color=black, fill=white, thick] ($(k2)-(\arista,\arista)$) rectangle ($(k2)+(\arista,\arista)$);
\filldraw[color=black, fill=white, thick] ($(k3)-(\arista,\arista)$) rectangle ($(k3)+(\arista,\arista)$);
\fill[white] (t1) circle (0.45ex);
\filldraw[pattern=north east lines, thick] (t1) circle (0.45ex);
\fill[white] (t2) circle (0.45ex);
\filldraw[pattern=north east lines, thick] (t2) circle (0.45ex);
}
}
\def\diagramextwoverticesSevenIm{\tikz[baseline=-1.4ex,scale=1.8, every node/.style={scale=1.4}]{
\coordinate (k1) at (-3ex,0ex);
\coordinate (k2) at (-1.ex,0ex);
\coordinate (k3) at (+3.ex,0ex);
\coordinate (t1) at (-2ex,-4ex);
\coordinate (t2) at (+2ex,-4ex);
\pgfmathsetmacro{\arista}{0.06}
\draw[thick] (-4ex,0ex) -- (4ex,0ex);
\draw[-,thick,double] (t1) -- (k1);
\draw[-,thick,dashed] (t1) -- (k2);
\draw[-,thick,double] (t2) -- (k3);
\draw[thick,scale=2,dashed] (t1) to [bend left=45] (t2);
\draw[thick,scale=2,double] (t1) to [bend left=-45] (t2);
\filldraw[color=black, fill=white, thick] ($(k1)-(\arista,\arista)$) rectangle ($(k1)+(\arista,\arista)$);
\filldraw[color=black, fill=white, thick] ($(k2)-(\arista,\arista)$) rectangle ($(k2)+(\arista,\arista)$);
\filldraw[color=black, fill=white, thick] ($(k3)-(\arista,\arista)$) rectangle ($(k3)+(\arista,\arista)$);
\fill[white] (t1) circle (0.45ex);
\filldraw[pattern=north east lines, thick] (t1) circle (0.45ex);
\fill[white] (t2) circle (0.45ex);
\filldraw[pattern=north east lines, thick] (t2) circle (0.45ex);
}
}
\def\diagramextwoverticesEightIm{\tikz[baseline=-1.4ex,scale=1.8, every node/.style={scale=1.4}]{
\coordinate (k1) at (-3ex,0ex);
\coordinate (k2) at (-1.ex,0ex);
\coordinate (k3) at (+3.ex,0ex);
\coordinate (t1) at (-2ex,-4ex);
\coordinate (t2) at (+2ex,-4ex);
\pgfmathsetmacro{\arista}{0.06}
\draw[thick] (-4ex,0ex) -- (4ex,0ex);
\draw[-,thick,dashed] (t1) -- (k1);
\draw[-,thick,double] (t1) -- (k2);
\draw[-,thick,double] (t2) -- (k3);
\draw[thick,scale=2,dashed] (t1) to [bend left=45] (t2);
\draw[thick,scale=2,double] (t1) to [bend left=-45] (t2);
\filldraw[color=black, fill=white, thick] ($(k1)-(\arista,\arista)$) rectangle ($(k1)+(\arista,\arista)$);
\filldraw[color=black, fill=white, thick] ($(k2)-(\arista,\arista)$) rectangle ($(k2)+(\arista,\arista)$);
\filldraw[color=black, fill=white, thick] ($(k3)-(\arista,\arista)$) rectangle ($(k3)+(\arista,\arista)$);
\fill[white] (t1) circle (0.45ex);
\filldraw[pattern=north east lines, thick] (t1) circle (0.45ex);
\fill[white] (t2) circle (0.45ex);
\filldraw[pattern=north east lines, thick] (t2) circle (0.45ex);
}
}
\def\diagramextwoverticesNineIm{\tikz[baseline=-1.4ex,scale=1.8, every node/.style={scale=1.4}]{
\coordinate (k1) at (-3ex,0ex);
\coordinate (k2) at (-1.ex,0ex);
\coordinate (k3) at (+3.ex,0ex);
\coordinate (t1) at (-2ex,-4ex);
\coordinate (t2) at (+2ex,-4ex);
\pgfmathsetmacro{\arista}{0.06}
\draw[thick] (-4ex,0ex) -- (4ex,0ex);
\draw[-,thick,dashed] (t1) -- (k1);
\draw[-,thick,dashed] (t1) -- (k2);
\draw[-,thick,dashed] (t2) -- (k3);
\draw[thick,scale=2,double] (t1) to [bend left=45] (t2);
\draw[thick,scale=2,dashed] (t1) to [bend left=-45] (t2);
\filldraw[color=black, fill=white, thick] ($(k1)-(\arista,\arista)$) rectangle ($(k1)+(\arista,\arista)$);
\filldraw[color=black, fill=white, thick] ($(k2)-(\arista,\arista)$) rectangle ($(k2)+(\arista,\arista)$);
\filldraw[color=black, fill=white, thick] ($(k3)-(\arista,\arista)$) rectangle ($(k3)+(\arista,\arista)$);
\fill[white] (t1) circle (0.45ex);
\filldraw[pattern=north east lines, thick] (t1) circle (0.45ex);
\fill[white] (t2) circle (0.45ex);
\filldraw[pattern=north east lines, thick] (t2) circle (0.45ex);
}
}
\def\diagramextwoverticesTenIm{\tikz[baseline=-1.4ex,scale=1.8, every node/.style={scale=1.4}]{
\coordinate (k1) at (-3ex,0ex);
\coordinate (k2) at (-1.ex,0ex);
\coordinate (k3) at (+3.ex,0ex);
\coordinate (t1) at (-2ex,-4ex);
\coordinate (t2) at (+2ex,-4ex);
\pgfmathsetmacro{\arista}{0.06}
\draw[thick] (-4ex,0ex) -- (4ex,0ex);
\draw[-,thick,dashed] (t1) -- (k1);
\draw[-,thick,dashed] (t1) -- (k2);
\draw[-,thick,dashed] (t2) -- (k3);
\draw[thick,scale=2,dashed] (t1) to [bend left=45] (t2);
\draw[thick,scale=2,double] (t1) to [bend left=-45] (t2);
\filldraw[color=black, fill=white, thick] ($(k1)-(\arista,\arista)$) rectangle ($(k1)+(\arista,\arista)$);
\filldraw[color=black, fill=white, thick] ($(k2)-(\arista,\arista)$) rectangle ($(k2)+(\arista,\arista)$);
\filldraw[color=black, fill=white, thick] ($(k3)-(\arista,\arista)$) rectangle ($(k3)+(\arista,\arista)$);
\fill[white] (t1) circle (0.45ex);
\filldraw[pattern=north east lines, thick] (t1) circle (0.45ex);
\fill[white] (t2) circle (0.45ex);
\filldraw[pattern=north east lines, thick] (t2) circle (0.45ex);
}
}
\bea
\diagramextwoverticesNoLabels &=& \diagramextwoverticesReImSplit + \diagramextwoverticesTwoImExt  \nn\\ &&  \hspace{-30pt}
+ \diagramextwoverticesThreeIm   + \diagramextwoverticesFourIm  \nn \\ && 
\hspace{-30pt}
+ \diagramextwoverticesFiveIm   + \diagramextwoverticesSixIm  \nn \\ &&
\hspace{-30pt}
+ \diagramextwoverticesSevenIm  + \diagramextwoverticesEightIm  \nn \\ &&
\hspace{-30pt}
+ \diagramextwoverticesNineIm + \diagramextwoverticesTenIm .
\eea
%
%
Given that the full result must be real, diagrams with an even number of vertices require an even number of imaginary propagators. Similarly, diagrams with an odd number of vertices require an odd number of imaginary propagators. In the following section we will exploit these rules in order to infer the number of infrared logarithms appearing in any given diagram.


\section{Loops and logarithms in momentum space}
\label{sec:logarithms}

In the previous section we learned that a connected $n$-point correlation function in momentum space $\tilde G^{(n)}_c ( {\k}_1 , \cdots , {\k}_n ; \tau )$ can be written as the following sum
\be
\tilde G^{(n)} ( {\k}_1 , \cdots , {\k}_n ; \tau ) =  \sum_{T}  D_{T} ( {\k}_1 , \cdots , {\k}_n ; \tau ) ,
\ee
where $D_{T} ( {\k}_1 , \cdots , {\k}_n ; \tau ) $ corresponds to the collection of all $n$-legged diagrams sharing the same topology $T$. Here, by topology we mean a specific shape for the diagram that remains indistinguishable after changing the color of vertices. For instance, both diagrams at the right hand side of (\ref{D-tree-level-1-vertex}) have vertices of different colors but share the same shape, and so they belong to the same category $T$ (diagrams with a different number of loops belong to different topologies). Similarly the two diagrams at the right hand side of (\ref{D-diagrams-2-vertices}) share the same topology. Thus, the contribution $ D_{T} ( {\k}_1 , \cdots , {\k}_n ; \tau ) $ consists in a collection of diagrams of the same shape but with different combinations of colors in their vertices.

In what follows we establish a general result regarding the time dependence of correlation functions in Fourier space in the superhorizon limit whereby every external comoving momentum $\k_i$ is restricted to satisfy $|\tau \k_i| \ll 1$. In this limit, a non-colored diagram $D_{T} (\k_1, \cdots , \k_n)$ of a given topology $T$, is found to be proportional to the product of time-valued logarithms of the form:
\bea \label{log-structure}
D_{T} (\k_1, \cdots , \k_n; \tau) &=&   \delta^{(3)} (\boldsymbol{K}) \sum_{s} A^T_{s}(\k_1, \cdots , \k_n) \nn \\
&& \!\!\!\!\!\!\!\!\!\!\!\!\!\!\! \times \prod_{a=1}^{V}  \ln  \big[ -\tau f^T_{s,a}(\k_1, \cdots , \k_n)  \big] , \quad
\eea
where $\boldsymbol{K} \equiv \k_1 + \cdots + \k_n$, $V$ is the number of vertices implied by the topology, while $A^T_{s}(\k_1, \cdots , \k_n)$ and $f^T_{s,a}(\k_1, \cdots , \k_n)$ are functions of momenta, that respect the scaling properties
\bea 
A^T_{s}(\alpha \k_1, \cdots , \alpha \k_n) &=& \alpha^{-3(n-1)} A^T_{s}(\k_1, \cdots , \k_n) ,  \quad \label{prop-g} \\
f^T_{s,a}( \alpha \k_1, \cdots , \alpha \k_n) &=& \alpha f^T_{s,a} (  \k_1, \cdots ,  \k_n) , \label{prop-f-log} 
\eea 
for some real parameter $\alpha>0$. This result will hold true independently of the number of loops participating in the diagram.

\subsection{Proof}
\label{sec:general-argument}

We would like to understand the time dependence of $ D_{T} ( {\k}_1 , \cdots , {\k}_n ; \tau ) $ in the limit $| k_i \tau | \ll 1$. In particular, we want to know how time-valued logarithms emerge. First of all, it should be clear that the only way a logarithm can emerge is as the consequence of the time integrals implied by vertices. Specifically, a logarithm of the form $\ln[-\tau \cdots ]$ (where the ellipses stand for combinations of external momenta) can only emerge from the behavior of integrals near their upper limit $\tau$. Thus, in order to deduce the number of logarithms implied by a collection of diagrams sharing the same topology, we must examine the behavior of $\tau_a$-integrals near the upper limits of the time-integrals, where $a=1, \cdots , V$ is the label running through the vertices of the diagram. Note that the integration variable $\tau_a$ belonging to a given vertex $a$ does not always appear multiplied by an external momentum $k_i$. The variable $\tau_a$ may also appear multiplied by an internal momentum $q$ having the role of a loop integration variable. In this case, it is necessary to notice that the only part of this loop-integral that can contribute to the appearance of a logarithm is that for which $|q \tau_a | \ll 1$. This is precisely the part from the loop that gives an integrand contribution near the upper limit $\tau$.

Having agreed on the need to examine the integrands of vertex-time-integrals near the upper limit $\tau$, let us determine the behavior of these integrands. According to Eq.~(\ref{Feynman-vertices}) a particular vertex $\tau_a$ comes with a time integral $\int^{\tau}_{-\infty} d \tau_a$ together with a factor ${\tau_a}^{-4}$. However, in Section~\ref{sec:split-propagators} we learned that there must be at least one imaginary propagator attached to each vertex. Such a propagator can be bulk-to-boundary or bulk-to-bulk. Thanks to Eq.~(\ref{prop-im}), we see that in the case of a bulk-to-boundary propagator $G_{\pm} (k,\tau_a,\tau)$, near the upper limit $\tau$, the imaginary part acquires the form
\bea
G^{\rm Im}_{\pm} (k,\tau_a,\tau) &=& \pm i \frac{H^2}{6} \Big[ (\tau_a^3 - \tau^3)  + \cdots \Big] , \qquad 
\eea
where the ellipses indicate terms suppressed in terms of the external momenta (which satisfy $|k_i \tau | \ll 1$). Similarly, the imaginary part of a bulk-to-bulk propagator $G_{\pm \pm} (k,\tau_a,\tau_b)$ connecting two vertices of the same color, near the upper limit $\tau$ becomes
\be
G^{\rm Im}_{\pm \pm} (k,\tau_a,\tau_b) = \mp i \frac{H^2}{6} \Big[ (\tau_a^3 - \tau_b^3)  + \cdots \Big] I (\tau_a , \tau_b), 
\ee
whereas the imaginary part of a bulk-to-bulk propagator $G_{\pm \mp} (k,\tau_a,\tau_b)$ connecting two vertices of opposite colors, has the form
\be
G^{\rm Im}_{\pm \mp} (k,\tau_a,\tau_b) = \pm i \frac{H^2}{6} \Big[ (\tau_a^3 - \tau_b^3)  + \cdots \Big] .
\ee
These expansions are not only valid for bulk-to-bulk propagators entering tree-level diagrams, but they are also valid for propagators participating in loops. Recall that the part of loop-integrals contributing to integrands near the upper limit has momenta $q$ satisfying $|q\tau | \ll 1$. On the other hand, the real part of any propagator, bulk-to-boundary or bulk-to-bulk, acquires the universal form (independent of the type of vertices they are connecting)
\be
G^{\rm Re}_{s_a s_b} (k,\tau_a,\tau_b) =  \frac{H^2}{2 k^3} \Big[1  + \cdots \Big] .
\ee
where $s_a$ stands for the sign of the vertices connected by the propagator.

The asymptotic behaviors just highlighted, imply that after performing all the time-integrals, the leading term must be proportional to an overall factor of the form $\ln[-\tau \cdots]^V$, where $V$ is the number of vertices in the diagram. To see this, one may simply split the time integrals as:
\be
\int_{-\infty}^{\tau} d\tau' = \int_{-\infty}^{\tau_0} d\tau'  +  \int_{\tau_0}^{\tau} d\tau' ,
\ee
for some arbitrary $\tau_0$. Given that we are only interested in the behavior near the upper limit $\tau$, the introduction of $\tau_0$ makes it easier to obtain the dependence of the integrals on $\tau$ with disregard of their behavior towards the lower limit $\tau' \to - \infty$. 

Now, the key point to notice is that the minimum number of imaginary propagators is precisely $V$, and that each one of them are attached to at least one vertex. This implies that the factor $\tau_a^{-4}$ accompanying the vertex $a$ will always encounter at least one factor of the form $(\tau_a^3-\tau^3)$ or $(\tau_a^3-\tau_b^3)$, lowering the degree of divergence of the integral to a logarithmic divergence. In fact, diagrams with a number $n_I$ of imaginary propagators larger than $V$ will necessarily end up being proportional to $\tau^{3(n_I - V)}$. 

It should be clear then that, after solving all $V$ time-integrals, the leading term will be a function with an overall factor $\propto \left[ \ln ( \tau / \tau_0) \right]^V$. The procedure must be independent of the choice $\tau_0$ and, because the entire diagram is a function of external momenta and time, the only way that a term proportional to $[\ln \tau]^V$ can appear is through a product of logarithms of the form (\ref{log-structure}) already advertised. Notice that, because the correlation function must respect the scaling property (\ref{scaling-props}), the functions $f$ and $g$ participating of (\ref{log-structure}) are forced to satisfy the scaling properties shown in Eqs.~(\ref{prop-g}) and (\ref{prop-f-log}).

We devote the rest of this section to explicitly show how (\ref{log-structure}) is realized in various relevant examples.

\subsection{Example 1: Tree-level diagram with 1-vertex }

Let us check the validity of (\ref{log-structure}) for the simplest case of a tree-level correlator built from a 1-vertex diagram. First, let us consider the case of an $n$-point function to first order in $\mathcal V(\varphi)$, depicted by the following 1-vertex diagram: 
\def\OconnectedVone{\tikz[baseline=0.5ex]{
\coordinate (P) at (0,-3ex);
\coordinate (C) at (-7ex,4ex);
\coordinate (k1) at (-5.0ex,4.0ex);
\coordinate (k2) at (-2.0ex,4.0ex);
\coordinate (kn) at (5.0ex,4.0ex);
\pgfmathsetmacro{\arista}{0.1}
\draw[thick] (-7ex,4ex) -- (7ex,4ex);
\draw[thick] (0,-3ex) -- (k1);
\draw[thick] (0,-3ex) -- (k2);
\draw[thick] (0,-3ex) -- (kn);
\filldraw[color=black, fill=black, thick] (0ex,2ex) circle (0.1ex);
\filldraw[color=black, fill=black, thick] (1.2ex,2ex) circle (0.1ex);
\filldraw[color=black, fill=black, thick] (2.4ex,2ex) circle (0.1ex);
\node at ($(P) + (-1.ex,0)$) [anchor=east]{{$\tau'$}};
\filldraw[color=black, fill=white, thick] ($(k1)-(\arista,\arista)$) rectangle ($(k1)+(\arista,\arista)$)
node[anchor=south]{\footnotesize{$k_1$}};
\filldraw[color=black, fill=white, thick] ($(k2)-(\arista,\arista)$) rectangle ($(k2)+(\arista,\arista)$)
node[anchor=south]{\footnotesize{$k_2$}};
\filldraw[color=black, fill=white, thick] ($(kn)-(\arista,\arista)$) rectangle ($(kn)+(\arista,\arista)$)
node[anchor=south]{\footnotesize{$k_n$}};
\fill[white] (P) circle (0.9ex);
\filldraw[pattern=north east lines, thick] (P) circle (0.9ex);
}
}
\bea \label{example-log-1-vertex}
\OconnectedVone &=&  (2\pi)^3 \delta^{(3)} (\boldsymbol{K}) \,  \frac{2 \lambda_{n}}{H^4} \frac{H^{2n}}{2^n k_1^3 \cdots k_n^3} \nn \\ 
[-10pt]
&& \times {\rm Im} \Big\{ I (\tau ; k_1 , \cdots , k_n) \Big\}.
\eea
where $I (\tau ; k_1 , \cdots , k_n)$ corresponds to the following time integral:
\be \label{vertex-integral-1}
I (\tau ; k_1 , \cdots , k_n) \equiv  \int_{-\infty}^{\tau}  \frac{d \tau'}{{ \tau' }^{4}} \prod_{j=1}^{n} (1 -  i k_j \tau ') e^{   i \tau' K} ,
\ee
with $K = k_1 + \cdots + k_n$. Notice that we are disregarding corrections of order $|k_i\tau|^2$ or higher in the bulk-to-boundary propagators used to obtain (\ref{example-log-1-vertex}). By performing integrations by part, the previous integral can be conveniently rewritten as:
\bea \label{vertex-integral-2}
I &=& \frac{e^{i K \tau}}{3} \bigg( -\frac{1}{\tau^3} + \frac{i K}{\tau^2} + \frac{K^2 - 3 \sum_{j=1}^{n} k_j^2  }{2\tau} \bigg)   \nn \\
&& 
+ \frac{i}{3} \bigg( \sum_{j=1}^{n} k_j^3 \bigg) \Big( i \pi + {\rm Ei}(-i K |\tau|) \Big)+ \mathcal O(\tau^0), \quad 
\eea
where ${\rm Ei}(z)$ is the usual exponential integral function, and $\mathcal O(\tau^0)$ represents terms of order $\tau^0$ or higher, and therefore suppressed with respect to those kept in the expression. In the limit $| k_i \tau| \ll 1$, Eq.~(\ref{vertex-integral-2}) becomes 
\bea \label{vertex-integral-3}
I (\tau ; k_1 , \cdots , k_n) &\simeq&  \frac{i}{3} \bigg( \sum_{j=1}^{n} k_j^3 \bigg) \ln (-\tau K) - \frac{1}{3\tau^3} \nn \\
&&  - \frac{   \sum_{j=1}^{n} k_j^2}{2\tau}  +  \mathcal O(\tau^0) . \qquad
\eea
Plugging the previous result back into (\ref{n-point-one-vertex}) we finally obtain that the leading contribution to the 1-vertex correlation function is:
\bea \label{D-n-point-single-vertex}
\OconnectedVone  \! &=&   (2\pi)^3 \delta^{(3)} (\boldsymbol{K}) \, \frac{2}{3}  \frac{\lambda_{n}}{H^4} \frac{H^{2n}}{2^n} \nn \\ [-15pt]
&& \times \frac{ k_1^3 + \cdots +k_n^3 }{k_1^3 \cdots k_n^3} \ln (-\tau K) . 
\eea
Thus, as anticipated by our analysis involving the splitting of propagators into real and imaginary components, we see that taking the imaginary part of the integral $I$ only keeps the logarithmic dependence of the correlator on $\tau$. 

\subsection{Example 2: Tree-level diagram with 2-vertices}
\label{sec:example-logs-2}

As a second example, let us consider a contribution to the $n$-point function given by a tree-level diagram where $n_1$ legs are attached to the first vertex $\tau_1$ and $n_2$ legs are attached to the second vertex $\tau_2$, such that $n_1 + n_2 = n$. As already discussed in Section~\ref{SK-examples}, in this case we have two distinct contributions; 2 diagrams with vertices of opposite colors, and 2 diagrams with vertices of the same color:
\begin{widetext}
\def\diagramextwoverticestree{\tikz[baseline=-1.4ex]{
\coordinate (k1) at (-9ex,0ex);
\coordinate (k2) at (-3.ex,0ex);
\coordinate (k3) at (+3.ex,0ex);
\coordinate (k4) at (+9.ex,0ex);
\coordinate (t1) at (-4.5ex,-7ex);
\coordinate (t2) at (+4.5ex,-7ex);
\pgfmathsetmacro{\arista}{0.1}
\draw[thick] (-11ex,0ex) -- (11ex,0ex);
\draw[-,thick] (t1) -- (k1);
\draw[-,thick] (t1) -- (k2);
\draw[-,thick] (t2) -- (k3);
\draw[-,thick] (t2) -- (k4);
\draw[thick,scale=2] (t1) to (t2);
\fill[white] (t1) circle (0.9ex);
\filldraw[pattern=north east lines, thick] (t1) circle (0.9ex);
\fill[white] (t2) circle (0.9ex);
\filldraw[pattern=north east lines, thick] (t2) circle (0.9ex);
\filldraw[color=black, fill=white, thick] ($(k1)-(\arista,\arista)$) rectangle ($(k1)+(\arista,\arista)$);
\filldraw[color=black, fill=white, thick] ($(k2)-(\arista,\arista)$) rectangle ($(k2)+(\arista,\arista)$);
\filldraw[color=black, fill=white, thick] ($(k3)-(\arista,\arista)$) rectangle ($(k3)+(\arista,\arista)$);
\filldraw[color=black, fill=white, thick] ($(k4)-(\arista,\arista)$) rectangle ($(k4)+(\arista,\arista)$);
\node at ($(k1)+(0ex,0.5ex)$) [anchor=south]{\footnotesize{$\k_1$}};
\node at ($(k2)+(0ex,0.5ex)$) [anchor=south]{\footnotesize{$\k_{n_1}$}};
\node at ($(k3)+(0ex,0.5ex)$) [anchor=south]{\footnotesize{$\k_{n_1+1}$}};
\node at ($(k4)+(0ex,0.5ex)$) [anchor=south]{\footnotesize{$\k_{n}$}};
}
}

\def\diagramextwoverticestreefirst{\tikz[baseline=-1.4ex]{
\coordinate (k1) at (-9ex,0ex);
\coordinate (k2) at (-3.ex,0ex);
\coordinate (k3) at (+3.ex,0ex);
\coordinate (k4) at (+9.ex,0ex);
\coordinate (t1) at (-4.5ex,-7ex);
\coordinate (t2) at (+4.5ex,-7ex);
\pgfmathsetmacro{\arista}{0.1}
\draw[thick] (-11ex,0ex) -- (11ex,0ex);
\draw[-,thick] (t1) -- (k1);
\draw[-,thick] (t1) -- (k2);
\draw[-,thick] (t2) -- (k3);
\draw[-,thick] (t2) -- (k4);
\draw[thick,scale=2] (t1) to (t2);
\fill[white] (t1) circle (0.9ex);
\filldraw[color=black, fill=black, thick]  (t1) circle (0.9ex);
\fill[white] (t2) circle (0.9ex);
\filldraw[color=black, fill=black, thick]  (t2) circle (0.9ex);
\filldraw[color=black, fill=white, thick] ($(k1)-(\arista,\arista)$) rectangle ($(k1)+(\arista,\arista)$);
\filldraw[color=black, fill=white, thick] ($(k2)-(\arista,\arista)$) rectangle ($(k2)+(\arista,\arista)$);
\filldraw[color=black, fill=white, thick] ($(k3)-(\arista,\arista)$) rectangle ($(k3)+(\arista,\arista)$);
\filldraw[color=black, fill=white, thick] ($(k4)-(\arista,\arista)$) rectangle ($(k4)+(\arista,\arista)$);
\node at ($(k1)+(0ex,0.5ex)$) [anchor=south]{\footnotesize{$\k_1$}};
\node at ($(k2)+(0ex,0.5ex)$) [anchor=south]{\footnotesize{$\k_{n_1}$}};
\node at ($(k3)+(0ex,0.5ex)$) [anchor=south]{\footnotesize{$\k_{n_1+1}$}};
\node at ($(k4)+(0ex,0.5ex)$) [anchor=south]{\footnotesize{$\k_{n}$}};
}
}

\def\diagramextwoverticestreesecond{\tikz[baseline=-1.4ex]{
\coordinate (k1) at (-9ex,0ex);
\coordinate (k2) at (-3.ex,0ex);
\coordinate (k3) at (+3.ex,0ex);
\coordinate (k4) at (+9.ex,0ex);
\coordinate (t1) at (-4.5ex,-7ex);
\coordinate (t2) at (+4.5ex,-7ex);
\pgfmathsetmacro{\arista}{0.1}
\draw[thick] (-11ex,0ex) -- (11ex,0ex);
\draw[-,thick] (t1) -- (k1);
\draw[-,thick] (t1) -- (k2);
\draw[-,thick] (t2) -- (k3);
\draw[-,thick] (t2) -- (k4);
\draw[thick,scale=2] (t1) to (t2);
\fill[white] (t1) circle (0.9ex);
\filldraw[color=black, fill=black, thick]  (t1) circle (0.9ex);
\fill[white] (t2) circle (0.9ex);
\filldraw[color=black, fill=white, thick]  (t2) circle (0.9ex);
\filldraw[color=black, fill=white, thick] ($(k1)-(\arista,\arista)$) rectangle ($(k1)+(\arista,\arista)$);
\filldraw[color=black, fill=white, thick] ($(k2)-(\arista,\arista)$) rectangle ($(k2)+(\arista,\arista)$);
\filldraw[color=black, fill=white, thick] ($(k3)-(\arista,\arista)$) rectangle ($(k3)+(\arista,\arista)$);
\filldraw[color=black, fill=white, thick] ($(k4)-(\arista,\arista)$) rectangle ($(k4)+(\arista,\arista)$);
\node at ($(k1)+(0ex,0.5ex)$) [anchor=south]{\footnotesize{$\k_1$}};
\node at ($(k2)+(0ex,0.5ex)$) [anchor=south]{\footnotesize{$\k_{n_1}$}};
\node at ($(k3)+(0ex,0.5ex)$) [anchor=south]{\footnotesize{$\k_{n_1+1}$}};
\node at ($(k4)+(0ex,0.5ex)$) [anchor=south]{\footnotesize{$\k_{n}$}};
}
}

\be \label{tree-level-two-vertices}
\diagramextwoverticestree =  2 {\rm Re} \Bigg\{  \diagramextwoverticestreesecond + \diagramextwoverticestreefirst \Bigg\} .
\ee
\end{widetext}
The first contribution at the right hand side, having vertices of opposite color, contains a bulk-to-bulk propagator that factorizes into two unnested integrals that can be solved just as in the case of our previous 1-vertex example. Therefore, repeating the procedure leading to (\ref{vertex-integral-3}), one finds:
\begin{widetext}
\def\diagramextwoverticestreesecond{\tikz[baseline=-1.4ex]{
\coordinate (k1) at (-9ex,0ex);
\coordinate (k2) at (-3.ex,0ex);
\coordinate (k3) at (+3.ex,0ex);
\coordinate (k4) at (+9.ex,0ex);
\coordinate (t1) at (-4.5ex,-7ex);
\coordinate (t2) at (+4.5ex,-7ex);
\pgfmathsetmacro{\arista}{0.1}
\draw[thick] (-11ex,0ex) -- (11ex,0ex);
\draw[-,thick] (t1) -- (k1);
\draw[-,thick] (t1) -- (k2);
\draw[-,thick] (t2) -- (k3);
\draw[-,thick] (t2) -- (k4);
\draw[thick,scale=2] (t1) to (t2);
\fill[white] (t1) circle (0.9ex);
\filldraw[color=black, fill=black, thick]  (t1) circle (0.9ex);
\fill[white] (t2) circle (0.9ex);
\filldraw[color=black, fill=white, thick]  (t2) circle (0.9ex);
\filldraw[color=black, fill=white, thick] ($(k1)-(\arista,\arista)$) rectangle ($(k1)+(\arista,\arista)$);
\filldraw[color=black, fill=white, thick] ($(k2)-(\arista,\arista)$) rectangle ($(k2)+(\arista,\arista)$);
\filldraw[color=black, fill=white, thick] ($(k3)-(\arista,\arista)$) rectangle ($(k3)+(\arista,\arista)$);
\filldraw[color=black, fill=white, thick] ($(k4)-(\arista,\arista)$) rectangle ($(k4)+(\arista,\arista)$);
\node at ($(k1)+(0ex,0.5ex)$) [anchor=south]{\footnotesize{$\k_1$}};
\node at ($(k2)+(0ex,0.5ex)$) [anchor=south]{\footnotesize{$\k_{n_1}$}};
\node at ($(k3)+(0ex,0.5ex)$) [anchor=south]{\footnotesize{$\k_{n_1+1}$}};
\node at ($(k4)+(0ex,0.5ex)$) [anchor=south]{\footnotesize{$\k_{n}$}};
}
}

\bea \label{two-vertex-diagram-1}
2 {\rm Re} \Bigg\{ \diagramextwoverticestreesecond \Bigg\}
&=&   (2\pi)^3 \delta^{(3)} ({\bf K}_1 + {\bf K}_2 ) \frac{\lambda_{n_1+1} \lambda_{n_2+1}}{ H^8}     \frac{H^2}{2 k_1^3} \cdots \frac{H^2}{2 k_{n_1}^3} \frac{H^2}{2 k_{n_1 + 1}^3} \cdots \frac{H^2}{2 k_{n}^3} \frac{H^2}{2 Q^3} \\   [-15pt]
 &&  \times \frac{2}{9}  \left( Q^3 + \sum_{j=1}^{n_1} k_{j}^3 \right)   \left( Q^3 + \sum_{j=n_1 + 1}^{n} k_{j}^3 \right) \ln [ - \tau (Q + K_1) ]  \ln [ - \tau (Q + K_2) ]   \nn \\
 && +  \mathcal O (\tau^{0}) + \sum_{n=1}^{6} \mathcal O (\tau^{-n})  , \nn
\eea
\end{widetext}
where $Q = |\k_1 + \cdots + \k_{n_1}|$ is the modulus of the total momentum interchanged by the two vertices. In addition, we have defined $K_1 = k_1 + \cdots + k_{n_1}$ and $K_2 = k_{n_1 + 1} + \cdots + k_{n}$. Notice that the third line of (\ref{two-vertex-diagram-1}) contains the contribution $\sum_{n=1}^{6} \mathcal O (\tau^{-n})$ which symbolizes divergent terms which grow much faster than the logarithmic contribution we are searching for. Nevertheless, recall that our analysis based on the splitting of propagators into real and imaginary part ensures that these terms cancel with similar contributions coming from the second diagram at the right hand side of (\ref{tree-level-two-vertices}).

The second contribution appearing at the right hand side of (\ref{tree-level-two-vertices}) involves solving nested integrals. However, these integrals can still be carried out near the boundary $\tau$ with the help of Eq.~(\ref{vertex-integral-3}). Keeping the leading logarithmic terms, one finally finds
\begin{widetext}

\def\diagramextwoverticestreefirst{\tikz[baseline=-1.4ex]{
\coordinate (k1) at (-9ex,0ex);
\coordinate (k2) at (-3.ex,0ex);
\coordinate (k3) at (+3.ex,0ex);
\coordinate (k4) at (+9.ex,0ex);
\coordinate (t1) at (-4.5ex,-7ex);
\coordinate (t2) at (+4.5ex,-7ex);
\pgfmathsetmacro{\arista}{0.1}
\draw[thick] (-11ex,0ex) -- (11ex,0ex);
\draw[-,thick] (t1) -- (k1);
\draw[-,thick] (t1) -- (k2);
\draw[-,thick] (t2) -- (k3);
\draw[-,thick] (t2) -- (k4);
\draw[thick,scale=2] (t1) to (t2);
\fill[white] (t1) circle (0.9ex);
\filldraw[color=black, fill=black, thick]  (t1) circle (0.9ex);
\fill[white] (t2) circle (0.9ex);
\filldraw[color=black, fill=black, thick]  (t2) circle (0.9ex);
\filldraw[color=black, fill=white, thick] ($(k1)-(\arista,\arista)$) rectangle ($(k1)+(\arista,\arista)$);
\filldraw[color=black, fill=white, thick] ($(k2)-(\arista,\arista)$) rectangle ($(k2)+(\arista,\arista)$);
\filldraw[color=black, fill=white, thick] ($(k3)-(\arista,\arista)$) rectangle ($(k3)+(\arista,\arista)$);
\filldraw[color=black, fill=white, thick] ($(k4)-(\arista,\arista)$) rectangle ($(k4)+(\arista,\arista)$);
\node at ($(k1)+(0ex,0.5ex)$) [anchor=south]{\footnotesize{$\k_1$}};
\node at ($(k2)+(0ex,0.5ex)$) [anchor=south]{\footnotesize{$\k_{n_1}$}};
\node at ($(k3)+(0ex,0.5ex)$) [anchor=south]{\footnotesize{$\k_{n_1+1}$}};
\node at ($(k4)+(0ex,0.5ex)$) [anchor=south]{\footnotesize{$\k_{n}$}};
}
}

\bea \label{two-vertex-diagram-2}
2 {\rm Re} \Bigg\{ \diagramextwoverticestreefirst \Bigg\} 
&=& - (2\pi)^3 \delta^{(3)} ({\bf K}_1 + {\bf K}_2 ) \frac{\lambda_{n_1+1} \lambda_{n_2+1}}{ H^8}     \frac{H^2}{2 k_1^3} \cdots \frac{H^2}{2 k_{n_1}^3} \frac{H^2}{2 k_{n_1 + 1}^3} \cdots \frac{H^2}{2 k_{n}^3}  \frac{H^2}{2 Q^3}  \\ [-15pt]
&& \Bigg[ \frac{1}{9}  \bigg(  Q^3 - \sum_{j=1}^{n_1} k_{j}^3  \bigg)   \bigg(  Q^3 + \sum_{j=n_1 + 1}^{n} k_{j}^3 \bigg) \bigg( \ln [ - \tau (Q + K_2)  ] \bigg)^2 \nn \\
&& + \frac{1}{9}  \bigg(  Q^3 + \sum_{j=1}^{n_1} k_{j}^3  \bigg)   \bigg(  Q^3 - \sum_{j=n_1 + 1}^{n} k_{j}^3  \bigg) \bigg( \ln [ - \tau (Q + K_1)  ] \bigg)^2 \Bigg]  \nn \\
 && +  \mathcal O (\tau^{0}) + \sum_{n=1}^{6} \mathcal O (\tau^{-n}) , \nn
\eea
\end{widetext}
where $Q$, $K_1$ and $K_2$ are defined as before. The term in the second line of (\ref{two-vertex-diagram-2}) comes from the nested integration whereby the second vertex is evaluated at earlier times than the first vertex. Conversely, the term at the third line comes from the nested integration whereby the second vertex is evaluated at later times than the first vertex. Also, notice again the presence of $\mathcal O (\tau^{-6})$, $\mathcal O (\tau^{-4})$ and $\mathcal O (\tau^{-2})$. These happen to have exactly the same form as those in (\ref{two-vertex-diagram-1}) but with opposite sign. In this way, the leading term contributing to (\ref{tree-level-two-vertices}) obtained after summing (\ref{two-vertex-diagram-1}) and (\ref{two-vertex-diagram-2}) consists only in those proportional to two powers of logarithms, as anticipated by our general formula (\ref{log-structure}).

Last but not least, it is noteworthy to appreciate that in all of these examples, the arguments of logarithms are strictly positive, making it impossible to have a divergent contribution for nonvanishing external momenta.

\subsection{Example 3: Daisy loops}
\label{sec:daisy-loops}

Daisy loops are loops formed by propagators starting and ending in the same vertex. From the examples offered in the previous section, we see that each daisy loop attached to a vertex of time $\tau_a$ contributes a factor $\frac{1}{2} G(0;\tau_a)$, where 
where $G(0;\tau)$ is nothing but the two point function $G(|\x - \x'|;\tau)$ evaluated at coincident point $\x = \x'$. Recall that in Section~\ref{sec:iso-vacua} we found that, in the case of light scalar fields, $G(0;\tau)$ is either a divergent constant, or just zero, a result encountered in the dimensional regularization scheme~(\ref{fk-t-d}). Moreover, it is worth noticing that if one were to write the interacting Hamiltonian of the theory in a scheme respecting normal ordering of its creation and annihilation operators, then these loops would also vanish~\footnote{Note that the constancy of daisy loops is in agreement with the treatment of Refs.~\cite{Lee:2023jby,Creminelli2024}.}. Thus, given that independently of the scheme, $G(0;\tau)$ is just a constant, let us momentarily adopt the more convenient notation 
\be \label{sigma-tot-def}
G(0;\tau)\equiv \sigma_{\rm tot}^2.
\ee

Now, the fact that these loops vanish in a particular scheme signals that, in other schemes, where they are divergent constant, it must be possible to renormalize them away by a trivial redefinition of parameters. Indeed, as shown in detail in Ref.~\cite{Huenupi:2024ksc}, the divergences implied by daisy loops via $G(0;\tau)=\sigma_{\rm tot}^2$ can be dealt with, to all orders in perturbation theory, by writing the bare coupling constants $\lambda_n$ of the potential as:
\be \label{lambda-lambda}
\lambda_n =\sum_{L} \frac{(-1)^L}{L!} \left( \frac{1}{2} \sigma_{\rm tot}^2 \right)^L \bar \lambda_{n + 2L} .
\ee
This redefinition of the bare coupling constants $\lambda_n$ in terms of new couplings $\bar \lambda_n$ removes the appearance of daisy loops in any given vertex after one sums over all possible configurations. That is, instead of computing diagrams with the couplings $\lambda_n$ and including in them daisy loops, one may as well work with the couplings $\bar \lambda_n$ and ignore daisy loops altogether. Both procedures are equivalent.

By recalling that the bare couplings $\lambda_n$ are the coefficients from the Taylor expansion of the bare potential $\mathcal V (\varphi)$ given in (\ref{taylor-bare-potential}), one can ask whether the couplings $\bar \lambda_n$ are the coefficients coming from the expansion of a certain potential $\bar {\mathcal V} (\varphi)$. From Eq.~(\ref{lambda-lambda}) one finds that the potential $\bar {\mathcal V} (\varphi)$ determining the coefficients $\bar \lambda_n$ is nothing but a Weierstrass transform of the potential $\mathcal V(\varphi)$:
\be \label{Weierstrass-V}
\bar {\mathcal V} (\varphi) = e^{\frac{1}{2} \sigma_{\rm tot}^2 \frac{\partial^2}{\partial \varphi^2}}   \mathcal V (\varphi) .
\ee
Dealing with loops in this way, simply corresponds to redefining the couplings $\lambda_n$, order by order, in terms of renormalized and counter-term couplings, adjusted to eliminate the divergences coming from daisy loops.

All in all, daisy loops, which do not carry external momenta, can be removed with a trivial redefinition of the bare potential, and cannot have any physical consequence on the computation of correlation functions. Consequentially, a correlation function to first order in the potential $\mathcal V (\varphi)$ and to all orders in loops, is indistinguishable from a tree-level correlation function.

\subsection{Example 4: Diagrams with 2 vertices and 1 loop}
\label{sec:example-logs-loop}

As already argued, loops cannot alter the structure of Eq.~(\ref{log-structure}), which is entirely dictated by the time dependence of the integrands near the boundary $\tau$. Loops will certainly imply divergences, but these cannot modify the time dependence involved in the integration of each vertex. This is obvious in the case of daisy loops, a case just analyzed.  

A less trivial example is provided by diagrams where loops carry some amount of external momentum. To check the validity of our claim for this instance, let us consider a $2$-point function with two vertices joined by one loop, as depicted in the following diagram:
\def\diagramextwoverticesNoLabels{\tikz[baseline=-1.4ex,scale=1.8, every node/.style={scale=1.4}]{
\coordinate (k1) at (-2.5ex,0ex);
\coordinate (k4) at (+2.5ex,0ex);
\coordinate (t1) at (-2ex,-4ex);
\coordinate (t2) at (+2ex,-4ex);
\pgfmathsetmacro{\arista}{0.06}
\draw[thick] (-3.5ex,0ex) -- (3.5ex,0ex);
\draw[-,thick] (t1) -- (k1);
\draw[-,thick] (t2) -- (k4);
\draw[thick,scale=2] (t1) to [bend left=60] (t2);
\draw[thick,scale=2] (t1) to [bend left=-60] (t2);
\fill[white] (t1) circle (0.45ex);
\filldraw[pattern=north east lines, thick] (t1) circle (0.45ex);
\fill[white] (t2) circle (0.45ex);
\filldraw[pattern=north east lines, thick] (t2) circle (0.45ex);
\filldraw[color=black, fill=white, thick] ($(k1)-(\arista,\arista)$) rectangle ($(k1)+(\arista,\arista)$);
\filldraw[color=black, fill=white, thick] ($(k4)-(\arista,\arista)$) rectangle ($(k4)+(\arista,\arista)$);
\node at ($(k1)+(0ex,0.15ex)$) [anchor=south]{\tiny{$\k_1$}};
\node at ($(k4)+(0ex,0.15ex)$) [anchor=south]{\tiny{$\k_{2}$}};
}
}
\def\diagramextwoverticesone{\tikz[baseline=-1.4ex,scale=1.8, every node/.style={scale=1.4}]{
\coordinate (k1) at (-2.5ex,0ex);
\coordinate (k4) at (+2.5ex,0ex);
\coordinate (t1) at (-2ex,-4ex);
\coordinate (t2) at (+2ex,-4ex);
\pgfmathsetmacro{\arista}{0.06}
\draw[thick] (-3.5ex,0ex) -- (3.5ex,0ex);
\draw[-,thick] (t1) -- (k1);
\draw[-,thick] (t2) -- (k4);
\draw[thick,scale=2] (t1) to [bend left=60] (t2);
\draw[thick,scale=2] (t1) to [bend left=-60] (t2);
\filldraw[color=black, fill=black, thick] (t1) circle (0.45ex);
\filldraw[color=black, fill=black, thick] (t2) circle (0.45ex);
\filldraw[color=black, fill=white, thick] ($(k1)-(\arista,\arista)$) rectangle ($(k1)+(\arista,\arista)$);
\filldraw[color=black, fill=white, thick] ($(k4)-(\arista,\arista)$) rectangle ($(k4)+(\arista,\arista)$);
\node at ($(k1)+(0ex,0.15ex)$) [anchor=south]{\tiny{$\k_1$}};
\node at ($(k4)+(0ex,0.15ex)$) [anchor=south]{\tiny{$\k_{2}$}};
}
}
\def\diagramextwoverticestwo{\tikz[baseline=-1.4ex,scale=1.8, every node/.style={scale=1.4}]{
\coordinate (k1) at (-2.5ex,0ex);
\coordinate (k4) at (+2.5ex,0ex);
\coordinate (t1) at (-2ex,-4ex);
\coordinate (t2) at (+2ex,-4ex);
\pgfmathsetmacro{\arista}{0.06}
\draw[thick] (-3.5ex,0ex) -- (3.5ex,0ex);
\draw[-,thick] (t1) -- (k1);
\draw[-,thick] (t2) -- (k4);
\draw[thick,scale=2] (t1) to [bend left=60] (t2);
\draw[thick,scale=2] (t1) to [bend left=-60] (t2);
\filldraw[color=black, fill=black, thick] (t1) circle (0.45ex);
\filldraw[color=black, fill=white, thick] (t2) circle (0.45ex);
\filldraw[color=black, fill=white, thick] ($(k1)-(\arista,\arista)$) rectangle ($(k1)+(\arista,\arista)$);
\filldraw[color=black, fill=white, thick] ($(k4)-(\arista,\arista)$) rectangle ($(k4)+(\arista,\arista)$);
\node at ($(k1)+(0ex,0.15ex)$) [anchor=south]{\tiny{$\k_1$}};
\node at ($(k4)+(0ex,0.15ex)$) [anchor=south]{\tiny{$\k_{2}$}};
}
}
\be \label{2-point-2-vertex-1-loop}
  \diagramextwoverticesNoLabels 
 =  2 {\rm Re} \Bigg\{  \diagramextwoverticestwo  +  \diagramextwoverticesone   \Bigg\} .  
\ee
The more general case with $n_1$ legs attached to the first vertex and $n_2$ legs attached to the second vertex is completely analogous. In addition, note that in the following discussion we will stick to $d=3$ dimensions. As we shall see, our claim regarding logarithms does not require us to resort to dimensional regularization.

To obtain analytical expressions for the diagrams at the right hand side of (\ref{2-point-2-vertex-1-loop}) it is convenient to define the following functions capturing the loop integrals:
\bea 
 F_{+-} (k , \tau_1 , \tau_2) &=& \int_{{\bf q}_1}  \int_{ {\bf q}_2 } (2\pi)^3 \delta^{(3)} (\k - \q_1 - \q_2 ) \nn \\
 && \!\!\!\!\!\!\!\!\!\!\!\! \times G_{+-} (q_1 , \tau_1 , \tau_2) G_{+-} (q_{2} , \tau_1 , \tau_2),  \label{def-F}
 \\
F_{++} (k , \tau_1 , \tau_2) &=& \int_{{\bf q}_1}  \int_{ {\bf q}_2 } (2\pi)^3 \delta^{(3)} (\k - \q_1 - \q_2 ) \nn \\
 && \!\!\!\!\!\!\!\!\!\!\!\! \times G_{++} (q_1 , \tau_1 , \tau_2) G_{++} (q_{2} , \tau_1 , \tau_2) .  \label{def-F-2}
\eea
One may also define $F_{-+} = F_{+-}^*$ and $F_{--}=F_{++}^*$. With these definitions, we can now write the two diagrams appearing on the right hand side of (\ref{2-point-2-vertex-1-loop}) as 
\bea \label{black-white-loop}
\diagramextwoverticestwo &=& \frac{ \lambda_{3}^2}{4  H^8} (2\pi)^3 \delta^{(3)} (\k_1 + \k_2)  \int^{\tau}_{-\infty} \!\! \frac{d\tau_1}{\tau_1^4} \nn \\ [-25pt]
&&    \int^{\tau}_{-\infty} \!\! \frac{d\tau_2}{\tau_2^4}    F_{+-} (k , \tau_1 , \tau_2) \nn \\
&&   G_{+} (k , \tau_1 , \tau)   G_{-} (k , \tau_2 , \tau)   ,
\eea
and 
\bea \label{black-black-loop}
 \diagramextwoverticesone   &=& - \frac{ \lambda_{3}^2}{4  H^8} (2\pi)^3 \delta^{(3)} (\k_1 + \k_2)  \int^{\tau}_{-\infty} \!\! \frac{d\tau_1}{\tau_1^4} \nn \\ [-25pt]
&&    \int^{\tau}_{-\infty} \!\! \frac{d\tau_2}{\tau_2^4}     F_{++} (k , \tau_1 , \tau_2) \nn \\
&&   G_{+} (k , \tau_1 , \tau)   G_{+} (k , \tau_2 , \tau)   ,
\eea
where $\k = \k_1 = - \k_2$. Notice the presence of the additional symmetry factor 2, due to the equal number of external legs attached to each vertex. By comparing these results with their tree-level counterparts studied in Section~\ref{sec:example-logs-2}, one can see that the only difference is that in these analytical expressions $F_{+-}$ and $F_{++}$ appear instead $G_{+-}$ and $G_{++}$. In fact, just as bulk-to-bulk propagators do, the functions $F_{+-}$ and $F_{++}$ satisfy the relation:
\bea
F_{++} (k , \tau_1 , \tau_2) &=& F_{-+}(k , \tau_1 , \tau_2) \theta(\tau_1 - \tau_2) \nn \\
&& +F_{+-} (k , \tau_1 , \tau_2)\theta(\tau_2 - \tau_1) , \\
F_{--} (k , \tau_1 , \tau_2) &=& F_{+-}(k , \tau_1 , \tau_2) \theta(\tau_1 - \tau_2) \nn \\
&& +F_{-+} (k , \tau_1 , \tau_2)\theta(\tau_2 - \tau_1) . \qquad
\eea
Moreover, $F_{+-}$ and $F_{++}$ satisfy the same transformation rules as $G_{+-}$ and $G_{++}$ under de Sitter transformations. Thanks to this, we see that the structure of the nested integrals in (\ref{black-black-loop}) is primarily determined by the form of $F_{+-}$. This being the case, let us focus our attention exclusively on the analytic form of $F_{+-}$ defined in (\ref{def-F}).

If one directly integrates in (\ref{def-F}) over one of the two internal momenta flowing through the loop, say $\q_2$, one obtains:
\be \label{Fpm-int}
 F_{+-} (k , \tau_1 , \tau_2) = \int_{{\bf q}}   G_{+-} (q , \tau_1 , \tau_2) G_{+-} ( |\k - \q| , \tau_1 , \tau_2) .
\ee
Thanks to the fact that $G_{+-} (k , \tau_1 , \tau_2) \sim k^{-3}$, it is possible to see that the integrand presents two IR divergent configuration in $q$-space: One at $\q = 0$ and the other at $\q = \k$.  Due to the third Wilson's axiom (\ref{translation-invariance}) one can see that both divergences are, in fact, equivalent. To make this property apparent, let us take a step back and integrate $F_{+-}$ in Eq.~(\ref{def-F}) with the help of the following identity involving a Dirac delta:
\bea \label{dirac}
&&\int_{{\bf q}_1}  \int_{{\bf q}_2}  (2\pi)^3  \delta^{(3)} ( \q_1 + \q_2 - \k) \cdots
\qquad  \nn \\
&& \qquad \qquad \qquad
 = \frac{k^3}{16 \pi^2}  \int_1^{\infty} \!\!\!\! d x  \int_{- 1}^{+1} \!\!\!\!\!\! d y  \, (x^2 -y^2)   \cdots   , \qquad
\eea
where we have defined $x = (q_1 + q_2)/k$ and $y = (q_1 - q_2)/k$. Note that we are employing Wilson's second axiom ensuring the scaling of the integrals. Using this relation back in~(\ref{def-F}), we get
\bea  \label{Fpm-x-y-tau}
 F_{+-} (k , \tau_1 , \tau_2) &=& \frac{ H^4}{ 2 \pi^{2}  k^3}  \int_1^{\infty} \!\!\!\! d x \, e^{ i x k (\tau_1 - \tau_2)} \int_{-1}^{+1} \!\!\!\!    \frac{d y}{ (x^2 - y^2)^2} \nn \\
 &&\!\!\!\!\!\! \left(1 - i \frac{x+y}{2} k \tau_1\right)\left(1 + i \frac{x+y}{2} k \tau_2\right) \nn \\
 &&\!\!\!\!\!\!
 \left(1 - i \frac{x-y}{2} k \tau_1\right)\left(1 + i \frac{x-y}{2} k \tau_2\right) . 
\eea
It can be seen that $ F_{+-} (k , \tau_1 , \tau_2)$ is a function only of $k\tau_1$ and $k\tau_2$. In addition, the $y$-integral must be a polynomial of $k \tau_1$ and $k \tau_2$ of order $4$ with coefficients that depend on the variable $x$. Performing the $y$ integral, one indeed finds:
\bea \label{Fpm-x-tau}
 F_{+-} (k , \tau_1 , \tau_2) &=& \frac{ H^4}{ 2 \pi^{2}  k^3}  \int_1^{\infty} \!\! d x e^{ i x k (\tau_1 - \tau_2)} \nn \\
&&  \!\!\!\!\!\!\!\!\!\!\!\!\!\!\!\!\!\!\!\! \times \Bigg[   \frac{(1-i x k \tau_1)(1+i x k \tau_2)}{x^2 (x^2-1)} + \frac{k^4 \tau_1^2 \tau_2^2 }{8} \nn \\ 
 && 
 \!\!\!\!\!\!\!\!\!\!\!\!\!\!\!\!\!\!\!\!
 + \frac{1}{4x^3} \Big(2 (1-i x k \tau_1)(1+i x k \tau_2)   - x^2 k^2(\tau_1^2 + \tau_2^2) \nn \\
 && 
 \!\!\!\!\!\!\!\!\!\!\!\!\!\!\!\!\!\!\!\!
 - i x^3 k^3 \tau_1 \tau_2( \tau_1 - \tau_2) \Big)  \ln \left[ \frac{x+1}{x-1} \right] \Bigg]
 .
\eea
Remarkably, the two divergences found in (\ref{Fpm-int}) have merged into a single divergence at $x=1$.

Only the first term of the second line in (\ref{Fpm-x-tau}) diverges towards the lower limit $x=1$. Thus, with the purpose of isolating the IR divergent contribution, let us define:
\bea \label{Fpm-x-tau-IR}
 F_{+-}^{\rm IR} (k , \tau_1 , \tau_2) &\equiv&   \frac{ H^2}{ 2   k^3}  (1-i  k \tau_1)(1+i  k \tau_2) e^{ i  k (\tau_1 - \tau_2)}  \nn \\ 
 && \times \frac{H^2}{\pi^{2}} \int_1^{\infty} \!\!  
 \frac{d x}{ x^2-1}  . \qquad
\eea
Notice that this corresponds to the divergent part of the full integral (\ref{Fpm-x-tau}), in such a way that the following combination
\be \label{Fpm-x-tau-fin}
F_{+-}^{\rm rest} (k , \tau_1 , \tau_2) = F_{+-} (k , \tau_1 , \tau_2) - F_{+-}^{\rm IR} (k , \tau_1 , \tau_2)
\ee
is IR-finite but still contains UV divergent parts that need to be dealt with separately. The outstanding aspect of this result is that $F_{+-}^{\rm IR} (k , \tau_1 , \tau_2)$ is just proportional to $G_{+-}(k,\tau_1 , \tau_2)$:
\be \label{Fpm-x-tau-IR-1}
 F_{+-}^{\rm IR} (k , \tau_1 , \tau_2) =   G_{+-}(k,\tau_1 , \tau_2)  \frac{H^2}{\pi^{2}} \int_1^{\infty} \!\!   
 \frac{d x}{ x^2 -1 }  .
\ee
In other words, the IR-divergent contribution coming from the loop under scrutiny is equivalent to a tree-level diagram multiplied by an overall infinite factor. We will examine how to deal with this divergence in the next subsection.

We can now assess the appearance of logarithms induced by the loop-integral $F_{+-} (k , \tau_1 , \tau_2)$. First, the divergent contribution (\ref{Fpm-x-tau-IR}), which is only proportional to a bulk-to-bulk propagator $G_{+-}(k,\tau_1 , \tau_2)$, reproduces the tree-level result already discussed in Section~\ref{sec:example-logs-2}, and so it does not lead to the appearance of extra logarithms. On the other hand, by expanding the integrand of the finite part $F_{+-}^{\rm rest} (k , \tau_1 , \tau_2)$ around $x=1$, we see that:
\bea \label{Fpm-x-tau-fin-2}
F_{+-}^{\rm rest}  &=& \frac{ H^4}{ 4 \pi^{2}  k^3}  \int_1^{\infty} \!\!\!\! d x e^{i k (\tau_1 - \tau_2)}\bigg[ \frac{k^4 \tau_1^2 \tau_2^2}{8}  \nn \\
&& + k^2 \tau_1^2 + k^2 \tau_2^2 + i k^3 \tau_1 \tau_2 (\tau_1 - \tau_2) \nn \\
&& + \Bigg(1 - \frac{k^2}{2} (\tau_1 - \tau_2)^2 - i k (\tau_1 - \tau_2) \nn \\
&& - \frac{1}{2} i k^3 \tau_1 \tau_2 (\tau_1 - \tau_2)\Bigg) \ln \left[ \frac{1}{x-1} \right] \nn \\ 
&& + \mathcal O (x-1) \bigg] .
\eea
One may now check explicitly that this form of $F_{+-}^{\rm rest} $ is unable to contribute logarithms beyond those already predicted in Section~\ref{sec:example-logs-2}. To verify this without the explicit integration of the time variable, note that the real and imaginary parts of $F_{+-}^{\rm rest} $ are given by:
\bea
{\rm Re} \Big\{ F_{+-}^{\rm rest} \Big\} &=& \frac{ H^4}{ 4 \pi^{2}  k^3}  \int_1^{\infty} \!\!\!\! d x \bigg[  \ln \left[ \frac{1}{x-1} \right]  \nn \\
&& +  k^2 (\tau_1^2 + \tau_2^2 ) + \cdots \bigg] , \\ 
{\rm Im} \Big\{ F_{+-}^{\rm rest} \Big\} &=& \frac{ H^4}{ 4 \pi^{2} }  \int_1^{\infty} \!\!\!\! d x  \bigg[ (\tau_1^3 - \tau_2^3)  \nn \\
&& \times  \Big(1 + \frac{1}{6} \ln \left[ \frac{1}{x-1} \right]  \Big)  + \cdots \bigg] .
\eea
By recalling our analysis regarding the splitting of propagators into real and imaginary parts, we see that these expressions precisely imply that a loop carrying external momentum will be unable to produce extra logarithms. Given that there is always an imaginary propagator attached to any vertex and each imaginary part is cubic with respect to the time variables, it follows that each time integral can at most contribute a single logarithm of the form (\ref{log-structure}).

\subsection{Confronting IR divergences}
\label{sec:IR-divergences}

With the exception of those appearing in daisy diagrams, IR divergences are fundamentally different from their UV counterparts in that they cannot be canceled by local counterterms. This is evident from the fact that the IR-divergent contribution to $F_{+-}$ is proportional to the bulk-to-bulk propagator $G_{+-}$. More generally, the IR-divergent part of any diagram consisting of two interaction vertices connected by two propagators forming a single loop is proportional to a tree-level diagram in which the same two vertices are connected by a single propagator. This is illustrated in the following example:
\def\diagramextwoverticesNoLabels{\tikz[baseline=-1.4ex,scale=1.8, every node/.style={scale=1.4}]{
\coordinate (k1) at (-4ex,0ex);
\coordinate (k2) at (-1ex,0ex);
\coordinate (k3) at (+1ex,0ex);
\coordinate (k4) at (+4ex,0ex);
\coordinate (t1) at (-2.5ex,-4ex);
\coordinate (t2) at (+2.5ex,-4ex);
\coordinate (P) at ($(-2.5ex,-1.0ex)$);
\coordinate (Q) at ($(2.5ex,-1.0ex)$);
\pgfmathsetmacro{\arista}{0.06}
\draw[thick] (-5ex,0ex) -- (5ex,0ex);
\draw[-,thick] (t1) -- (k1);
\draw[-,thick] (t1) -- (k2);
\draw[-,thick] (t2) -- (k3);
\draw[-,thick] (t2) -- (k4);
\draw[thick,scale=2] (t1) to [bend left=45] (t2);
\draw[thick,scale=2] (t1) to [bend left=-45] (t2);
\fill[white] (t1) circle (0.45ex);
\filldraw[pattern=north east lines, thick] (t1) circle (0.45ex);
\fill[white] (t2) circle (0.45ex);
\filldraw[pattern=north east lines, thick] (t2) circle (0.45ex);
\filldraw[color=black, fill=white, thick] ($(k1)-(\arista,\arista)$) rectangle ($(k1)+(\arista,\arista)$);
\filldraw[color=black, fill=white, thick] ($(k2)-(\arista,\arista)$) rectangle ($(k2)+(\arista,\arista)$);
\filldraw[color=black, fill=white, thick] ($(k3)-(\arista,\arista)$) rectangle ($(k3)+(\arista,\arista)$);
\filldraw[color=black, fill=white, thick] ($(k4)-(\arista,\arista)$) rectangle ($(k4)+(\arista,\arista)$);
\filldraw[color=black, fill=black, thick] ($(Q)+(-0.5ex,0ex)$) circle (0.05ex);
\filldraw[color=black, fill=black, thick] (Q) circle (0.05ex);
\filldraw[color=black, fill=black, thick] ($(Q)+(0.5ex,0ex)$) circle (0.05ex);
\filldraw[color=black, fill=black, thick] ($(P)+(-0.5ex,0ex)$) circle (0.05ex);
\filldraw[color=black, fill=black, thick] (P) circle (0.05ex);
\filldraw[color=black, fill=black, thick] ($(P)+(0.5ex,0ex)$) circle (0.05ex);
}
}
\def\diagramextwovertextree{\tikz[baseline=-1.4ex,scale=1.8, every node/.style={scale=1.4}]{
\coordinate (k1) at (-4ex,0ex);
\coordinate (k2) at (-1ex,0ex);
\coordinate (k3) at (+1ex,0ex);
\coordinate (k4) at (+4ex,0ex);
\coordinate (t1) at (-2.5ex,-4ex);
\coordinate (t2) at (+2.5ex,-4ex);
\coordinate (P) at ($(-2.5ex,-1.0ex)$);
\coordinate (Q) at ($(2.5ex,-1.0ex)$);
\pgfmathsetmacro{\arista}{0.06}
\draw[thick] (-5ex,0ex) -- (5ex,0ex);
\draw[-,thick] (t1) -- (k1);
\draw[-,thick] (t1) -- (k2);
\draw[-,thick] (t2) -- (k3);
\draw[-,thick] (t2) -- (k4);
\draw[-,thick] (t1) -- (t2);
\fill[white] (t1) circle (0.45ex);
\filldraw[pattern=north east lines, thick] (t1) circle (0.45ex);
\fill[white] (t2) circle (0.45ex);
\filldraw[pattern=north east lines, thick] (t2) circle (0.45ex);
\filldraw[color=black, fill=white, thick] ($(k1)-(\arista,\arista)$) rectangle ($(k1)+(\arista,\arista)$);
\filldraw[color=black, fill=white, thick] ($(k2)-(\arista,\arista)$) rectangle ($(k2)+(\arista,\arista)$);
\filldraw[color=black, fill=white, thick] ($(k3)-(\arista,\arista)$) rectangle ($(k3)+(\arista,\arista)$);
\filldraw[color=black, fill=white, thick] ($(k4)-(\arista,\arista)$) rectangle ($(k4)+(\arista,\arista)$);
\filldraw[color=black, fill=black, thick] ($(Q)+(-0.5ex,0ex)$) circle (0.05ex);
\filldraw[color=black, fill=black, thick] (Q) circle (0.05ex);
\filldraw[color=black, fill=black, thick] ($(Q)+(0.5ex,0ex)$) circle (0.05ex);
\filldraw[color=black, fill=black, thick] ($(P)+(-0.5ex,0ex)$) circle (0.05ex);
\filldraw[color=black, fill=black, thick] (P) circle (0.05ex);
\filldraw[color=black, fill=black, thick] ($(P)+(0.5ex,0ex)$) circle (0.05ex);
}
}
\be \label{IR-div-comp-tree}
\Bigg[ \diagramextwoverticesNoLabels \Bigg]_{\rm IR \,\, div}  \propto \diagramextwovertextree .
\ee
This relation holds regardless of the number of external legs attached to the interaction vertices. Crucially, no local operator can generate a single-vertex diagram proportional to the right-hand side of Eq.~(\ref{IR-div-comp-tree}), confirming that the associated divergence is inherently nonlocal and cannot be removed through conventional UV renormalization techniques.

Of course, as already emphasized elsewhere, IR divergences are expected to be resolved upon resummation of contributions order by order in the number of interaction vertices. For instance, a concrete way to address the divergence in the diagram of Eq.~(\ref{2-point-2-vertex-1-loop}) is to consider the insertion of mass counterterms proportional to $\lambda_2$, as shown in the following sum:
\def\resummationone{\tikz[baseline=-1.4ex,scale=1.8, every node/.style={scale=1.4}]{
\coordinate (k1) at (-4ex,0ex);
\coordinate (k2) at (-1ex,0ex);
\coordinate (k3) at (+1ex,0ex);
\coordinate (k4) at (+4ex,0ex);
\coordinate (t1) at (-2.5ex,-4ex);
\coordinate (t2) at (+2.5ex,-4ex);
\coordinate (i1) at (+0ex,-5ex);
\coordinate (P) at ($(-2.5ex,-1.0ex)$);
\coordinate (Q) at ($(2.5ex,-1.0ex)$);
\pgfmathsetmacro{\arista}{0.06}
\draw[thick] (-5ex,0ex) -- (5ex,0ex);
\draw[-,thick] (t1) -- (k1);
\draw[-,thick] (t1) -- (k2);
\draw[-,thick] (t2) -- (k3);
\draw[-,thick] (t2) -- (k4);
\draw[thick,scale=2] (t1) to [bend left=45] (t2);
\draw[thick,scale=2] (t1) to [bend left=-45] (t2);
\fill[white] (t1) circle (0.45ex);
\filldraw[pattern=north east lines, thick] (t1) circle (0.45ex);
\fill[white] (t2) circle (0.45ex);
\filldraw[pattern=north east lines, thick] (t2) circle (0.45ex);
\fill[white] (i1) circle (0.45ex);
\filldraw[pattern=north east lines, thick] (i1) circle (0.45ex);
\filldraw[color=black, fill=white, thick] ($(k1)-(\arista,\arista)$) rectangle ($(k1)+(\arista,\arista)$);
\filldraw[color=black, fill=white, thick] ($(k2)-(\arista,\arista)$) rectangle ($(k2)+(\arista,\arista)$);
\filldraw[color=black, fill=white, thick] ($(k3)-(\arista,\arista)$) rectangle ($(k3)+(\arista,\arista)$);
\filldraw[color=black, fill=white, thick] ($(k4)-(\arista,\arista)$) rectangle ($(k4)+(\arista,\arista)$);
\filldraw[color=black, fill=black, thick] ($(Q)+(-0.5ex,0ex)$) circle (0.05ex);
\filldraw[color=black, fill=black, thick] (Q) circle (0.05ex);
\filldraw[color=black, fill=black, thick] ($(Q)+(0.5ex,0ex)$) circle (0.05ex);
\filldraw[color=black, fill=black, thick] ($(P)+(-0.5ex,0ex)$) circle (0.05ex);
\filldraw[color=black, fill=black, thick] (P) circle (0.05ex);
\filldraw[color=black, fill=black, thick] ($(P)+(0.5ex,0ex)$) circle (0.05ex);
}
}
\def\resummationtwo{\tikz[baseline=-1.4ex,scale=1.8, every node/.style={scale=1.4}]{
\coordinate (k1) at (-4ex,0ex);
\coordinate (k2) at (-1ex,0ex);
\coordinate (k3) at (+1ex,0ex);
\coordinate (k4) at (+4ex,0ex);
\coordinate (t1) at (-2.5ex,-4ex);
\coordinate (t2) at (+2.5ex,-4ex);
\coordinate (i1) at (+0ex,-3ex);
\coordinate (P) at ($(-2.5ex,-1.0ex)$);
\coordinate (Q) at ($(2.5ex,-1.0ex)$);
\pgfmathsetmacro{\arista}{0.06}
\draw[thick] (-5ex,0ex) -- (5ex,0ex);
\draw[-,thick] (t1) -- (k1);
\draw[-,thick] (t1) -- (k2);
\draw[-,thick] (t2) -- (k3);
\draw[-,thick] (t2) -- (k4);
\draw[thick,scale=2] (t1) to [bend left=45] (t2);
\draw[thick,scale=2] (t1) to [bend left=-45] (t2);
\fill[white] (t1) circle (0.45ex);
\filldraw[pattern=north east lines, thick] (t1) circle (0.45ex);
\fill[white] (t2) circle (0.45ex);
\filldraw[pattern=north east lines, thick] (t2) circle (0.45ex);
\fill[white] (i1) circle (0.45ex);
\filldraw[pattern=north east lines, thick] (i1) circle (0.45ex);
\filldraw[color=black, fill=white, thick] ($(k1)-(\arista,\arista)$) rectangle ($(k1)+(\arista,\arista)$);
\filldraw[color=black, fill=white, thick] ($(k2)-(\arista,\arista)$) rectangle ($(k2)+(\arista,\arista)$);
\filldraw[color=black, fill=white, thick] ($(k3)-(\arista,\arista)$) rectangle ($(k3)+(\arista,\arista)$);
\filldraw[color=black, fill=white, thick] ($(k4)-(\arista,\arista)$) rectangle ($(k4)+(\arista,\arista)$);
\filldraw[color=black, fill=black, thick] ($(Q)+(-0.5ex,0ex)$) circle (0.05ex);
\filldraw[color=black, fill=black, thick] (Q) circle (0.05ex);
\filldraw[color=black, fill=black, thick] ($(Q)+(0.5ex,0ex)$) circle (0.05ex);
\filldraw[color=black, fill=black, thick] ($(P)+(-0.5ex,0ex)$) circle (0.05ex);
\filldraw[color=black, fill=black, thick] (P) circle (0.05ex);
\filldraw[color=black, fill=black, thick] ($(P)+(0.5ex,0ex)$) circle (0.05ex);
}
}
\def\resummationthree{\tikz[baseline=-1.4ex,scale=1.8, every node/.style={scale=1.4}]{
\coordinate (k1) at (-4ex,0ex);
\coordinate (k2) at (-1ex,0ex);
\coordinate (k3) at (+1ex,0ex);
\coordinate (k4) at (+4ex,0ex);
\coordinate (t1) at (-2.5ex,-4ex);
\coordinate (t2) at (+2.5ex,-4ex);
\coordinate (i1) at (-0.8ex,-4.9ex);
\coordinate (i2) at (+0.8ex,-4.9ex);
\coordinate (P) at ($(-2.5ex,-1.0ex)$);
\coordinate (Q) at ($(2.5ex,-1.0ex)$);
\pgfmathsetmacro{\arista}{0.06}
\draw[thick] (-5ex,0ex) -- (5ex,0ex);
\draw[-,thick] (t1) -- (k1);
\draw[-,thick] (t1) -- (k2);
\draw[-,thick] (t2) -- (k3);
\draw[-,thick] (t2) -- (k4);
\draw[thick,scale=2] (t1) to [bend left=45] (t2);
\draw[thick,scale=2] (t1) to [bend left=-45] (t2);
\fill[white] (t1) circle (0.45ex);
\filldraw[pattern=north east lines, thick] (t1) circle (0.45ex);
\fill[white] (t2) circle (0.45ex);
\filldraw[pattern=north east lines, thick] (t2) circle (0.45ex);
\fill[white] (i1) circle (0.45ex);
\filldraw[pattern=north east lines, thick] (i1) circle (0.45ex);
\fill[white] (i2) circle (0.45ex);
\filldraw[pattern=north east lines, thick] (i2) circle (0.45ex);
\filldraw[color=black, fill=white, thick] ($(k1)-(\arista,\arista)$) rectangle ($(k1)+(\arista,\arista)$);
\filldraw[color=black, fill=white, thick] ($(k2)-(\arista,\arista)$) rectangle ($(k2)+(\arista,\arista)$);
\filldraw[color=black, fill=white, thick] ($(k3)-(\arista,\arista)$) rectangle ($(k3)+(\arista,\arista)$);
\filldraw[color=black, fill=white, thick] ($(k4)-(\arista,\arista)$) rectangle ($(k4)+(\arista,\arista)$);
\filldraw[color=black, fill=black, thick] ($(Q)+(-0.5ex,0ex)$) circle (0.05ex);
\filldraw[color=black, fill=black, thick] (Q) circle (0.05ex);
\filldraw[color=black, fill=black, thick] ($(Q)+(0.5ex,0ex)$) circle (0.05ex);
\filldraw[color=black, fill=black, thick] ($(P)+(-0.5ex,0ex)$) circle (0.05ex);
\filldraw[color=black, fill=black, thick] (P) circle (0.05ex);
\filldraw[color=black, fill=black, thick] ($(P)+(0.5ex,0ex)$) circle (0.05ex);
}
}
\def\resummationfour{\tikz[baseline=-1.4ex,scale=1.8, every node/.style={scale=1.4}]{
\coordinate (k1) at (-4ex,0ex);
\coordinate (k2) at (-1ex,0ex);
\coordinate (k3) at (+1ex,0ex);
\coordinate (k4) at (+4ex,0ex);
\coordinate (t1) at (-2.5ex,-4ex);
\coordinate (t2) at (+2.5ex,-4ex);
\coordinate (i1) at (-0.8ex,-3.1ex);
\coordinate (i2) at (+0.8ex,-3.1ex);
\coordinate (P) at ($(-2.5ex,-1.0ex)$);
\coordinate (Q) at ($(2.5ex,-1.0ex)$);
\pgfmathsetmacro{\arista}{0.06}
\draw[thick] (-5ex,0ex) -- (5ex,0ex);
\draw[-,thick] (t1) -- (k1);
\draw[-,thick] (t1) -- (k2);
\draw[-,thick] (t2) -- (k3);
\draw[-,thick] (t2) -- (k4);
\draw[thick,scale=2] (t1) to [bend left=45] (t2);
\draw[thick,scale=2] (t1) to [bend left=-45] (t2);
\fill[white] (t1) circle (0.45ex);
\filldraw[pattern=north east lines, thick] (t1) circle (0.45ex);
\fill[white] (t2) circle (0.45ex);
\filldraw[pattern=north east lines, thick] (t2) circle (0.45ex);
\fill[white] (i1) circle (0.45ex);
\filldraw[pattern=north east lines, thick] (i1) circle (0.45ex);
\fill[white] (i2) circle (0.45ex);
\filldraw[pattern=north east lines, thick] (i2) circle (0.45ex);
\filldraw[color=black, fill=white, thick] ($(k1)-(\arista,\arista)$) rectangle ($(k1)+(\arista,\arista)$);
\filldraw[color=black, fill=white, thick] ($(k2)-(\arista,\arista)$) rectangle ($(k2)+(\arista,\arista)$);
\filldraw[color=black, fill=white, thick] ($(k3)-(\arista,\arista)$) rectangle ($(k3)+(\arista,\arista)$);
\filldraw[color=black, fill=white, thick] ($(k4)-(\arista,\arista)$) rectangle ($(k4)+(\arista,\arista)$);
\filldraw[color=black, fill=black, thick] ($(Q)+(-0.5ex,0ex)$) circle (0.05ex);
\filldraw[color=black, fill=black, thick] (Q) circle (0.05ex);
\filldraw[color=black, fill=black, thick] ($(Q)+(0.5ex,0ex)$) circle (0.05ex);
\filldraw[color=black, fill=black, thick] ($(P)+(-0.5ex,0ex)$) circle (0.05ex);
\filldraw[color=black, fill=black, thick] (P) circle (0.05ex);
\filldraw[color=black, fill=black, thick] ($(P)+(0.5ex,0ex)$) circle (0.05ex);
}
}
\def\resummationfive{\tikz[baseline=-1.4ex,scale=1.8, every node/.style={scale=1.4}]{
\coordinate (k1) at (-4ex,0ex);
\coordinate (k2) at (-1ex,0ex);
\coordinate (k3) at (+1ex,0ex);
\coordinate (k4) at (+4ex,0ex);
\coordinate (t1) at (-2.5ex,-4ex);
\coordinate (t2) at (+2.5ex,-4ex);
\coordinate (i1) at (0ex,-5ex);
\coordinate (i2) at (0ex,-3ex);
\coordinate (P) at ($(-2.5ex,-1.0ex)$);
\coordinate (Q) at ($(2.5ex,-1.0ex)$);
\pgfmathsetmacro{\arista}{0.06}
\draw[thick] (-5ex,0ex) -- (5ex,0ex);
\draw[-,thick] (t1) -- (k1);
\draw[-,thick] (t1) -- (k2);
\draw[-,thick] (t2) -- (k3);
\draw[-,thick] (t2) -- (k4);
\draw[thick,scale=2] (t1) to [bend left=45] (t2);
\draw[thick,scale=2] (t1) to [bend left=-45] (t2);
\fill[white] (t1) circle (0.45ex);
\filldraw[pattern=north east lines, thick] (t1) circle (0.45ex);
\fill[white] (t2) circle (0.45ex);
\filldraw[pattern=north east lines, thick] (t2) circle (0.45ex);
\fill[white] (i1) circle (0.45ex);
\filldraw[pattern=north east lines, thick] (i1) circle (0.45ex);
\fill[white] (i2) circle (0.45ex);
\filldraw[pattern=north east lines, thick] (i2) circle (0.45ex);
\filldraw[color=black, fill=white, thick] ($(k1)-(\arista,\arista)$) rectangle ($(k1)+(\arista,\arista)$);
\filldraw[color=black, fill=white, thick] ($(k2)-(\arista,\arista)$) rectangle ($(k2)+(\arista,\arista)$);
\filldraw[color=black, fill=white, thick] ($(k3)-(\arista,\arista)$) rectangle ($(k3)+(\arista,\arista)$);
\filldraw[color=black, fill=white, thick] ($(k4)-(\arista,\arista)$) rectangle ($(k4)+(\arista,\arista)$);
\filldraw[color=black, fill=black, thick] ($(Q)+(-0.5ex,0ex)$) circle (0.05ex);
\filldraw[color=black, fill=black, thick] (Q) circle (0.05ex);
\filldraw[color=black, fill=black, thick] ($(Q)+(0.5ex,0ex)$) circle (0.05ex);
\filldraw[color=black, fill=black, thick] ($(P)+(-0.5ex,0ex)$) circle (0.05ex);
\filldraw[color=black, fill=black, thick] (P) circle (0.05ex);
\filldraw[color=black, fill=black, thick] ($(P)+(0.5ex,0ex)$) circle (0.05ex);
}
}
\bea \label{lambda-2-resummation}
&& \diagramextwoverticesNoLabels + \resummationone \nn  \\  
&& + \resummationtwo  + \resummationthree  \nn \\ 
&& + \resummationfour + \resummationfive + \cdots \qquad
\eea
In this case, within the relevant IR limit under consideration, each $\lambda_2$ insertion contributes a factor of $\lambda_2 \ln(q|\tau|)$ after performing the time integrals, where $q$ denotes the comoving momentum flowing through the vertex. As a result, the resummed expression in Eq.~(\ref{lambda-2-resummation}) is equivalent to a loop diagram in which the two internal lines correspond to propagators of a massive free theory with mass parameter $\lambda_2$. Naturally, these massive propagators yield IR-convergent loop integrals.

However, the resummation presented in Eq.~(\ref{lambda-2-resummation}) omits other contributions of the same order in the number of vertices and therefore does not capture the full loop-corrected result at second order in the vertex expansion. It is important to recall that, in the case of light scalar fields, all interaction vertices must be treated on equal footing. This is because the time dependence of correlation functions is governed primarily by the total number of vertices, regardless of the number of legs attached to each one.

This state of affairs should not prevent us from identifying the form of the final loop-corrected contribution to the diagram in Eq.~(\ref{2-point-2-vertex-1-loop}) that dominates in the infrared. First, we recall that in the class of interactions considered here, integrals cannot be extended all the way down to $k = 0$, including those arising in loop diagrams. This restriction stems from the existence of a strongly nonlinear scale $\Lambda_{\rm IR}$, which acts as a physical cutoff for momentum integrals. Beyond this scale, nonlinear effects become dominant and must be resummed—a task that currently remains intractable. In the specific case under consideration, this implies that the IR divergence in Eq.~(\ref{Fpm-x-tau-IR-1}) can be equivalently recast as
\be \label{Fpm-x-tau-IR-2}
F_{+-}^{\rm IR} (k , \tau_1 , \tau_2) = G_{+-}(k,\tau_1 , \tau_2) \frac{H^2}{2 \pi^{2}} \ln \frac{H}{\Lambda_{\rm IR}} .
\ee
With this expression in hand, it becomes straightforward to construct nonlocal operators proportional to $\ln(H/\Lambda_{\rm IR})$, which generate Feynman rules such that the resulting nonlocal vertex reproduces the diagram on the right-hand side of Eq.~(\ref{IR-div-comp-tree}). Such an operator would act as a counterterm, subtracting the divergent contribution in Eq.~(\ref{Fpm-x-tau-IR-2}), which is itself proportional to $\ln(H/\Lambda_{\rm IR})$. Although we do not present the explicit form of this operator—since it is not particularly illuminating—we emphasize that it would encapsulate the nonlinear physics associated with modes of wavelength longer than $\Lambda_{\rm IR}^{-1}$.

Now, it is important to point out that the procedure used to isolate the IR divergence in $F_{+-}$—in order to extract the IR-finite part given in Eq.~(\ref{Fpm-x-tau-fin})—involves a degree of arbitrariness, which is typical in perturbative calculations. In defining $F_{+-}^{\rm IR}$ in Eq.~(\ref{Fpm-x-tau-IR}), nothing prevents us from adding additional terms proportional to the bulk-to-bulk propagator $G_{+-}$ without altering the conclusions of the present discussion. However, since physical observables must be independent of such ambiguities, we are led to conclude that there exists a unique de Sitter invariant form that captures the full loop-corrected structure of $F_{+-}$. That is, the complete result must take the form
\be \label{Fpm-x-tau-IR-3}
F_{+-} (k , \tau_1 , \tau_2) = F^{\rm rest}_{+-} (k , \tau_1 , \tau_2) + c_{1} \, G_{+-}(k,\tau_1 , \tau_2) ,
\ee
where $F^{\rm rest}_{+-}$ is the finite part defined in Eq.~(\ref{Fpm-x-tau-fin-2}), and $c_1$ is a finite constant capturing the contribution from the infrared sector of the loop integral.

Let us conclude this section by speculating on the structure of higher-order loop corrections to two-vertex diagrams. Having established that the leading IR contribution from a one-loop correction yields a term proportional to the corresponding tree-level diagram, it is tempting to conjecture that this pattern persists at higher orders. That is, the IR-divergent part of multi-loop diagrams involving two vertices may also reduce to a tree-level diagram with the same vertex structure, as suggested by the following expression:
\def\diagramextwoverticesNoLabels{\tikz[baseline=-1.4ex,scale=1.8, every node/.style={scale=1.4}]{
\coordinate (k1) at (-4ex,0ex);
\coordinate (k2) at (-1ex,0ex);
\coordinate (k3) at (+1ex,0ex);
\coordinate (k4) at (+4ex,0ex);
\coordinate (t1) at (-2.5ex,-4ex);
\coordinate (t2) at (+2.5ex,-4ex);
\coordinate (P) at ($(-2.5ex,-1.0ex)$);
\coordinate (Q) at ($(2.5ex,-1.0ex)$);
\coordinate (R) at ($(0ex,-3.8ex)$);
\pgfmathsetmacro{\arista}{0.06}
\draw[thick] (-5ex,0ex) -- (5ex,0ex);
\draw[-,thick] (t1) -- (k1);
\draw[-,thick] (t1) -- (k2);
\draw[-,thick] (t2) -- (k3);
\draw[-,thick] (t2) -- (k4);
\draw[thick,scale=2] (t1) to [bend left=50] (t2);
\draw[thick,scale=2] (t1) to [bend left=-30] (t2);
\draw[thick,scale=2] (t1) to [bend left=-60] (t2);
\fill[white] (t1) circle (0.45ex);
\filldraw[pattern=north east lines, thick] (t1) circle (0.45ex);
\fill[white] (t2) circle (0.45ex);
\filldraw[pattern=north east lines, thick] (t2) circle (0.45ex);
\filldraw[color=black, fill=white, thick] ($(k1)-(\arista,\arista)$) rectangle ($(k1)+(\arista,\arista)$);
\filldraw[color=black, fill=white, thick] ($(k2)-(\arista,\arista)$) rectangle ($(k2)+(\arista,\arista)$);
\filldraw[color=black, fill=white, thick] ($(k3)-(\arista,\arista)$) rectangle ($(k3)+(\arista,\arista)$);
\filldraw[color=black, fill=white, thick] ($(k4)-(\arista,\arista)$) rectangle ($(k4)+(\arista,\arista)$);
\filldraw[color=black, fill=black, thick] ($(Q)+(-0.5ex,0ex)$) circle (0.05ex);
\filldraw[color=black, fill=black, thick] (Q) circle (0.05ex);
\filldraw[color=black, fill=black, thick] ($(Q)+(0.5ex,0ex)$) circle (0.05ex);
\filldraw[color=black, fill=black, thick] ($(P)+(-0.5ex,0ex)$) circle (0.05ex);
\filldraw[color=black, fill=black, thick] (P) circle (0.05ex);
\filldraw[color=black, fill=black, thick] ($(P)+(0.5ex,0ex)$) circle (0.05ex);
\filldraw[color=black, fill=black, thick] ($(R)+(0ex,0.5ex)$) circle (0.05ex);
\filldraw[color=black, fill=black, thick] (R) circle (0.05ex);
\filldraw[color=black, fill=black, thick]($(R)+(0ex,-0.5ex)$) circle (0.05ex);
}
}
\be \label{IR-div-comp-speculation}
\Bigg[ \diagramextwoverticesNoLabels \Bigg]_{\rm IR \,\, div}  \propto \diagramextwovertextree .
\ee
We leave the task of establishing this result more rigorously to future work.


\section{Stochastic approach}
\label{sec:stoch}

The stochastic approach is widely regarded as the appropriate framework for resumming the divergent secular growth that arises in the correlation functions of light scalar fields in de Sitter space, including contributions from loop corrections~\cite{Adshead:2008gk,Enqvist:2008kt,Fujita:2013cna,Fujita:2014tja,PerreaultLevasseur:2013kfq,PerreaultLevasseur:2014ziv,Mirbabayi:2020vyt,Cohen:2020php,Green:2022ovz,Cohen:2021fzf,Cohen:2022clv,Moss:2016uix,Ailiga:2024wdx,Choudhury:2018bcf,Li:2025azq,Cruces:2022imf,Bhattacharya:2025eqf,Tokuda:2017fdh,Cruces:2018cvq,Cruces:2024pni}. However, as established in the previous sections, loop corrections do not, in fact, generate genuine secular growth. The goal of this section is to revisit the stochastic formalism and clarify the connection between loop effects and the classical Fokker--Planck equation, which underpins the stochastic description of infrared dynamics.

Recall that the Fokker--Planck equation governs the time evolution of the probability density function $\rho(\varphi, t)$, which describes the coincident-point statistics of the field fluctuations $\varphi$ over superhorizon scales in de Sitter space. The original form, first derived by Starobinsky in Ref.~\cite{Starobinsky:1986fx}, is given by
\be \label{Fokker-Planck-1}
\frac{d \rho }{dt} = \frac{H^3}{8 \pi^2} \rho'' + \frac{1}{3 H} \left( \rho {\mathcal V'} \right)' ,
\ee
where primes denote derivatives with respect to the field $\varphi$. Time derivatives are taken with respect to cosmic time $t$, which is related to conformal time via $\tau = - \frac{1}{H} e^{-H t}$. The first term on the right-hand side of Eq.~(\ref{Fokker-Planck-1}) represents diffusion, while the second term, determined by the shape of the potential $\mathcal{V}(\varphi)$, corresponds to drift.

In the absence of drift, i.e., when $\mathcal{V}' = 0$, Eq.~(\ref{Fokker-Planck-1}) implies that an initially Gaussian probability distribution $\rho(\varphi)$ remains Gaussian, though it evolves in time:
\be
\rho (\varphi , t) = \frac{1}{\sqrt{2 \pi} \sigma (t)}
\exp \left[- \frac{1}{2} \frac{\varphi^2}{\sigma^2 (t)}\right] ,
\ee
with a time-dependent variance given by
\be
\sigma^2 (t) \propto t - t_0 ,
\ee
for some initial time $t_0$. Noting that $t = \frac{1}{H} \ln a(\tau)$, this behavior coincides with that of Eq.~(\ref{secular-growth}), which describes the secular growth of the two-point function at coincident points, as discussed in Section~\ref{sec:iso-vacua}.

When drift is present, with $\mathcal{V}' \neq 0$, the initially Gaussian distribution begins to develop non-Gaussian features. These deviations are initially small, but grow with time, eventually dominating the dynamics. In this regime, one expects $\rho(\varphi, t)$ to approach a non-Gaussian equilibrium configuration. Indeed, Eq.~(\ref{Fokker-Planck-1}) admits a stationary solution $\dot{\rho}_{\rm eq} = 0$, given by~\cite{Starobinsky:1994bd}
\be
\rho_{\rm eq} (\varphi) \propto e^{ - \frac{8 \pi^2}{3 H^4} \mathcal{V}(\varphi)} . \label{rho-equil}
\ee
In this equilibrium regime, nonlinear effects dominate the dynamics of $\varphi$, rendering perturbative computations, such as those of previous sections, invalid. 

In what follows, we examine the stochastic description within the perturbative regime, where the nonlinear effects associated with the potential $\mathcal{V}(\varphi)$ remain small. In this regime, the probability density function remains close to Gaussian and can be systematically expanded in terms of time-dependent cumulants $\langle \varphi^n(t) \rangle_c$ as~\cite{Palma:2023idj}
\bea \label{expansion-rho-hermite}
\rho (\varphi , t) &=& \frac{e^{- \frac{1}{2} \frac{\varphi^2}{\sigma^2(t)}}}{\sqrt{2 \pi} \sigma(t)} \sum_{N=0}^{\infty} \frac{1}{N!} \sum_{n_1=0}^{\infty} \cdots \sum_{n_N=0}^{\infty}
\nn \\ 
&&
\frac{\langle \varphi^{n_1} (t) \rangle_c }{n_1! \sigma^{n_1} (t)} \cdots \frac{ \langle \varphi^{n_N} (t) \rangle_c }{n_N! \sigma^{n_N} (t)} \nn \\ 
&& {\rm He}_{n_1 + \cdots + n_N} \big[ \varphi / \sigma(t) \big] ,
\eea
where ${\rm He}_n(x) = (-1)^n e^{\frac{x^2}{2}} \frac{d^n}{dx^n} e^{-\frac{x^2}{2}}$ are the probabilistic Hermite polynomials. For convenience, we set $\langle \varphi^0 \rangle_c = 1$. The cumulants $\langle \varphi^n(t) \rangle_c$ correspond to connected $n$-point correlation functions evaluated at a single spacetime point. In particular, within a perturbative scheme in which the cumulants are expanded as $\langle \varphi^n \rangle_c = \langle \varphi^n \rangle_1 + \langle \varphi^n \rangle_2 + \cdots$, the leading non-Gaussian correction to the probability density takes the form
\be \label{expansion-rho-hermite-first}
\!\!\rho (\varphi , t) = \frac{e^{- \frac{1}{2} \frac{\varphi^2}{\sigma^2(t)}}}{\sqrt{2 \pi} \sigma(t)} \!\left[ 1 \!+ \!\sum_{n=0}^{\infty}
\frac{\langle \varphi^{n} (t) \rangle_c }{n! \sigma^{n} (t)} {\rm He}_{n} \big[ \varphi / \sigma(t) \big] \!\right] \!,
\ee
which is simply an Edgeworth expansion.
Equation~(\ref{expansion-rho-hermite-first}) provides a concrete bridge between the stochastic formalism and the correlation functions computed in the previous sections. As we will demonstrate, the standard derivation of the stochastic framework implicitly relies on a specific scheme for regularizing loop integrals. Crucially, if this scheme is replaced with one that preserves de Sitter invariance—as advocated throughout this work—then the resulting Fokker--Planck equation acquires corrections that depart from its classical form~\eqref{Fokker-Planck-1}.

\subsection{Smoothed fields over comoving patches}
\label{sec:smoothed-fields-comoving-patch}

The stochastic formalism is designed to describe the statistics of the field $\varphi$ in configuration space rather than in momentum space. However, correlation functions in configuration space are generally divergent, irrespective of whether secular growth is present. This situation requires the introduction of a window function to regulate the field by selecting a finite range of scales over which the observable field is defined. To this end, we define a smoothed scalar field $\varphi_w(\x, \tau)$ as
\bea \label{field-Fourier-2}
\varphi_w(\x, \tau) &\equiv&  \hat W  \Big\{ \varphi (\x, \tau) \Big\} \nn \\ 
&=&  \int \frac{d^3 k}{(2\pi)^3} W(k, \tau) \, \tilde{\varphi}(\k, \tau) \, e^{-i \k \cdot \x},
\eea
where $\tilde{\varphi}(\k, \tau)$ denotes the Fourier transform of the full nonlinear field, and $W(k, \tau)$ is a window function that selects a finite band of momenta. The first line in Eq.~(\ref{field-Fourier-2}) introduces the integral operator $\hat{W}$ acting on the field in coordinate space to implement the smoothing.

In the stochastic formalism, the field is defined over a comoving patch of size $L$, representing a spatial region that expands with time. This choice implies that $\varphi_w(\x, \tau)$ includes modes with comoving wavelengths no larger than $L$, or equivalently, modes with comoving momentum $k \geq k_L = 2\pi / L$. Simultaneously, one introduces a physical smoothing scale $R_S \equiv \Lambda_S^{-1}$ to exclude subhorizon modes. It is customary to take $R_S \gg R_H$ by choosing $\Lambda_S = \varepsilon H$ with $\varepsilon \ll 1$, ensuring that smoothing always occurs over superhorizon scales. These choices can be encoded in the following form for the window function:
\be \label{W}
W(k, \tau) = \theta(k - k_{L}) \, \theta(\Lambda_{S} a(\tau) - k),
\ee
which selects the comoving momentum range $k_L < k < \Lambda_S a(\tau)$. Note that $\Lambda_S$ is a physical UV cutoff, while $k_L$ acts as a comoving IR cutoff. As time evolves, the smoothed field~\eqref{field-Fourier-2} includes more and more modes, as physical wavelengths redshift from the sub-UV regime ($k / a(\tau) > \Lambda_S$) to the super-UV regime ($k / a(\tau) < \Lambda_S$). This construction implies the existence of a minimum time $\tau_0$ defined by
\be
\Lambda_S a(\tau_0) = k_L,
\ee 
corresponding to the moment when the physical size of the comoving patch coincides with the smoothing scale $R_S$.

It is important to emphasize that, in principle, these smoothing scales have nothing to do with the cutoff scales used to regularize loops in momentum space. The scales in~\eqref{W} determine the range of external momenta contributing to observables, not the internal loop momenta. Of course, if one is convinced that loop integrals should also be regulated using a comoving IR cutoff $k_{\rm IR}$—with the goal of explicitly inducing de Sitter-breaking secular growth—then one is free to identify $k_{\rm IR} = k_L$.

\subsection{Free theory 2-point correlation function}

Before examining the general computation of $n$-point correlation functions, let us consider the 2-point function of the smoothed field, defined as
\be
G_{w} (|\x - \x' |, \tau ) = \langle \varphi_{w} (\x , \tau) \varphi_{w} (\x' , \tau) \rangle 
\ee
in the free-theory case. Upon quantization, the field takes the form
\be \label{field-Fourier-3}
\varphi_{w} (\x , \tau) \equiv \!\! \int \!\!\! \frac{d^3 k}{(2 \pi)^3} W(k,\tau) \!\! \left[  f_{k} (\tau) \hat a_{\k} + f_{k}^* (\tau) \hat a_{-\k}^\dag  \right] e^{ i \k \cdot \x} \!\! .
\ee
Then, a direct computation leads to:
\bea
G_{w} (|\x - \x' |, \tau ) &=& \frac{H^2}{4 \pi^2} \!\! \int_{k_{L}/ a(\tau)}^{\Lambda_{S}} \! \frac{dp}{p}  \left( 1 + \frac{p^2}{H^2}\right)  \nn \\ 
&& \times \frac{\sin ( p | (\x - \x')/H\tau |) }{p | (\x - \x')/H\tau |} .
\eea
Now, taking the limit $\Lambda_{S} / k_{L} \gg H |\tau|$, the previous integral can be easily solved in two relevant regimes. First, for distances $k^{-1}_{L} \gg |\x - \x' |  \gg H |\tau| \Lambda^{-1}_{S} $ one finds
\be \label{smoothed-two-point-1}
\!\!G_{w} (|\x - \x' |, \tau ) \simeq \frac{H^2}{4\pi^2} \left( \frac{ \tau^2}{ |\x - \x' |^2} - \ln \frac{ \left| \x - \x' \right|}{ k_{L}^{-1}}    \right) .
\ee
Notice that this result coincides with the equal-time limit of the expression~\eqref{2point-1}. On the other hand, for distances $|\x - \x' | / H |\tau| \ll  \Lambda^{-1}_{S}$ (which effectively corresponds to evaluating the 2-point function at a coincident point), one obtains Eq.~\eqref{secular-growth}, which we rewrite here:
\be \label{smoothed-two-point-2}
G_{w} (|\x - \x' |, \tau ) \simeq \frac{H^2}{4 \pi^2} \ln \frac{ a(\tau)}{a(\tau_0)} ,
\ee
where $a(\tau_0) = k_{L} / \Lambda_{S}$. In other words, the 2-point function of the smoothed field displays secular growth, which is to be expected due to the introduction of the comoving IR cutoff scale $k_{L}$ over external momenta. It should be emphasized that, in this case, the appearance of the factor $\log[a(\tau)/a(\tau_0)]$ is enforced by the very definition of the smoothed field $\varphi_{w} (\x , \tau)$ and is unrelated to the shift symmetry discussed in Section~\ref{sec:iso-vacua}.

\subsection{Cumulants}
\label{sec:cumulants}

The field $\varphi_w$ defined in (\ref{field-Fourier-3}) can be used to compute $n$-cumulants, denoted by $\big\langle \varphi_w^n (\tau) \big\rangle_c$. These correspond to the connected $n$-point correlation functions in coordinate space evaluated at a single coincident point, and are given by
\bea \label{G-n-coordinate-2}
\big \langle \varphi_w^n (\tau) \big \rangle_c &=&  \! \int^{\Lambda_{S} a (\tau)}_{k_{L}} \!\!  \frac{dk_1}{k_1}  \!\! \int \!\! \frac{d\Omega_1}{(2 \pi)^3}  \cdots \!\! \int^{\Lambda_{S} a (\tau)}_{k_{L}}  \!\! \frac{dk_n}{k_n}  \!\! \int \!\! \frac{d\Omega_n}{(2 \pi)^3}  \nn \\
 && \times ( k_1^3 \cdots k_n^3) \, \tilde G_c^{(n)} (\k_1, \cdots \! , \k_n ; \tau) ,
\eea
where $d\Omega_i$ denotes the solid angle integration measure associated with $\k_i$. Recall that the smoothing is applied to the external momenta, thereby defining the range of scales over which the observable is defined. It does not apply to the internal momenta involved in virtual processes, such as loops. In other words, the smoothing procedure does not affect the computation of $\tilde G^{(n)}_c ( {\k}_1 , \cdots  , {\k}_n , \tau )$, which is fully determined by the underlying theory.

We aim to determine the time dependence of $\big \langle \varphi_w^n (\tau) \big \rangle_c$. To this end, let us recall the scaling property (\ref{n-point-momentum-scaling}), which holds for any correlation function in momentum space. By choosing the scaling factor as $e^{\theta} = 1/a(\tau)$, we find that (\ref{G-n-coordinate-2}) takes the form
\bea  \label{G-n-smoothed-p}
\big \langle \varphi_w^n (\tau) \big \rangle_c &=&  \int^{\Lambda_S}_{k_{\rm IR}/ a(\tau)}  \frac{dp_1}{p_1}  \!\! \int \!\! \frac{d\Omega_1}{(2 \pi)^3}  \cdots \int^{\Lambda_S}_{k_{\rm IR} / a(\tau)}  \frac{dp_n}{p_n}  \!\! \int \!\! \frac{d\Omega_n}{(2 \pi)^3}  \nn \\
 && \times ( p_1^3 \cdots p_n^3) \, G^{(n)} (\p_1, \cdots \! , \p_n ; - H^{-1}) .
\eea
Note that the time dependence of $ \langle \varphi_w^n (\tau) \rangle_c$ arises solely through the lower limits of integration. Consequently, to determine its time dependence, it is sufficient to analyze the asymptotic behavior of the integrand, $(p_1^3 \cdots p_n^3) G^{(n)} (\p_1, \cdots , \p_n , - H^{-1})$, in the regime $p_i \ll H$.

Next, recall that the connected $n$-point correlation function $\tilde G_c^{(n)} (\k_1, \cdots , \k_n , \tau)$ appearing in (\ref{G-n-coordinate-2}) can be expressed as a sum over diagrams:
\be
\tilde G_c^{(n)} ( {\k}_1 , \cdots , {\k}_n ; \tau ) = \sum_{T}  D_{T} ( {\k}_1 , \cdots , {\k}_n ; \tau ) ,
\ee
where $T$ denotes the topology of the diagram $D_T$. Since the external momenta entering $\tilde G^{(n)} (\k_1, \cdots  , \k_n ; \tau)$ satisfy $|k_i \tau| \ll 1$, each contributing diagram $D_T$ exhibits the asymptotic behavior described by (\ref{log-structure}). Inserting this structure into (\ref{G-n-smoothed-p}) and using the Dirac delta function to integrate over one of the momenta (e.g., $\p_n$), we obtain
\bea  \label{G-n-smoothed-p-2}
\big \langle \varphi_w^n (\tau) \big \rangle_c &=&  \int^{\Lambda_{\rm UV} }_{k_{\rm IR}/ a(\tau)}  \frac{dp_1}{p_1}    \cdots \int^{\Lambda_{\rm UV} }_{k_{\rm IR} / a(\tau)}  \frac{dp_{n-1}}{p_{n-1}}   \nn \\
 && \times \sum_s B^T_s (\p_1, \cdots , \p_{n-1})\nn \\ 
&& \times \prod_{a=1}^{V}  \ln  \big[ \frac{1}{H}  g^T_{s,a}(\p_1, \cdots , \p_{n-1})  \big] ,
\eea
where $B^T_s (\p_1, \cdots , \p_{n-1})$ is a function invariant under rescalings of the momenta, and the functions $g^T_{s,a}$ satisfy the homogeneity condition
\be
g^T_{s,a}(\alpha \p_1, \cdots , \alpha \p_{n-1}) = \alpha g^T_{s,a}(\p_1, \cdots , \p_{n-1}) .
\ee
Equation (\ref{G-n-smoothed-p-2}) shows that the cumulants $\big \langle \varphi_w^n (\tau) \big \rangle_c$ can be written as a sum of contributions classified by the number of vertices $V$ involved in the corresponding diagrams:
\be \label{sum-G-V}
\big \langle \varphi_w^n (\tau) \big \rangle_c = \sum_V \big \langle \varphi_w^n (\tau) \big \rangle_V ,
\ee
The scaling properties of the integrand imply that one can take up to $n-1 + V$ logarithmic derivatives $- \tau \frac{d}{d\tau}$ of each term $\big \langle \varphi_w^n (\tau) \big \rangle_V$ without the result vanishing in the limit $\tau \to 0$. This leads to the following time dependence:
\be \label{G-n-smoothed-1}
\big \langle \varphi_w^n (\tau) \big \rangle_V \propto \left[ \ln \frac{a(\tau)}{ a(\tau_0)} \right]^{n-1+V} .
\ee
Remarkably, the time evolution of each contribution depends solely on the number of vertices $V$, and not on the number of loops present in the corresponding diagrams.

It is instructive to verify the form of (\ref{G-n-smoothed-1}) in the simplest case, namely that of single-vertex contributions to connected $n$-point correlation functions. Inserting the tree-level single-vertex result (\ref{D-n-point-single-vertex}) into (\ref{G-n-coordinate-2}), one finds
\be \label{cum-1-vertex}
\big \langle \varphi_w^n (\tau) \big \rangle_1 = - \frac{4 \pi^2 n}{3 H^4} \sigma^{2 n} (\tau) \lambda_n ,
\ee
where, for notational convenience, we have introduced
\be \label{sigma}
\sigma^{2} (\tau) \equiv \frac{H^2}{4 \pi^2} \ln \frac{ a(\tau)}{a(\tau_0)}.
\ee
One may also consider the inclusion of daisy-loop corrections. In this case, the expression becomes~\cite{Palma:2023idj}

\be \label{cum-1-vertex-2}
\!\!\big \langle \varphi_w^n (\tau) \big \rangle_1 = - \frac{4 \pi^2 n}{3 H^4} \sigma^{2 n} (\tau) \sum_{L} \frac{1}{L!} \left( \frac{\sigma_{\rm tot}^2}{2} \right)^L  \lambda_{n + 2L} , 
\ee
where $\sigma_{\rm tot}^2 \equiv G(0;\tau)$ denotes the constant loop integral defined in (\ref{G-0-p}). Recall from Section~\ref{sec:daisy-loops} that this class of diagrams leads to divergences, which can be absorbed by redefining the bare couplings $\lambda_n$ in terms of finite, renormalized couplings $\bar \lambda_n$ via the relation (\ref{lambda-lambda}). This renormalization leads to the final result
\be  \label{cum-1-vertex-3}
\big \langle \varphi_w^n (\tau) \big \rangle_1 = - \frac{4 \pi^2 n}{3 H^4} \sigma^{2 n} (\tau) \bar \lambda_n  ,
\ee
which is structurally identical to the tree-level expression in (\ref{cum-1-vertex}).

\subsection{Other schemes to compute cumulants}
\label{sec:other-scheme}

In the previous section, we derived expression (\ref{cum-1-vertex-2}), or equivalently (\ref{cum-1-vertex-3}), as the general form of the cumulants characterizing the statistics of the field $\varphi$ over a smoothed patch of comoving size $L$, taking into account interactions to first order in the potential. It is instructive, however, to consider an alternative scheme for computing single-vertex cumulants that leads to a qualitatively different result.

To begin, let us return to expression (\ref{n-point-one-vertex}). Recall that, after performing the time integral, one arrives at (\ref{D-n-point-single-vertex}), which provides an accurate expression for the $n$-point function in momentum space valid in the superhorizon limit for external momenta. In this alternative approach, we refrain from evaluating the time integral in (\ref{n-point-one-vertex}) and instead compute the cumulant directly, without specifying momentum cutoffs a priori. The formal result, including loop contributions to all orders, is
\bea
\big \langle \varphi_w^n (\tau) \big \rangle_1 &=&   \sum_{L=0}^{\infty} 2\,\text{Im}\Bigg\{  \frac{\lambda_{n+2L}}{2^L L! H^4} 
     \int_{-\infty}^{\tau}  \frac{d \tau'}{{ \tau' }^{4}} \int_{\k_1} \cdots \int_{\k_n} \nn\\
     && (2\pi)^3 \delta^{(3)} (\boldsymbol{K}) G_+(\tau', \tau, k_1)\cdots G_+(\tau', \tau ,k_n) 
     \nn \\ 
     &&  \left[\int_{\bf k} G( \tau', \tau',k)\right]^L  (k_1^3 \cdots k_n^3)  \Bigg\}.
\eea
We now impose that all integrated momenta—both external and internal—lie within the range $k \in (k_L, \Lambda_S a(\tau'))$, where $\Lambda_S$ is the physical cutoff scale introduced in Section~\ref{sec:smoothed-fields-comoving-patch}. This corresponds to a constant comoving infrared cutoff regulating loops. This prescription yields
\bea \label{alt-scheme-cum}
\big \langle \varphi_w^n (\tau) \big \rangle_1 &=&   \sum_{L=0}^{\infty} 2\,\text{Im}\Bigg\{  \frac{\lambda_{n+2L}}{2^L L! H^4} 
     \int_{-\infty}^{\tau}  \frac{d \tau'}{{ \tau' }^{4}}  \nn\\
     &&  \int^{\Lambda_{S} a (\tau')}_{k_{L}} \!\!  \frac{dk_1}{k_1}  \!\! \int \!\! \frac{d\Omega_1}{(2 \pi)^3}  \cdots \!\! \int^{\Lambda_{S} a (\tau')}_{k_{L}}  \!\! \frac{dk_n}{k_n} \nn \\ 
     && (2\pi)^3 \delta^{(3)} (\boldsymbol{K}) G_+(\tau', \tau, k_1)\cdots G_+(\tau', \tau ,k_n) 
     \nn \\ 
     && \left[ \int^{\Lambda_{S} a (\tau')}_{k_{L}} \frac{dk_1}{k_1}  \!\! \int \!\! \frac{d\Omega_1}{(2 \pi)^3}  k^3 G( \tau', \tau',k)\right]^L\! \nn\\
     && (k_1^3 \cdots k_n^3)  \Bigg\}. 
\eea
Notice that this choice of momentum cutoff introduces an explicit time dependence into the integrals that was not present in the original time integral. It also alters the way interactions affect modes of different momentum. Performing the momentum integrals first, followed by the time integral, one obtains the following expression for the cumulant in the leading-log approximation:
\be \label{G-com-loops}
\big \langle \varphi_w^n (\tau) \big \rangle_1 = - \frac{4 \pi^2 n}{3 H^4} \sigma^{2 n} (\tau) \sum_{L=0}^{\infty}  \frac{\lambda_{n + 2 L}}{(n+L)L!} \left( \frac{\sigma^2(\tau)}{2} \right)^L .
\ee
This result clearly differs from (\ref{cum-1-vertex-2}), even at tree level (i.e., for $L = 0$). One key difference lies in the time dependence of the loop contributions, which arises here due to the use of time-dependent momentum cutoffs involving a mixture of physical and comoving scales. Another important distinction is the altered structure of interactions across different wavelengths. The expression in (\ref{D-n-point-single-vertex}) allows for unrestricted momentum exchange between external legs in momentum space, whereas the structure in (\ref{alt-scheme-cum}) imposes a restriction on the momentum flowing through the vertex, dictated by the size of the observable patch.

In other words, under this scheme, no universal expression for correlation functions in momentum space exists, as the form of the interaction becomes dependent on the comoving size $L$ of the region being probed and the smoothing scale $\Lambda_S$. Naturally, we do not agree with this approach to computing cumulants.

\subsection{An intriguing connection}

Having discussed the computation of cumulants from correlation functions, let us now return to the stochastic formalism. By multiplying both sides of the Fokker-Planck equation (\ref{Fokker-Planck-1}) by $\varphi^n$, expanding the potential $\mathcal V(\varphi)$ in a Taylor series as in (\ref{taylor-bare-potential}), and integrating over the full range $\varphi \in (-\infty, +\infty)$, one obtains
\bea \label{moments-fokker}
 \frac{d}{dt}  \big\langle \varphi^n  \big\rangle &=&  n (n-1) \frac{H^3 }{8 \pi^2} \big\langle \varphi^{n-2}  \big\rangle  \nn \\
 &&  
 - \frac{n}{3 H} \sum_{m=2}^\infty \frac{\lambda_m}{(m-1)!} \big \langle \varphi^{m+n-2} \big \rangle , 
\eea
where $\big\langle \varphi^n \big\rangle$ denotes the $n$-th moment of the probability distribution $\rho(\varphi, t)$, defined by
\be \label{n-moment-def}
 \big \langle \varphi^n (t)  \big\rangle \equiv \int d\varphi \, \rho (\varphi, t) \, \varphi^n .
\ee
Equation (\ref{moments-fokker}) encodes a non-trivial relation among moments of different powers and highlights the role of the stochastic formalism in resumming $n$-point functions. 

It is important to stress that the moment $\big \langle \varphi^n \big\rangle$, as defined in (\ref{n-moment-def}), should not be confused with the cumulant $\big \langle \varphi^n \big\rangle_c$. While the moment includes both connected and disconnected contributions, the cumulant retains only the connected part. If one computes cumulants perturbatively (for instance, by expanding cumulants as $\langle \varphi^n \rangle_c = \langle \varphi^n \rangle_1 + \langle \varphi^n \rangle_2 + \cdots$) then, using Eq.~(\ref{expansion-rho-hermite-first}), the relation between an $n$-th moment and cumulants takes the form
\be
 \big \langle \varphi^n (t) \big \rangle = \sum_{m=0}^{\lfloor n/2 \rfloor} \frac{n!}{m! (n-2m)! 2^m} \sigma^{2m} (t) \big  \langle \varphi^{n-2m} (t) \big \rangle_c .
 \ee
This expression can, in principle, be substituted back into Eq.~(\ref{moments-fokker}) to derive a corresponding relation among cumulants of various orders. However, such a step will not be necessary for our purposes here.

Now, is Eq.~(\ref{moments-fokker}) satisfied by the cumulants computed via perturbation theory, as discussed in the previous sections? It turns out that Eq.~(\ref{moments-fokker}) is indeed satisfied by the cumulants in Eq.~(\ref{G-com-loops}), computed using the scheme where interactions are conditioned by the size of the observable patch. This correspondence was first observed in Ref.~\cite{Tsamis:2005hd} in the context of a $\lambda \varphi^4$ theory. The authors of that work noted that Eq.~(\ref{moments-fokker}) entangles contributions from loops of different orders—a nontrivial feature they interpreted as a double validation: both of the use of comoving, de Sitter-breaking cutoffs and of the stochastic formalism as a means to resum infrared-divergent terms in correlation functions. In the next subsections, we show that this interpretation is premature.

\subsection{Perturbative derivation of the Fokker--Planck equation}

In the previous subsection, we saw that the Fokker--Planck equation (\ref{Fokker-Planck-1}) is consistent with perturbation theory, provided that interactions and loop corrections explicitly introduce de Sitter-breaking secular growth. How can this be the case? To clarify the situation, it is instructive to revisit the derivation of the Fokker--Planck equation starting from the full quantum equation obtained from the action (\ref{action-cosmo-coord}). Expressed in cosmic time, the equation reads:
\be \label{full-quantum}
\frac{d^2}{dt^2} \varphi + 3 H \frac{d}{dt} \varphi + \frac{1}{a^2(t)} \nabla^2 \varphi + \mathcal V ' (\varphi) = 0 .
\ee
Since we are interested in coincident-point computations, we may evaluate fields at $\x = 0$ in what follows. The first step is to apply the operator $\hat W$, defined in (\ref{field-Fourier-2}), to Eq.~(\ref{full-quantum}). This operator suppresses both second-order time derivatives and spatial derivatives, yielding a first-order equation:
\be \label{pre-langevin}
  \frac{d}{dt} \varphi_w + \frac{1}{ 3 H}\hat W  \Big\{ \mathcal V ' (\varphi) \Big\} = H \hat \xi (t) ,
\ee
where $\hat \xi(t)$ is a Gaussian noise term defined as:
\be
\hat \xi (\tau) \equiv H^{-1} \int \!\! \frac{d^3 k}{(2 \pi)^3} \left[ \frac{d}{dt} W(k,\tau) \right] \tilde \varphi (\k , \tau) .
\ee
Thanks to the definition of the window function adopted in (\ref{W}), the noise $\hat \xi(t)$ is sensitive only to fluctuations evaluated at the physical scale $\Lambda_S$. Moreover, it satisfies the normalization condition
\be \label{xi-norm}
\big \langle \hat \xi (\tau) \hat \xi (\tau') \big \rangle = \frac{H}{4 \pi^2} \delta (t - t') .
\ee
To derive the Fokker-Planck equation (\ref{Fokker-Planck-1}) from Eq.~(\ref{pre-langevin}), a final step is required: we assume that
\be \label{assumption-W-V}
\hat W \Big\{  \mathcal V ' (\varphi) \Big\}=  \mathcal V ' (\varphi_w) .
\ee
This critical assumption leads to the Langevin equation
\be \label{langevin}
  \frac{d}{dt} \varphi_w + \frac{1}{ 3 H} \mathcal V ' (\varphi_w) = H \hat \xi (t) ,
\ee
which is known to imply the Fokker--Planck equation (\ref{Fokker-Planck-1}) through standard nonperturbative methods. However, because our interest is in assessing the perturbative results developed in previous sections, we now provide an alternative derivation of (\ref{Fokker-Planck-1}) valid to first order in perturbation theory~\cite{Palma:2023uwo}. 

Integrating Eq.~(\ref{langevin}) over time gives:
\be \label{soch-varphi-w-0}
\varphi_w (t) =  \varphi_G (t) - \frac{1}{ 3 H} \int^{t}_{t_0} \!\!\! dt'   \frac{ d \mathcal V}{d \varphi} \big( \varphi_w (t') \big)  ,
\ee
where $\varphi_G(t)$ is a Gaussian field defined from $\hat \xi(t)$ as
\be \label{Gaussian-field-def-xi}
\varphi_G (t) \equiv H \int_{t_0}^t dt' \hat \xi (t').
\ee
The variance of this Gaussian field is
\be
\big \langle \varphi_G^2 (t) \big \rangle = \frac{H^3}{4 \pi^2} (t - t_0),
\ee
which matches $\sigma^2(t)$ from Eq.~(\ref{sigma}) when expressed in conformal time. 

We can now compute cumulants of $\varphi_w(t)$ using Eq.~(\ref{soch-varphi-w-0}). Iterating once, we find:
\be \label{soch-varphi-w-0-1}
\varphi_w (t) =  \varphi_G (t) - \frac{1}{ 3 H} \int^{t}_{t_0} \!\!\! dt'   \frac{ d \mathcal V}{d \varphi} \big( \varphi_G (t') \big)  .
\ee
Expanding the potential $\mathcal V(\varphi)$ in a Taylor series and evaluating cumulants to first order in the potential yields:
\bea \label{soch-varphi-w-0-2}
\big\langle \varphi_w^n (t) \big\rangle_c &=& - \frac{n}{3H} \int^{t}_{t_0} \!\!\! dt' \, \Big( \big \langle \varphi_G (t) \varphi_G (t')  \big \rangle  \Big) ^{n-1} \nn \\
&& \sum_{L=0}^{\infty} \frac{\lambda_{n+2L}}{L!} \Big( \frac{1}{2} \big \langle \varphi_G (t') \varphi_G (t')  \big \rangle  \Big)^L  .
\eea
Here, the expectation values $\big\langle \varphi_G(t) \varphi_G(t') \big\rangle$ and $\big\langle \varphi_G (t') \varphi_G (t') \big\rangle$ arise from Wick contractions of the Gaussian field. The sum over $L$ corresponds to loop corrections from self-contractions of the interaction term.

Using Eq.~(\ref{Gaussian-field-def-xi}) and the noise normalization (\ref{xi-norm}), we find:
\be \label{wrong-corr}
\big \langle \varphi_G(t') \varphi_G(t') \big \rangle =
\big \langle \varphi_G(t) \varphi_G(t') \big \rangle =  \frac{H^3}{4 \pi^2} (t' - t_0) , 
\ee
where we used $t > t'$. Plugging these expressions into Eq.~(\ref{soch-varphi-w-0-2}) and performing the time integral yields:
\be \label{Langegin-cumulants-FP}
\big\langle \varphi_w^n (t) \big\rangle_c = - \frac{4 \pi^2 n}{3 H^4} \sigma^{2 n} (t) \sum_{L=0}^{\infty}  \frac{\lambda_{n + 2 L}}{(n+L)L!} \left( \frac{\sigma^2(t)}{2} \right)^L .
\ee
Not surprisingly, this result agrees with Eq.~(\ref{G-com-loops}) obtained in the scheme analyzed in Section\ref{sec:other-scheme}. As we have emphasized, we regard that scheme as unjustified.

To conclude, one may use Eq.~(\ref{Langegin-cumulants-FP}) together with the Hermite expansion (\ref{expansion-rho-hermite-first}) to reconstruct the probability density function $\rho(\varphi, t)$. Taking the time derivative of the resulting expression and using standard recurrence relations satisfied by Hermite polynomials, one recovers Eq.~(\ref{Fokker-Planck-1}).

\subsection{Clarifying the connection}

Clearly, the assumption in Eq.~(\ref{assumption-W-V}) is invalid and underlies the apparent agreement between the stochastic formalism and the unjustified scheme that led to the analytical expression for cumulants in Eq.~(\ref{G-com-loops}). To understand the implications of adopting Eq.~(\ref{assumption-W-V}), it is instructive to derive the Fokker--Planck equation directly from Eq.~(\ref{pre-langevin}), without invoking this assumption. Integrating Eq.~(\ref{pre-langevin}) in time gives:
\be \label{soch-varphi-w}
\varphi_w (t) =  \varphi_G (t) - \frac{1}{ 3 H} \int^{t}_{-\infty} \!\!\!\! dt' \, W  \bigg\{ \frac{ d \mathcal V}{d \varphi} \big( \varphi (t') \big) \bigg\} ,
\ee
where $\varphi_G(t)$ is the Gaussian field defined in Eq.~(\ref{Gaussian-field-def-xi}). Using Eq.~(\ref{soch-varphi-w}), we can compute the cumulants of $\varphi_w(t)$ to first order in the potential. These take the form:
\be \label{soch-varphi-w-2}
\big\langle \varphi_w^n (t) \big\rangle_c = - \frac{n}{3H} \int^{t}_{t_{0}} \!\!\! dt' \, \Big \langle \varphi^{n-1}_G(t)  \hat W  \bigg\{ \frac{d \mathcal V}{d \varphi} [\varphi (t')] \bigg\} \Big \rangle .
\ee
To evaluate this expression, we expand the potential in a Taylor series and apply the window function explicitly, yielding:
\bea \label{expansion-W-V}
&& \hat W  \bigg\{ \frac{d \mathcal V}{d \varphi} [\varphi (t')] \bigg\} = (2 \pi)^3 \sum_n \frac{\lambda_n}{(n-1)!} \int_{\k}   W(k,\tau)  \nn \\ 
 && \qquad
 \int_{\k_1} \cdots  \int_{\k_{n-1}}  \!\!\! \delta^{(3)} (\k-{\bf K}) \tilde \varphi_{\k_1} (t') \cdots \tilde \varphi_{\k_{n-1}} (t') , \quad
\eea
where $\boldsymbol{K} = \k_1 + \cdots + \k_{n-1}$. Since we are working to first order in the potential, we treat the field $\varphi(t')$ as a free field inside the potential. Substituting Eq.~(\ref{expansion-W-V}) into Eq.~(\ref{soch-varphi-w-2}) gives:
\bea \label{soch-varphi-w-3}
\big\langle \varphi_w^n (t) \big\rangle_c &=& - \frac{n}{3H} \int^{t}_{t_0} \!\!\! dt' \, \Big( \big \langle \varphi_G (t) \varphi (t')  \big \rangle  \Big) ^{n-1} \nn \\
&& \sum_{L=0}^{\infty} \frac{\lambda_{n+2L}}{L!} \Big( \frac{1}{2} \big \langle \varphi (t') \varphi (t')  \big \rangle  \Big)^L  .
\eea
While this expression resembles Eq.~(\ref{soch-varphi-w-0-2}), the contractions now involve different fields, leading to a markedly different result after time integration. Specifically, the contraction between the Gaussian field $\varphi_G(t)$ and the full field $\varphi(t')$ is given by
\be 
\big \langle \varphi_G (t) \varphi (t')  \big \rangle =  \frac{H^3}{4 \pi^2} (t - t_0) ,
\ee
which is independent of $t'$. In contrast, the self-contraction of the full field,
\be
\big \langle \varphi (t') \varphi (t')  \big \rangle = \sigma_{\rm tot}^2 ,
\ee
is a daisy loop contribution, where $\sigma_{\rm tot}^2$ denotes the full two-point function at coincident point, as defined in Eq.~(\ref{sigma-tot-def}). Note the difference from Eq.~\eqref{wrong-corr}. Substituting these results into Eq.~(\ref{soch-varphi-w-3}), one finds
\be \label{cum-phi-w-correct}
\big\langle \varphi_w^n (t) \big\rangle_c = - \frac{4 \pi^2 n}{3 H^4} \sigma^{2 n} (\tau) \sum_{L} \frac{1}{L!} \left( \frac{\sigma_{\rm tot}^2}{2}  \right)^L  \lambda_{n + 2L} , 
\ee
which matches the result obtained earlier in Eq.~(\ref{cum-1-vertex-2}) from the diagrammatic computation in Section~\ref{sec:cumulants}.

Equation~(\ref{cum-phi-w-correct}) is significant for two reasons. First, it agrees with the correct cumulant computation established in Section~\ref{sec:cumulants}. Second, it highlights the critical role of the assumption in Eq.~(\ref{assumption-W-V}) in deriving the original Fokker--Planck equation of Ref.~\cite{Starobinsky:1986fx}.

As a final step, one can insert Eq.~(\ref{cum-phi-w-correct}) into the Hermite expansion (\ref{expansion-rho-hermite-first}) to reconstruct the probability density function $\rho(\varphi, t)$ describing the statistics of $\varphi$ within a comoving patch. Taking the time derivative of $\rho(\varphi, t)$ and applying standard recurrence relations for Hermite polynomials yields the following modified Fokker--Planck equation~\cite{Palma:2023idj}:
\be \label{Fokker-Planck-2}
\frac{d \rho }{dt} = \frac{H^3}{8 \pi^2}  \left[  \left( 1 - \frac{8 \pi^2}{3H^4 } \sigma^2 {\mathcal V''_{\rm patch}} \right) \rho \right] ''  + \frac{1}{3 H}  \Big( \rho  \mathcal V{\,}'_{\!\! \rm patch}  \Big)' ,
\ee
where we have defined the effective potential over the patch as
\be \label{V-patch}
\mathcal V_{\rm patch} (\varphi) = e^{- \frac{\sigma^2 (t)}{2} \frac{\partial^2}{\partial \varphi^2}} \bar {\mathcal V} (\varphi) ,
\ee
with $\bar{\mathcal V}(\varphi)$ being the renormalized potential introduced in Eq.~(\ref{Weierstrass-V}) during our discussion of daisy loops in Section~\ref{sec:daisy-loops}. The potential $\mathcal V_{\rm patch}(\varphi)$ can thus be interpreted as an effective potential defined over the expanding comoving patch, and is explicitly time-dependent due to the growing variance $\sigma^2(t)$.

Beyond the time dependence in the potential, note that the diffusion term in Eq.~(\ref{Fokker-Planck-2}) also acquires a correction proportional to $\sigma^2(t)$. As a result, this equation does not describe the nonperturbative evolution of $\rho$ for arbitrarily late times, since it eventually breaks down when
\be
\frac{8\pi^2}{3H^4} \sigma^2(t) \langle \mathcal V''_{\rm patch} \rangle \sim 1.
\ee
This breakdown is a straightforward reminder that Eq.~(\ref{Fokker-Planck-2}) is valid only to first order in the potential~\footnote{Similar statements have been made in Ref.~\cite{Cruces:2018cvq} for the case of a quasi-de Sitter background.}. Extending its applicability to longer timescales would require resumming higher-order contributions.


\section{Summary and conclusions}
\label{sec:conclusions}

In this work, we have examined the widely held view that light scalar fields in de Sitter space inevitably lead to secular growth in correlation functions due to infrared divergences. Focusing on theories with non-derivative interactions, we have shown that such growth is not a genuine physical effect but rather an artifact of regularization schemes that break de Sitter invariance. Our analysis distinguishes between theories that preserve shift symmetry and those that do not. In the shift-symmetric case, physical observables involve derivatives of the field and are insensitive to the logarithmic growth of two-point functions. On the other hand, when the shift symmetry is broken, interactions introduce a physical infrared scale $\Lambda_{\rm IR}$, which sets the onset of strong nonlinear behavior and replaces the role commonly assigned to a comoving infrared cutoff.

By employing a de Sitter-invariant renormalization scheme based on Wilson’s axioms for momentum-space integration, we have demonstrated that infrared divergences can be systematically removed within perturbation theory. As a result, correlation functions remain finite and de Sitter invariant for all external momenta above the threshold $\Lambda_{\rm IR}$. Moreover, we have shown that the time dependence of equal-time $n$-point correlation functions in the superhorizon regime is determined entirely by the number of interaction vertices, not by the loop order. This means that loop corrections do not generate additional time dependence relative to tree-level contributions and therefore do not lead to secular growth.

These findings challenge the standard use of the stochastic formalism, which typically relies on comoving infrared cutoffs and, in doing so, explicitly breaks de Sitter invariance. We have argued that if loop integrals are regularized in a de Sitter-invariant way, the resulting stochastic dynamics deviates from the conventional picture and requires modifications to the Fokker–Planck equation governing the evolution of long-wavelength fluctuations.

Altogether, our results support a consistent and symmetry-preserving treatment of light scalar fields in de Sitter space. Although infrared divergences do arise, they can be renormalized without introducing spurious time dependence, and the framework of de Sitter-invariant effective field theory remains valid. The secular growth seen in other approaches originates from symmetry-breaking choices and should not be regarded as a physical prediction.

The abundance of light perturbations during inflation~\cite{Baumann:2014nda} renders the current study both theoretically and phenomenologically relevant, and hence, there are compelling reasons to dedicate time and effort to reexamine how to compute statistics of interacting fields in the massless limit. For instance, the absence of secular growth has implications for the stability of de Sitter space~\cite{Ford:1984hs,Antoniadis:1985pj,Polyakov:2012uc,Anderson:2013zia} and eternal inflation~\cite{Weinberg:1987dv,Creminelli:2008es,Senatore:2009cf}, and the nonperturbative statistics of curvature fluctuations and structure formation~\cite{Ezquiaga:2019ftu,Celoria:2021vjw,Hooshangi:2023kss,Cai:2022erk,Cai:2021zsp,Hooshangi:2021ubn,Pi:2022ysn,Vennin:2020kng,Achucarro:2021pdh, Gow:2022jfb,Cruces:2025typ}.

\begin{acknowledgments}

\vspace{3pt}
\noindent We are grateful to Ana Achúcarro, Ignatios Antoniadis, Sebastián Céspedes, Xingang Chen, Thomas Colas, Francisco Colipí, Diego Cruces, Jinn-Ouk Gong, Harry Goodhew, Javier Huenupi, Ellie Hughes, Gabriel Marín, Sonia Paban, Enrico Pajer, Subodh Patil, Sébastien Renaux-Petel, Koenraad Schalm, Bruno Scheihing, Xi Tong, Yuko Urakawa, Dong-Gang Wang, Richard Woodard and Vicharit Yingcharoenrat for useful discussions and comments on various aspects surrounding this work. We acknowledge support from the Fondecyt Regular projects 1210876 and 1251511 (ANID).

\end{acknowledgments}

\appendix

\section{Prevalence of imaginary propagators} 
\label{app:split-prop}

For completeness, in this appendix we provide a proof of the statements I and II offered regarding the appearance of imaginary propagators analyzed in Section~\ref{sec:split-propagators}. Any propagator, bulk-to-bulk or bull-to-boundary, can be split between real and imaginary parts. Let us denote the real and imaginary parts of the function $G (k , \tau_a , \tau_b)$ defined in (\ref{G-basic-prop}), as $G_R (k , \tau_a , \tau_b)$ and $G_I (k , \tau_a , \tau_b)$ respectively:
\be
G (k , \tau_a , \tau_b) = G_R (k , \tau_a , \tau_b) + i G_I (k , \tau_a , \tau_b) .
\ee
These functions take the form
\bea
G_R (k , \tau_a , \tau_b)  &=&  \frac{H^2}{2 k^3} \bigg( k ( \tau_a - \tau_b) \sin \Big[ k (\tau_a - \tau_b) \Big]  \nn \\
&& + (1 + k^2 \tau_a \tau_b) \cos \Big[ k (\tau_a - \tau_b) \Big]  \bigg), \qquad \\
G_I (k , \tau_a , \tau_b)  &=&   \frac{H^2}{2 k^3} \bigg( k ( \tau_a - \tau_b) \cos \Big[ k (\tau_a - \tau_b) \Big] \nn \\
&&- (1 + k^2 \tau_a \tau_b) \sin \Big[ k (\tau_a - \tau_b) \Big]  \bigg). \qquad
\eea
Notice that $G_R (k , \tau_a , \tau_b)$ is an even function under the interchange of $\tau_a$ and $\tau_b$ whereas $G_I (k , \tau_a , \tau_b)$ is found to be odd. One can now split the various propagators of the theory into real and imaginary contributions. These take the form:
\bea
G_{++} (k, \tau_a , \tau_b)  &=& G_R (k , \tau_a , \tau_b) \nn \\
&& + i G_I (k , \tau_a , \tau_b) I ( \tau_a , \tau_b)  , \\
G_{--} (k, \tau_a , \tau_b)  &=& G_R (k , \tau_a , \tau_b) \nn \\
&& - i G_I (k , \tau_a , \tau_b) I ( \tau_a , \tau_b)  , \\
G_{+-} ( k, \tau_a , \tau_b)  &=& G_R (k , \tau_a , \tau_b) -  i G_I (k , \tau_a , \tau_b)  , \\
G_{-+} (k, \tau_a , \tau_b)  &=& G_R (k , \tau_a , \tau_b) + i G_I (k , \tau_a , \tau_b)  , \quad
\eea
where we have defined
\be
I ( \tau_a , \tau_b)  \equiv  \theta (\tau_a - \tau_b) - \theta (\tau_b - \tau_a)  . \label{def-I}
\ee
Notice that $I ( \tau_a , \tau_b)$ is an odd function under the interchange of $\tau_a$ and $\tau_b$. With these definitions in mind, let us decompose the diagrammatic representation of propagators as
\def\propbb{\tikz[baseline=-0.6ex,scale=1.8, every node/.style={scale=1.4}]{
\coordinate (tau1) at (-2.5ex,0ex);
\coordinate (tau2) at (2.5ex,0ex);
\draw[thick] (tau1) -- (tau2);
\filldraw[color=black, fill=black, thick] (tau1) circle (0.5ex);
\node[anchor=south] at ($(tau1)+(0,0.5ex)$) {\scriptsize{$\tau_a$}};
\filldraw[color=black, fill=black, thick] (tau2) circle (0.5ex);
\node[anchor=south] at ($(tau2)+(0,0.5ex)$) {\scriptsize{$\tau_b$}};
}
}
\def\propww{\tikz[baseline=-0.6ex,scale=1.8, every node/.style={scale=1.4}]{
\coordinate (tau1) at (-2.5ex,0ex);
\coordinate (tau2) at (2.5ex,0ex);
\draw[thick] (tau1) -- (tau2);
\filldraw[color=black, fill=white, thick] (tau1) circle (0.5ex);
\node[anchor=south] at ($(tau1)+(0,0.5ex)$) {\scriptsize{$\tau_a$}};
\filldraw[color=black, fill=white, thick] (tau2) circle (0.5ex);
\node[anchor=south] at ($(tau2)+(0,0.5ex)$) {\scriptsize{$\tau_b$}};
}
}
\def\propbw{\tikz[baseline=-0.6ex,scale=1.8, every node/.style={scale=1.4}]{
\coordinate (tau1) at (-2.5ex,0ex);
\coordinate (tau2) at (2.5ex,0ex);
\draw[thick] (tau1) -- (tau2);
\filldraw[color=black, fill=black, thick] (tau1) circle (0.5ex);
\node[anchor=south] at ($(tau1)+(0,0.5ex)$) {\scriptsize{$\tau_a$}};
\filldraw[color=black, fill=white, thick] (tau2) circle (0.5ex);
\node[anchor=south] at ($(tau2)+(0,0.5ex)$) {\scriptsize{$\tau_b$}};
}
}
\def\propwb{\tikz[baseline=-0.6ex,scale=1.8, every node/.style={scale=1.4}]{
\coordinate (tau1) at (-2.5ex,0ex);
\coordinate (tau2) at (2.5ex,0ex);
\draw[thick] (tau1) -- (tau2);
\filldraw[color=black, fill=white, thick] (tau1) circle (0.5ex);
\node[anchor=south] at ($(tau1)+(0,0.5ex)$) {\scriptsize{$\tau_a$}};
\filldraw[color=black, fill=black, thick] (tau2) circle (0.5ex);
\node[anchor=south] at ($(tau2)+(0,0.5ex)$) {\scriptsize{$\tau_b$}};
}
}
\def\imbb{\tikz[baseline=-0.6ex,scale=1.8, every node/.style={scale=1.4}]{
\coordinate (tau1) at (-2.5ex,0ex);
\coordinate (tau2) at (2.5ex,0ex);
\draw[thick, dashed] (tau1) -- (tau2);
\filldraw[color=black, fill=black, thick] (tau1) circle (0.5ex);
\node[anchor=south] at ($(tau1)+(0,0.5ex)$) {\scriptsize{$\tau_a$}};
\filldraw[color=black, fill=black, thick] (tau2) circle (0.5ex);
\node[anchor=south] at ($(tau2)+(0,0.5ex)$) {\scriptsize{$\tau_b$}};
}
}
\def\rebb{\tikz[baseline=-0.6ex,scale=1.8, every node/.style={scale=1.4}]{
\coordinate (tau1) at (-2.5ex,0ex);
\coordinate (tau2) at (2.5ex,0ex);
\draw[thick, double] (tau1) -- (tau2);
\filldraw[color=black, fill=black, thick] (tau1) circle (0.5ex);
\node[anchor=south] at ($(tau1)+(0,0.5ex)$) {\scriptsize{$\tau_a$}};
\filldraw[color=black, fill=black, thick] (tau2) circle (0.5ex);
\node[anchor=south] at ($(tau2)+(0,0.5ex)$) {\scriptsize{$\tau_b$}};
}
}
\def\imww{\tikz[baseline=-0.6ex,scale=1.8, every node/.style={scale=1.4}]{
\coordinate (tau1) at (-2.5ex,0ex);
\coordinate (tau2) at (2.5ex,0ex);
\draw[thick, dashed] (tau1) -- (tau2);
\filldraw[color=black, fill=white, thick] (tau1) circle (0.5ex);
\node[anchor=south] at ($(tau1)+(0,0.5ex)$) {\scriptsize{$\tau_a$}};
\filldraw[color=black, fill=white, thick] (tau2) circle (0.5ex);
\node[anchor=south] at ($(tau2)+(0,0.5ex)$) {\scriptsize{$\tau_b$}};
}
}
\def\reww{\tikz[baseline=-0.6ex,scale=1.8, every node/.style={scale=1.4}]{
\coordinate (tau1) at (-2.5ex,0ex);
\coordinate (tau2) at (2.5ex,0ex);
\draw[thick, double] (tau1) -- (tau2);
\filldraw[color=black, fill=white, thick] (tau1) circle (0.5ex);
\node[anchor=south] at ($(tau1)+(0,0.5ex)$) {\scriptsize{$\tau_a$}};
\filldraw[color=black, fill=white, thick] (tau2) circle (0.5ex);
\node[anchor=south] at ($(tau2)+(0,0.5ex)$) {\scriptsize{$\tau_b$}};
}
}
\def\imbw{\tikz[baseline=-0.6ex,scale=1.8, every node/.style={scale=1.4}]{
\coordinate (tau1) at (-2.5ex,0ex);
\coordinate (tau2) at (2.5ex,0ex);
\draw[thick, dashed] (tau1) -- (tau2);
\filldraw[color=black, fill=black, thick] (tau1) circle (0.5ex);
\node[anchor=south] at ($(tau1)+(0,0.5ex)$) {\scriptsize{$\tau_a$}};
\filldraw[color=black, fill=white, thick] (tau2) circle (0.5ex);
\node[anchor=south] at ($(tau2)+(0,0.5ex)$) {\scriptsize{$\tau_b$}};
}
}
\def\rebw{\tikz[baseline=-0.6ex,scale=1.8, every node/.style={scale=1.4}]{
\coordinate (tau1) at (-2.5ex,0ex);
\coordinate (tau2) at (2.5ex,0ex);
\draw[thick, double] (tau1) -- (tau2);
\filldraw[color=black, fill=black, thick] (tau1) circle (0.5ex);
\node[anchor=south] at ($(tau1)+(0,0.5ex)$) {\scriptsize{$\tau_a$}};
\filldraw[color=black, fill=white, thick] (tau2) circle (0.5ex);
\node[anchor=south] at ($(tau2)+(0,0.5ex)$) {\scriptsize{$\tau_b$}};
}
}
\def\imwb{\tikz[baseline=-0.6ex,scale=1.8, every node/.style={scale=1.4}]{
\coordinate (tau1) at (-2.5ex,0ex);
\coordinate (tau2) at (2.5ex,0ex);
\draw[thick, dashed] (tau1) -- (tau2);
\filldraw[color=black, fill=white, thick] (tau1) circle (0.5ex);
\node[anchor=south] at ($(tau1)+(0,0.5ex)$) {\scriptsize{$\tau_a$}};
\filldraw[color=black, fill=black, thick] (tau2) circle (0.5ex);
\node[anchor=south] at ($(tau2)+(0,0.5ex)$) {\scriptsize{$\tau_b$}};
}
}
\def\rewb{\tikz[baseline=-0.6ex,scale=1.8, every node/.style={scale=1.4}]{
\coordinate (tau1) at (-2.5ex,0ex);
\coordinate (tau2) at (2.5ex,0ex);
\draw[thick, double] (tau1) -- (tau2);
\filldraw[color=black, fill=white, thick] (tau1) circle (0.5ex);
\node[anchor=south] at ($(tau1)+(0,0.5ex)$) {\scriptsize{$\tau_a$}};
\filldraw[color=black, fill=black, thick] (tau2) circle (0.5ex);
\node[anchor=south] at ($(tau2)+(0,0.5ex)$) {\scriptsize{$\tau_b$}};
}
}
\bea
\propbb &=& \rebb + \imbb , \qquad \quad \\
\propww &=& \reww + \imww ,  \qquad \quad \\
\propbw &=& \rebw + \imbw , \qquad \quad\\
\propwb &=& \rewb + \imwb , \qquad \quad
\eea
where double lines stand for the real part and dashed lines denote their imaginary part. 

The previous splitting is also to be performed for bulk-to-boundary propagators. In this case we have:
\bea
G_{+} ( k, \tau_a , \tau)  &=& G_R (k , \tau_a , \tau) -  i G_I (k , \tau_a , \tau)  , \\
G_{-} (k, \tau_a , \tau)  &=& G_R (k , \tau_a , \tau) + i G_I (k , \tau_a , \tau)  , \quad
\eea

\def\propbf{\tikz[baseline=-0.6ex,scale=1.8, every node/.style={scale=1.4}]{
\coordinate (tau) at (-2.5ex,0ex);
\coordinate (phi) at (2.5ex,0ex);
\draw[thick] (tau) -- (phi);
\filldraw[color=black, fill=black, thick] (tau) circle (0.5ex);
\node[anchor=south] at ($(tau)+(0,0.5ex)$) {\scriptsize{$\tau_a$}};
\pgfmathsetmacro{\arista}{0.06}
\filldraw[color=black, fill=white, thick] ($(phi)-(\arista,\arista)$) rectangle ($(phi)+(\arista,\arista)$);
\node[anchor=south] at ($(phi)+(0,0.5ex)$){\scriptsize{$\tau$}};
}
}
\def\propwf{\tikz[baseline=-0.6ex,scale=1.8, every node/.style={scale=1.4}]{
\coordinate (tau) at (-2.5ex,0ex);
\coordinate (phi) at (2.5ex,0ex);
\draw[thick] (tau) -- (phi);
\filldraw[color=black, fill=white, thick] (tau) circle (0.5ex);
\node[anchor=south] at ($(tau)+(0,0.5ex)$) {\scriptsize{$\tau_a$}};
\pgfmathsetmacro{\arista}{0.06}
\filldraw[color=black, fill=white, thick] ($(phi)-(\arista,\arista)$) rectangle ($(phi)+(\arista,\arista)$);
\node[anchor=south] at ($(phi)+(0,0.5ex)$){\scriptsize{$\tau$}};
}
}
\def\propbfim{\tikz[baseline=-0.6ex,scale=1.8, every node/.style={scale=1.4}]{
\coordinate (tau) at (-2.5ex,0ex);
\coordinate (phi) at (2.5ex,0ex);
\draw[thick, dashed] (tau) -- (phi);
\filldraw[color=black, fill=black, thick] (tau) circle (0.5ex);
\node[anchor=south] at ($(tau)+(0,0.5ex)$) {\scriptsize{$\tau_a$}};
\pgfmathsetmacro{\arista}{0.06}
\filldraw[color=black, fill=white, thick] ($(phi)-(\arista,\arista)$) rectangle ($(phi)+(\arista,\arista)$);
\node[anchor=south] at ($(phi)+(0,0.5ex)$){\scriptsize{$\tau$}};
}
}
\def\propwfim{\tikz[baseline=-0.6ex,scale=1.8, every node/.style={scale=1.4}]{
\coordinate (tau) at (-2.5ex,0ex);
\coordinate (phi) at (2.5ex,0ex);
\draw[thick, dashed] (tau) -- (phi);
\filldraw[color=black, fill=white, thick] (tau) circle (0.5ex);
\node[anchor=south] at ($(tau)+(0,0.5ex)$) {\scriptsize{$\tau_a$}};
\pgfmathsetmacro{\arista}{0.06}
\filldraw[color=black, fill=white, thick] ($(phi)-(\arista,\arista)$) rectangle ($(phi)+(\arista,\arista)$);
\node[anchor=south] at ($(phi)+(0,0.5ex)$){\scriptsize{$\tau$}};
}
}
\def\propbfre{\tikz[baseline=-0.6ex,scale=1.8, every node/.style={scale=1.4}]{
\coordinate (tau) at (-2.5ex,0ex);
\coordinate (phi) at (2.5ex,0ex);
\draw[thick, double] (tau) -- (phi);
\filldraw[color=black, fill=black, thick] (tau) circle (0.5ex);
\node[anchor=south] at ($(tau)+(0,0.5ex)$) {\scriptsize{$\tau_a$}};
\pgfmathsetmacro{\arista}{0.06}
\filldraw[color=black, fill=white, thick] ($(phi)-(\arista,\arista)$) rectangle ($(phi)+(\arista,\arista)$);
\node[anchor=south] at ($(phi)+(0,0.5ex)$){\scriptsize{$\tau$}};
}
}
\def\propwfre{\tikz[baseline=-0.6ex,scale=1.8, every node/.style={scale=1.4}]{
\coordinate (tau) at (-2.5ex,0ex);
\coordinate (phi) at (2.5ex,0ex);
\draw[thick, double] (tau) -- (phi);
\filldraw[color=black, fill=white, thick] (tau) circle (0.5ex);
\node[anchor=south] at ($(tau)+(0,0.5ex)$) {\scriptsize{$\tau_a$}};
\pgfmathsetmacro{\arista}{0.06}
\filldraw[color=black, fill=white, thick] ($(phi)-(\arista,\arista)$) rectangle ($(phi)+(\arista,\arista)$);
\node[anchor=south] at ($(phi)+(0,0.5ex)$){\scriptsize{$\tau$}};
}
}
The diagrammatic representation of this splitting for bulk-to-boundary propagators is
\bea
\propbf &=& \propbfre + \propbfim , \qquad \quad \\
\propwf &=& \propbfre + \propwfim ,  \qquad \quad 
\eea
Notice that real propagators (represented by double lines) are all identical, independently of the type of vertex to which they are attached to, whereas the imaginary propagators (dashed lines) change according to the type of vertices to which they are attached.

Let us now consider an arbitrary diagram with every propagator, internal or external, decomposed into real and imaginary parts. A diagram with only real propagators (double lines) will necessarily vanish. Indeed, real propagators are vertex-independent, and so the only difference between different diagrams comes from vertices; but given that for every black vertex there is a white one which contributes with the opposite sign, the sum of all diagrams will thus vanish. 

The previous reasoning can be made more precise without the need of restricting our attention only to diagrams where every propagator is real. Consider an arbitrary diagram with a given arbitrary vertex labelled by $\tau$:
\def\arbdiagram{\tikz[baseline=-1.4ex,scale=1.8, every node/.style={scale=1.4}]{
\coordinate (k1) at (-3ex,0ex);
\coordinate (k2) at (-1.5ex,0ex);
\coordinate (kn) at (3ex,0ex);
\coordinate (tau) at (0,-7ex);
\coordinate (dots1) at (1ex,-1.7ex);
\pgfmathsetmacro{\arista}{0.06}
\draw[thick] (-4ex,0ex) -- (4ex,0ex);
\draw[-,thick] (-2.5ex,-3ex) -- (k1);
\draw[-,thick] (-1.0ex,-3ex) -- (k2);
\draw[-,thick] (2.5ex,-3ex) -- (kn);
\draw[-,thick] (-2.5ex,-3ex) -- (tau);
\draw[-,thick] (0ex,-3ex) -- (tau);
\draw[-,thick] (2.5ex,-3ex) -- (tau);
\node[anchor=south] at ($(dots1)$) {\scriptsize{$\cdots$}};
\filldraw[color=black, fill=gray, thick] (0.0ex,-3.0ex) ellipse (3ex and 1.5ex); 
\node[anchor=north] at ($(tau)+(0.0ex,-0.1ex)$) {\scriptsize{$\tau$}};
\filldraw[color=black, fill=white, thick] ($(k1)-(\arista,\arista)$) rectangle ($(k1)+(\arista,\arista)$);
\filldraw[color=black, fill=white, thick] ($(k2)-(\arista,\arista)$) rectangle ($(k2)+(\arista,\arista)$);
\filldraw[color=black, fill=white, thick] ($(kn)-(\arista,\arista)$) rectangle ($(kn)+(\arista,\arista)$);
\fill[white] (tau) circle (0.5ex);
\filldraw[pattern=north east lines, thick] (tau) circle (0.5ex);
}
}
\def\arbdiagramb{\tikz[baseline=-1.4ex,scale=1.8, every node/.style={scale=1.4}]{
\coordinate (k1) at (-3ex,0ex);
\coordinate (k2) at (-1.5ex,0ex);
\coordinate (kn) at (3ex,0ex);
\coordinate (tau) at (0,-7ex);
\coordinate (dots1) at (1ex,-1.7ex);
\pgfmathsetmacro{\arista}{0.06}
\draw[thick] (-4ex,0ex) -- (4ex,0ex);
\draw[-,thick] (-2.5ex,-3ex) -- (k1);
\draw[-,thick] (-1.0ex,-3ex) -- (k2);
\draw[-,thick] (2.5ex,-3ex) -- (kn);
\draw[-,thick] (-2.5ex,-3ex) -- (tau);
\draw[-,thick] (0ex,-3ex) -- (tau);
\draw[-,thick] (2.5ex,-3ex) -- (tau);
\node[anchor=south] at ($(dots1)$) {\scriptsize{$\cdots$}};
\filldraw[color=black, fill=gray, thick] (0.0ex,-3.0ex) ellipse (3ex and 1.5ex); 
\node[anchor=north] at ($(tau)+(0.0ex,-0.1ex)$) {\scriptsize{$\tau$}};
\filldraw[color=black, fill=white, thick] ($(k1)-(\arista,\arista)$) rectangle ($(k1)+(\arista,\arista)$);
\filldraw[color=black, fill=white, thick] ($(k2)-(\arista,\arista)$) rectangle ($(k2)+(\arista,\arista)$);
\filldraw[color=black, fill=white, thick] ($(kn)-(\arista,\arista)$) rectangle ($(kn)+(\arista,\arista)$);
\filldraw[color=black, fill=black, thick] (tau) circle (0.5ex);
}
}
\def\arbdiagramw{\tikz[baseline=-1.4ex,scale=1.8, every node/.style={scale=1.4}]{
\coordinate (k1) at (-3ex,0ex);
\coordinate (k2) at (-1.5ex,0ex);
\coordinate (kn) at (3ex,0ex);
\coordinate (tau) at (0,-7ex);
\coordinate (dots1) at (1ex,-1.7ex);
\pgfmathsetmacro{\arista}{0.06}
\draw[thick] (-4ex,0ex) -- (4ex,0ex);
\draw[-,thick] (-2.5ex,-3ex) -- (k1);
\draw[-,thick] (-1.0ex,-3ex) -- (k2);
\draw[-,thick] (2.5ex,-3ex) -- (kn);
\draw[-,thick] (-2.5ex,-3ex) -- (tau);
\draw[-,thick] (0ex,-3ex) -- (tau);
\draw[-,thick] (2.5ex,-3ex) -- (tau);
\node[anchor=south] at ($(dots1)$) {\scriptsize{$\cdots$}};
\filldraw[color=black, fill=gray, thick] (0.0ex,-3.0ex) ellipse (3ex and 1.5ex); 
\node[anchor=north] at ($(tau)+(0.0ex,-0.1ex)$) {\scriptsize{$\tau$}};
\filldraw[color=black, fill=white, thick] ($(k1)-(\arista,\arista)$) rectangle ($(k1)+(\arista,\arista)$);
\filldraw[color=black, fill=white, thick] ($(k2)-(\arista,\arista)$) rectangle ($(k2)+(\arista,\arista)$);
\filldraw[color=black, fill=white, thick] ($(kn)-(\arista,\arista)$) rectangle ($(kn)+(\arista,\arista)$);
\fill[white] (tau) circle (0.5ex);
\filldraw[color=black, fill=white, thick] (tau) circle (0.5ex);
}
}
\be
\arbdiagram  = \arbdiagramb  +  \arbdiagramw .
\ee
If we now expand only the propagators attached to the $\tau$-vertex into real and imaginary parts, we see that the contributions that contain only the real parts will necessarily cancel each other, as the following diagrammatic relation shows:
\def\arbdiagramb{\tikz[baseline=-1.4ex,scale=1.8, every node/.style={scale=1.4}]{
\coordinate (k1) at (-3ex,0ex);
\coordinate (k2) at (-1.5ex,0ex);
\coordinate (kn) at (3ex,0ex);
\coordinate (tau) at (0,-7ex);
\coordinate (dots1) at (1ex,-1.7ex);
\pgfmathsetmacro{\arista}{0.06}
\draw[thick] (-4ex,0ex) -- (4ex,0ex);
\draw[-,thick] (-2.5ex,-3ex) -- (k1);
\draw[-,thick] (-1.0ex,-3ex) -- (k2);
\draw[-,thick] (2.5ex,-3ex) -- (kn);
\draw[-,thick, double] (-2.5ex,-3ex) -- (tau);
\draw[-,thick, double] (0ex,-3ex) -- (tau);
\draw[-,thick, double] (2.5ex,-3ex) -- (tau);
\node[anchor=south] at ($(dots1)$) {\scriptsize{$\cdots$}};
\filldraw[color=black, fill=gray, thick] (0.0ex,-3.0ex) ellipse (3ex and 1.5ex); 
\node[anchor=north] at ($(tau)+(0.0ex,-0.1ex)$) {\scriptsize{$\tau$}};
\filldraw[color=black, fill=white, thick] ($(k1)-(\arista,\arista)$) rectangle ($(k1)+(\arista,\arista)$);
\filldraw[color=black, fill=white, thick] ($(k2)-(\arista,\arista)$) rectangle ($(k2)+(\arista,\arista)$);
\filldraw[color=black, fill=white, thick] ($(kn)-(\arista,\arista)$) rectangle ($(kn)+(\arista,\arista)$);
\filldraw[color=black, fill=black, thick] (tau) circle (0.5ex);
}
}
\def\arbdiagramw{\tikz[baseline=-1.4ex,scale=1.8, every node/.style={scale=1.4}]{
\coordinate (k1) at (-3ex,0ex);
\coordinate (k2) at (-1.5ex,0ex);
\coordinate (kn) at (3ex,0ex);
\coordinate (tau) at (0,-7ex);
\coordinate (dots1) at (1ex,-1.7ex);
\pgfmathsetmacro{\arista}{0.06}
\draw[thick] (-4ex,0ex) -- (4ex,0ex);
\draw[-,thick] (-2.5ex,-3ex) -- (k1);
\draw[-,thick] (-1.0ex,-3ex) -- (k2);
\draw[-,thick] (2.5ex,-3ex) -- (kn);
\draw[-,thick, double] (-2.5ex,-3ex) -- (tau);
\draw[-,thick, double] (0ex,-3ex) -- (tau);
\draw[-,thick, double] (2.5ex,-3ex) -- (tau);
\node[anchor=south] at ($(dots1)$) {\scriptsize{$\cdots$}};
\filldraw[color=black, fill=gray, thick] (0.0ex,-3.0ex) ellipse (3ex and 1.5ex); 
\node[anchor=north] at ($(tau)+(0.0ex,-0.1ex)$) {\scriptsize{$\tau$}};
\filldraw[color=black, fill=white, thick] ($(k1)-(\arista,\arista)$) rectangle ($(k1)+(\arista,\arista)$);
\filldraw[color=black, fill=white, thick] ($(k2)-(\arista,\arista)$) rectangle ($(k2)+(\arista,\arista)$);
\filldraw[color=black, fill=white, thick] ($(kn)-(\arista,\arista)$) rectangle ($(kn)+(\arista,\arista)$);
\fill[white] (tau) circle (0.5ex);
\filldraw[color=black, fill=white, thick] (tau) circle (0.5ex);
}
}
\be
\arbdiagramb \quad + \quad \arbdiagramw \quad = \quad  0 \,. \nn  
\ee
This is because the only difference between the two diagrams of the previous sum is given by the opposite sign of the two vertices. 
This will not be the case if at least one imaginary propagator remains attached to the vertex $\tau$. This result allows us to conclude that there must be at least one imaginary propagator attached to every vertex for a given diagram not to vanish. 

Taken on face value, this statement implies that for a diagram with $V$ vertices, the minimum number of imaginary (bulk-to-bulk) propagators appears to be $V-1$, which would ensure that every vertex has at least one imaginary propagator attached to it. However, a single diagram constructed with $V$ vertices and $V-1$ imaginary propagators is necessarily imaginary, from where it follows that the sum of subdiagrams constructed in this way must vanish. This finally implies that, for a diagram with $V$ vertices, the minimum number of imaginary propagators must be $V$.

Furthermore, in a diagram with $V$ imaginary propagators (the rest being real), it is possible to show that at least one of them must always correspond to an external leg. If this was not the case, then for a diagram with $V$ vertices and $V$ imaginary propagators, there would necessarily be a closed loop formed entirely by imaginary propagators. But a loop formed by only imaginary propagators must vanish. To see this, consider an arbitrary undotted diagram $D_n$ with a subdiagram consisting of a loop formed only by imaginary internal propagators (denoted by dashed lines), with $n$ vertices attached to it. The rest of the diagram is constructed with full propagators (solid lines). The undotted diagram can then be written as the sum of diagrams with a dotted imaginary loop (that is, with every other vertex undotted, except for those attached to the loop):
\begin{widetext}
\def\loopy{\tikz[baseline=-1.4ex,scale=1.8, every node/.style={scale=1.4}]{
\coordinate (loop) at (0ex,-2ex);
\coordinate (t1) at (-3ex,0ex);
\coordinate (t2) at (-2.121ex,2.121ex);
\coordinate (t3) at (0ex,3ex);
\coordinate (t4) at (2.121ex,2.121ex);
\coordinate (t5) at (3ex,0ex);
\draw[thick, fill=white, dashed](loop) circle (3.0ex);
\fill[white] ($(loop) + (t1)$) circle (0.5ex);
\filldraw[pattern=north east lines, thick] ($(loop) + (t1)$) circle (0.4ex);
\fill[white] ($(loop) + (t2)$) circle (0.5ex);
\filldraw[pattern=north east lines, thick] ($(loop) + (t2)$) circle (0.4ex);
\fill[white] ($(loop) + (t3)$) circle (0.5ex);
\filldraw[pattern=north east lines, thick] ($(loop) + (t3)$) circle (0.4ex);
\fill[white] ($(loop) + (t4)$) circle (0.5ex);
\filldraw[pattern=north east lines, thick] ($(loop) + (t4)$) circle (0.4ex);
\fill[white] ($(loop) + (t5)$) circle (0.5ex);
\filldraw[pattern=north east lines, thick] ($(loop) + (t5)$) circle (0.4ex);
\node[anchor=west] at ($(loop) + (t1) + (-0.2ex,-0.7ex)$) {\scriptsize{$\tau_n$}};
\node[anchor=west] at ($(loop) + (t2) + (-0.7ex,-1.0ex)$) {\scriptsize{$\tau_1$}};
\node[anchor=north] at ($(loop) + (t3) + (0.0ex,0.0ex)$) {\scriptsize{$\tau_2$}};
\node[anchor=east] at ($(loop) + (t4) + (+0.7ex,-1.0ex)$) {\scriptsize{$\tau_3$}};
\node[anchor=east] at ($(loop) + (t5) + (+0.2ex,-0.7ex)$) {\scriptsize{$\tau_4$}};
\filldraw[color=black, fill=black, thick] (-0.7ex,-4.3ex) circle (0.035ex);
\filldraw[color=black, fill=black, thick] (0ex,-4.4ex) circle (0.035ex);
\filldraw[color=black, fill=black, thick] (0.7ex,-4.3ex) circle (0.035ex);
}
}
\def\loopone{\tikz[baseline=-1.4ex,scale=1.8, every node/.style={scale=1.4}]{
\coordinate (loop) at (0ex,-2ex);
\coordinate (t1) at (-3ex,0ex);
\coordinate (t2) at (-2.121ex,2.121ex);
\coordinate (t3) at (0ex,3ex);
\coordinate (t4) at (2.121ex,2.121ex);
\coordinate (t5) at (3ex,0ex);
\draw[thick, fill=white, dashed](loop) circle (3.0ex);
\filldraw[color=black, fill=black, thick] ($(loop) + (t1)$) circle (0.4ex);
\filldraw[color=black, fill=black, thick] ($(loop) + (t2)$) circle (0.4ex);
\filldraw[color=black, fill=black, thick] ($(loop) + (t3)$) circle (0.4ex);
\filldraw[color=black, fill=black, thick] ($(loop) + (t4)$) circle (0.4ex);
\filldraw[color=black, fill=black, thick] ($(loop) + (t5)$) circle (0.4ex);
\node[anchor=west] at ($(loop) + (t1) + (-0.2ex,-0.7ex)$) {\scriptsize{$\tau_n$}};
\node[anchor=west] at ($(loop) + (t2) + (-0.7ex,-1.0ex)$) {\scriptsize{$\tau_1$}};
\node[anchor=north] at ($(loop) + (t3) + (0.0ex,0.0ex)$) {\scriptsize{$\tau_2$}};
\node[anchor=east] at ($(loop) + (t4) + (+0.7ex,-1.0ex)$) {\scriptsize{$\tau_3$}};
\node[anchor=east] at ($(loop) + (t5) + (+0.2ex,-0.7ex)$) {\scriptsize{$\tau_4$}};
\filldraw[color=black, fill=black, thick] (-0.7ex,-4.3ex) circle (0.035ex);
\filldraw[color=black, fill=black, thick] (0ex,-4.4ex) circle (0.035ex);
\filldraw[color=black, fill=black, thick] (0.7ex,-4.3ex) circle (0.035ex);
}
}
\def\looptwo{\tikz[baseline=-1.4ex,scale=1.8, every node/.style={scale=1.4}]{
\coordinate (loop) at (0ex,-2ex);
\coordinate (t1) at (-3ex,0ex);
\coordinate (t2) at (-2.121ex,2.121ex);
\coordinate (t3) at (0ex,3ex);
\coordinate (t4) at (2.121ex,2.121ex);
\coordinate (t5) at (3ex,0ex);
\draw[thick, fill=white, dashed](loop) circle (3.0ex);
\filldraw[color=black, fill=black, thick] ($(loop) + (t1)$) circle (0.4ex);
\filldraw[color=black, fill=white, thick] ($(loop) + (t2)$) circle (0.4ex);
\filldraw[color=black, fill=black, thick] ($(loop) + (t3)$) circle (0.4ex);
\filldraw[color=black, fill=black, thick] ($(loop) + (t4)$) circle (0.4ex);
\filldraw[color=black, fill=black, thick] ($(loop) + (t5)$) circle (0.4ex);
\node[anchor=west] at ($(loop) + (t1) + (-0.2ex,-0.7ex)$) {\scriptsize{$\tau_n$}};
\node[anchor=west] at ($(loop) + (t2) + (-0.7ex,-1.0ex)$) {\scriptsize{$\tau_1$}};
\node[anchor=north] at ($(loop) + (t3) + (0.0ex,0.0ex)$) {\scriptsize{$\tau_2$}};
\node[anchor=east] at ($(loop) + (t4) + (+0.7ex,-1.0ex)$) {\scriptsize{$\tau_3$}};
\node[anchor=east] at ($(loop) + (t5) + (+0.2ex,-0.7ex)$) {\scriptsize{$\tau_4$}};
\filldraw[color=black, fill=black, thick] (-0.7ex,-4.3ex) circle (0.035ex);
\filldraw[color=black, fill=black, thick] (0ex,-4.4ex) circle (0.035ex);
\filldraw[color=black, fill=black, thick] (0.7ex,-4.3ex) circle (0.035ex);
}
}
\def\loopthree{\tikz[baseline=-1.4ex,scale=1.8, every node/.style={scale=1.4}]{
\coordinate (loop) at (0ex,-2ex);
\coordinate (t1) at (-3ex,0ex);
\coordinate (t2) at (-2.121ex,2.121ex);
\coordinate (t3) at (0ex,3ex);
\coordinate (t4) at (2.121ex,2.121ex);
\coordinate (t5) at (3ex,0ex);
\draw[thick, fill=white, dashed](loop) circle (3.0ex);
\filldraw[color=black, fill=black, thick] ($(loop) + (t1)$) circle (0.4ex);
\filldraw[color=black, fill=black, thick] ($(loop) + (t2)$) circle (0.4ex);
\filldraw[color=black, fill=white, thick] ($(loop) + (t3)$) circle (0.4ex);
\filldraw[color=black, fill=black, thick] ($(loop) + (t4)$) circle (0.4ex);
\filldraw[color=black, fill=black, thick] ($(loop) + (t5)$) circle (0.4ex);
\node[anchor=west] at ($(loop) + (t1) + (-0.2ex,-0.7ex)$) {\scriptsize{$\tau_n$}};
\node[anchor=west] at ($(loop) + (t2) + (-0.7ex,-1.0ex)$) {\scriptsize{$\tau_1$}};
\node[anchor=north] at ($(loop) + (t3) + (0.0ex,0.0ex)$) {\scriptsize{$\tau_2$}};
\node[anchor=east] at ($(loop) + (t4) + (+0.7ex,-1.0ex)$) {\scriptsize{$\tau_3$}};
\node[anchor=east] at ($(loop) + (t5) + (+0.2ex,-0.7ex)$) {\scriptsize{$\tau_4$}};
\filldraw[color=black, fill=black, thick] (-0.7ex,-4.3ex) circle (0.035ex);
\filldraw[color=black, fill=black, thick] (0ex,-4.4ex) circle (0.035ex);
\filldraw[color=black, fill=black, thick] (0.7ex,-4.3ex) circle (0.035ex);
}
}
\def\loopfour{\tikz[baseline=-1.4ex,scale=1.8, every node/.style={scale=1.4}]{
\coordinate (loop) at (0ex,-2ex);
\coordinate (t1) at (-3ex,0ex);
\coordinate (t2) at (-2.121ex,2.121ex);
\coordinate (t3) at (0ex,3ex);
\coordinate (t4) at (2.121ex,2.121ex);
\coordinate (t5) at (3ex,0ex);
\draw[thick, fill=white, dashed](loop) circle (3.0ex);
\filldraw[color=black, fill=black, thick] ($(loop) + (t1)$) circle (0.4ex);
\filldraw[color=black, fill=white, thick] ($(loop) + (t2)$) circle (0.4ex);
\filldraw[color=black, fill=white, thick] ($(loop) + (t3)$) circle (0.4ex);
\filldraw[color=black, fill=black, thick] ($(loop) + (t4)$) circle (0.4ex);
\filldraw[color=black, fill=black, thick] ($(loop) + (t5)$) circle (0.4ex);
\node[anchor=west] at ($(loop) + (t1) + (-0.2ex,-0.7ex)$) {\scriptsize{$\tau_n$}};
\node[anchor=west] at ($(loop) + (t2) + (-0.7ex,-1.0ex)$) {\scriptsize{$\tau_1$}};
\node[anchor=north] at ($(loop) + (t3) + (0.0ex,0.0ex)$) {\scriptsize{$\tau_2$}};
\node[anchor=east] at ($(loop) + (t4) + (+0.7ex,-1.0ex)$) {\scriptsize{$\tau_3$}};
\node[anchor=east] at ($(loop) + (t5) + (+0.2ex,-0.7ex)$) {\scriptsize{$\tau_4$}};
\filldraw[color=black, fill=black, thick] (-0.7ex,-4.3ex) circle (0.035ex);
\filldraw[color=black, fill=black, thick] (0ex,-4.4ex) circle (0.035ex);
\filldraw[color=black, fill=black, thick] (0.7ex,-4.3ex) circle (0.035ex);
}
}
\def\loopfive{\tikz[baseline=-1.4ex,scale=1.8, every node/.style={scale=1.4}]{
\coordinate (loop) at (0ex,-2ex);
\coordinate (t1) at (-3ex,0ex);
\coordinate (t2) at (-2.121ex,2.121ex);
\coordinate (t3) at (0ex,3ex);
\coordinate (t4) at (2.121ex,2.121ex);
\coordinate (t5) at (3ex,0ex);
\draw[thick, fill=white, dashed](loop) circle (3.0ex);
\filldraw[color=black, fill=black, thick] ($(loop) + (t1)$) circle (0.4ex);
\filldraw[color=black, fill=black, thick] ($(loop) + (t2)$) circle (0.4ex);
\filldraw[color=black, fill=white, thick] ($(loop) + (t3)$) circle (0.4ex);
\filldraw[color=black, fill=white, thick] ($(loop) + (t4)$) circle (0.4ex);
\filldraw[color=black, fill=black, thick] ($(loop) + (t5)$) circle (0.4ex);
\node[anchor=west] at ($(loop) + (t1) + (-0.2ex,-0.7ex)$) {\scriptsize{$\tau_n$}};
\node[anchor=west] at ($(loop) + (t2) + (-0.7ex,-1.0ex)$) {\scriptsize{$\tau_1$}};
\node[anchor=north] at ($(loop) + (t3) + (0.0ex,0.0ex)$) {\scriptsize{$\tau_2$}};
\node[anchor=east] at ($(loop) + (t4) + (+0.7ex,-1.0ex)$) {\scriptsize{$\tau_3$}};
\node[anchor=east] at ($(loop) + (t5) + (+0.2ex,-0.7ex)$) {\scriptsize{$\tau_4$}};
\filldraw[color=black, fill=black, thick] (-0.7ex,-4.3ex) circle (0.035ex);
\filldraw[color=black, fill=black, thick] (0ex,-4.4ex) circle (0.035ex);
\filldraw[color=black, fill=black, thick] (0.7ex,-4.3ex) circle (0.035ex);
}
}
\bea \
\loopy \,\, &=& \,\, 2 {\rm Re} \Bigg[ \quad \loopone \quad \Bigg] 
 + 2 {\rm Re} \Bigg[ \quad \looptwo \quad + \quad \loopthree \quad + \quad \cdots \quad \Bigg] \nn \\
&& + 2 {\rm Re} \Bigg[ \quad \loopfour \quad + \quad \loopfive \quad + \quad \cdots \quad \Bigg]  + \cdots  . \label{imagine-a-loop}
\eea
\end{widetext}
In translating the sum of these diagrams into Feynman rules, one will find that the imaginary loop will contribute to the entire sum as an overall factor equal to the sum of products of the function $I(\tau_i , \tau_j)$ defined in equation~(\ref{def-I}). 

For instance, in the particular case whereby the number $n$ of vertices attached to the loop is even, the first line in the previous diagrammatic relation would be proportional to the product $I(\tau_1, \tau_2) I(\tau_2, \tau_3) \cdots I(\tau_n, \tau_1)$. Similarly, the  second line, which contains diagrams with two consecutive white vertices and every other vertex black, would be proportional to the product $I(\tau_2, \tau_3) I(\tau_4, \tau_5) \cdots I(\tau_{n-1}, \tau_n)$ plus terms obtained by means of cyclic permutations.

More generally, the number of functions $I(\tau_i , \tau_j)$ participating in these products will depend on the number of consecutive vertices of the same color. It is then straightforward to find that if $n$ is even, then the diagram under examination must be proportional to the following combination:
\bea
D_n &\propto& I (\tau_1 , \tau_2) + I (\tau_2 , \tau_3) + \cdots + I (\tau_n , \tau_1) \nn \\
&& + I (\tau_1 , \tau_2) I (\tau_2 , \tau_3) I (\tau_3 , \tau_4) + {\rm perms .} \nn \\
&& + I (\tau_1 , \tau_2) \times \cdots \times I (\tau_5 , \tau_6 ) + {\rm perms .} \nn \\
&&
+ \cdots  \nn \\
&& + I (\tau_1 , \tau_2) \times \cdots \times I (\tau_n , \tau_1 ).
\eea
On the other hand, if $n$ is odd, then one obtains
\bea
D_n &\propto& 1  + I (\tau_1 , \tau_2) I (\tau_2 , \tau_3)  + {\rm perms .} \nn \\
&& + I (\tau_1 , \tau_2) \times \cdots \times I (\tau_4 , \tau_5 ) + {\rm perms .} \nn \\
&&
+ \cdots  \nn \\
&& + I (\tau_1 , \tau_2) \times \cdots \times I (\tau_{n-1} , \tau_n ) + {\rm perms .}
\eea
One can readily verify that in both cases the total sum adds up to zero, rendering the diagram above null. To see this, it is enough to evaluate the sum for the particular ordered sequence $\tau_1 > \tau_2 > \cdots > \tau_n$, which is easily verified to vanish. Then, one can interchange the order of any pair of times (for example $\tau_1$ and $\tau_2$) and check that a change of sign in any one term is always compensated by a change of sing of another term (which originally had the opposite sign), leaving the conclusion unchanged.

Notice that the previous argument is independent of the class of propagators connecting the loop with the rest of the diagram. To see this, first notice that every loop contributing to (\ref{imagine-a-loop}) is purely real by construction (there is an equal number of imaginary propagators and vertices). This implies that the real part of the rest of the diagram (to which loops are attached) must have a non-vanishing real part. Now, consider as a baseline the situation where the rest of the diagram is constructed only with real propagators (which implies an even number of extra vertices). In this case, the previous argument follows trivially, as every subdiagram appearing in (\ref{imagine-a-loop}) would be part of the same analytical expression. Next, one can consider the effect of replacing real propagators by imaginary propagators. If we replace them by an odd number of imaginary propagators the result is a purely imaginary expression that doesn't contribute to the correlation function. If we do so by an even number of imaginary propagators one easily sees that each loop in (\ref{imagine-a-loop}) which recall, is real, ends up multiplied by the same expression (the same argument applies in the case where the rest of the diagram has an odd number of vertices).


%

\end{document}